\documentclass[12pt]{article}

\pdfoutput=1

\usepackage{amsmath,amssymb,amsfonts}
\usepackage{color}
\definecolor{darkblue}{rgb}{0.1,0.1,.7}
\usepackage[colorlinks, linkcolor=darkblue, citecolor=darkblue, urlcolor=darkblue, linktocpage]{hyperref} 
\usepackage[square, comma, compress,numbers]{natbib}
\usepackage[]{amsmath}
\usepackage[]{graphicx}
\usepackage[]{latexsym}
\usepackage{geometry}
\usepackage{amscd}
\usepackage[all,cmtip]{xy}
\usepackage{mathrsfs}
\usepackage{dsfont}
\usepackage{dsdshorthand}
\usepackage[margin=10pt,font=small,labelfont=bf]{caption}
\usepackage{float}

\numberwithin{equation}{section}

\newcommand{\reef}[1]{(\ref{#1})}
\renewcommand{\be}{\begin{equation}}
\renewcommand{\ee}{\end{equation}}
\newcommand{\bea}{\begin{eqnarray}}
\newcommand{\eea}{\end{eqnarray}}

\newcommand{\eps}{\epsilon}
\def\beq{\begin{equation}} 
\def\eeq{\end{equation}} 

\def\del {\partial} 
\def\nn{\nonumber} 
\def\bZ {\mathbb{Z}} 
\def\bbR {\mathbb{R}}

\def\bZ {\mathbb{Z}}

\newcommand\bh{\mathbf{h}}
\newcommand\bd{\mathbf{d}}
\newcommand\bc{\mathbf{c}}
\newcommand\bb{\mathbf{b}}

\newcommand{\CT}{c}

\begin{document}

\vspace*{-.6in} \thispagestyle{empty}
\begin{flushright}
CERN--PH-TH/2014-038\\
NSF-KITP-14-022 
\end{flushright}
\vspace{.2in} {\large
\begin{center}
{\bf Solving the 3d Ising Model with the Conformal Bootstrap\\
II. $\CT$-Minimization and Precise Critical Exponents}
\end{center}
}
\vspace{.2in}
\begin{center}
{\bf Sheer El-Showk$^{a}$, 
Miguel F. Paulos$^{b}$, 
David Poland$^{c}$,}\\\vspace{.2in} 
{\bf Slava Rychkov$^{a,d}$, 
David Simmons-Duffin$^{e}$, 
Alessandro Vichi$^{f}$}
\\
\vspace{.2in} 
$^a$ {\it Theory Division, CERN, Geneva, Switzerland}
\\\vspace{.2in} 
$^b$ {\it Department of Physics, Brown University, Box 1843, Providence, RI 02912-1843, USA}
\\\vspace{.2in} 
$^c$ {\it Department of Physics, Yale University, New Haven, CT 06520, USA}
\\\vspace{.2in} 
$^d$ {\it  Facult\'{e} de Physique, Universit\'{e} Pierre et Marie Curie\\
\& Laboratoire de Physique Th\'{e}orique, \'{E}cole Normale Sup\'{e}rieure, Paris, France}
\\\vspace{.2in} 
$^e$ {\it School of Natural Sciences, Institute for Advanced Study, \\Princeton, New Jersey 08540, USA}
\\\vspace{.2in} 
$^f$ {\it Theoretical Physics Group, Ernest Orlando Lawrence Berkeley National Laboratory\\
 \& Center for Theoretical Physics, University of California, Berkeley, CA 94720, USA}
\end{center}

\vspace{.2in}

\begin{abstract}
We use the conformal bootstrap to perform a precision study of the operator spectrum of the critical 3d Ising model. We conjecture that the 3d Ising spectrum minimizes the central charge $c$ in the space of unitary solutions to crossing symmetry. Because extremal solutions to crossing symmetry are uniquely determined, we are able to precisely reconstruct the first several $\bZ_2$-even operator dimensions and their OPE coefficients. We observe that a sharp transition in the operator spectrum occurs at the 3d Ising dimension $\Delta_\s = 0.518154(15)$, and find strong numerical evidence that operators decouple from the spectrum as one approaches the 3d Ising point. We compare this behavior to the analogous situation in 2d, where the disappearance of operators can be understood in terms of degenerate Virasoro representations.
\end{abstract}
\vspace{.2in}
\vspace{.3in}
\begin{center}
March 2014\\[5pt]
\emph{To appear in a special issue of J.Stat.Phys.~in memory of Kenneth Wilson} 
\end{center}

\newpage
\setcounter{page}{1}

\tableofcontents



\section{Introduction}

In \cite{ElShowk:2012ht}, we initiated the conformal bootstrap approach to studying the ``3d Ising CFT''---the CFT describing the 3d Ising model at criticality. This CFT seems to occupy a very special place in the space of unitary $\bZ_2$-symmetric 3d CFTs. In the plane parametrized by the leading $\bZ_2$-odd and $\bZ_2$-even scalar dimensions $\De_\sigma$ and $\De_\eps$, it seems to live at a ``kink" on the boundary of the region allowed by the constraints of crossing symmetry and unitarity. Turning this around, the position of this kink can be used to determine $\De_\sigma$ and $\De_\eps$, giving results in agreement with the renormalization group and lattice methods. Moreover, compared to these other techniques our method has two important advantages. Firstly, the needed computational time scales favorably with the final accuracy (the kink localization), promising high-precision results for the critical exponents. Secondly, due to the fact that the many parameters characterizing the CFT talk to each other in the bootstrap equation, the results show interesting structure and interrelations. The very existence of the kink tying $\De_\sigma$ to $\De_\eps$ is one such interrelation, and there are many more \cite{ElShowk:2012ht}. For example, previously we found that the subleading $\bZ_2$-even scalar and spin 2 operator dimensions show interesting variation near the kink, while the stress tensor central charge $\CT$ seems to take a minimal value there. 

Exploiting these advantages of the bootstrap approach will be a major theme of this second work in the series. Our primary goal will be to use the conformal bootstrap to make a high precision determination of all low-lying operator dimensions and OPE coefficients in the 3d Ising CFT. In doing so, we would also like to gain some insights into why the 3d Ising solution is special. We will make extensive use of the fact that solutions living on the boundary of the region allowed by crossing symmetry and unitarity can be uniquely reconstructed~\cite{ElShowk:2012hu}. In order to reach the boundary, we will work with the conjecture that the central charge $c$ is minimized for the 3d Ising CFT. This conjecture combined with uniqueness of the boundary solution allows us to numerically reconstruct the solution to crossing symmetry for different values of $\Delta_{\s}$, using a modification of Dantzig's simplex algorithm.

We find that the resulting $\bZ_2$-even spectrum shows a dramatic transition in the vicinity of $\Delta_\s = 0.518154(15)$, giving a high precision determination of the leading critical exponent $\eta$. Focusing on the transition region, we are able to extract precise values of the first several $\bZ_2$-even operator dimensions and of their OPE coefficients, see Table \ref{tab:dims}.
We also give reasonable estimates for the locations of all low dimension ($\Delta \lesssim 13$) scalar and spin 2 operators in the $\bZ_2$-even spectrum.
\begin{table}[htdp]
\begin{center}
\begin{tabular}{|l|l|l|l|}
\hline
spin \& $\mathbb{Z}_2$ & name & $\Delta$ & OPE coefficient \\
\hline
\hline
$\ell=0$, $\mathbb{Z}_2=-$  & $\s$ &0.518154(15) & \\
\hline
$\ell=0$, $\mathbb{Z}_2=+$  & $\eps$ &1.41267(13) & $f_{\s\s\e}^2 = 1.10636(9)$ \\
 & $\eps'$ & 3.8303(18)& $f_{\s\s\e'}^2 = 0.002810(6)$ \\
   \hline
$\ell=2$, $\mathbb{Z}_2=+$  & $T$ & 3 & $\CT/\CT_{\text{free}} = 0.946534(11)$ \\
 & $T'$ & 5.500(15) & $f_{\s\s T'}^2=2.97(2) \times 10^{-4}$ \\
\hline
\end{tabular}
\end{center}
\caption{Precision information about the low-lying 3d Ising CFT spectrum and OPE coefficients extracted in this work. 
See sections \ref{sec:spin0} and \ref{sec:spin2} for preliminary information about higher-dimension $\ell=0$ and $\ell=2$ operators. See also section \ref{sec:comp} for a comparison to results by other techniques.}
\label{tab:dims}
\end{table}%

The transition also shows the highly intriguing feature that certain operators {\it disappear} from the spectrum as one approaches the 3d Ising point. This decoupling of states gives an important characterization of the 3d Ising CFT. This is similar to what occurs in the 2d Ising model, where the decoupling of operators can be rigorously understood in terms of degenerate representations of the Virasoro symmetry. To better understand this connection, we give a detailed comparison to the application of our $c$-minimization algorithm in 2d, where the exact spectrum of the 2d Ising CFT and its interpolation through the minimal models is known. We conclude with a discussion of important directions for future research.

\section{A Conjecture for the 3d Ising Spectrum}

\begin{figure}[htbp]
\begin{center}
\raisebox{0.5cm}{\includegraphics[scale=0.6]{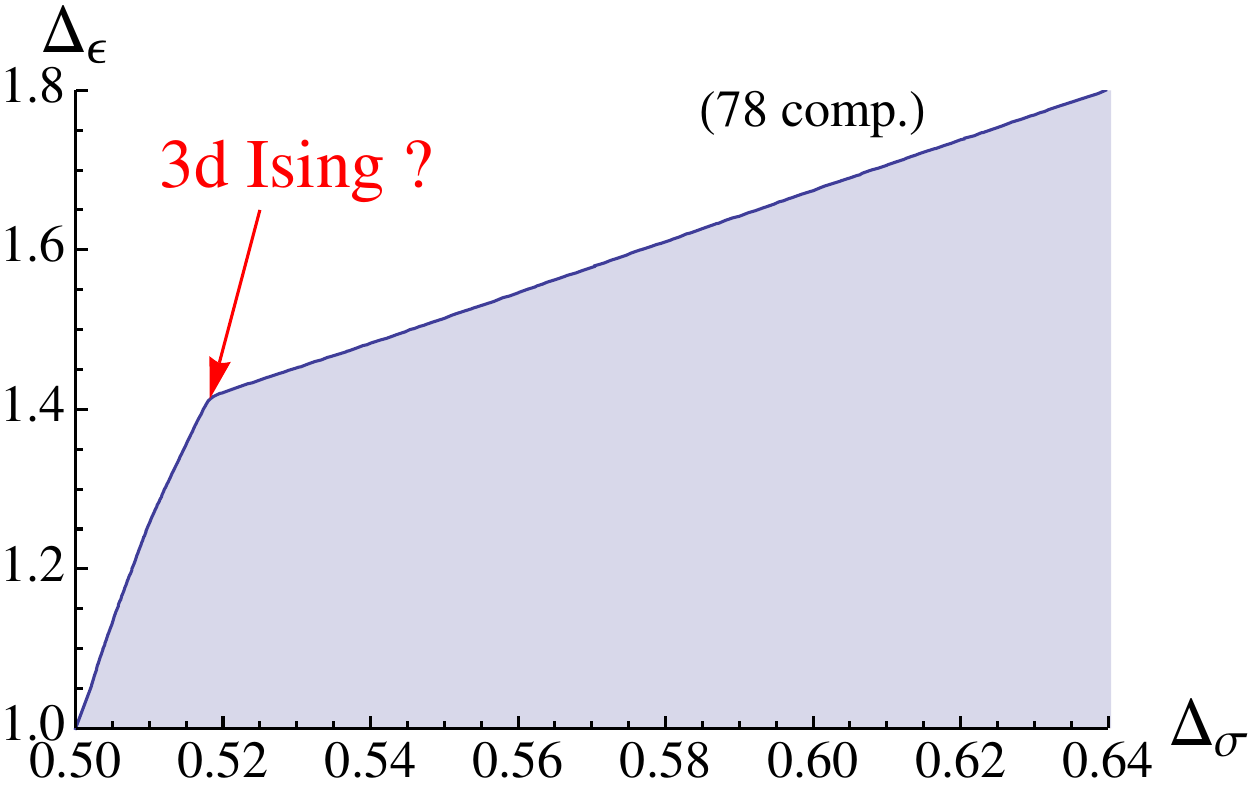}}
\includegraphics[scale=0.6]{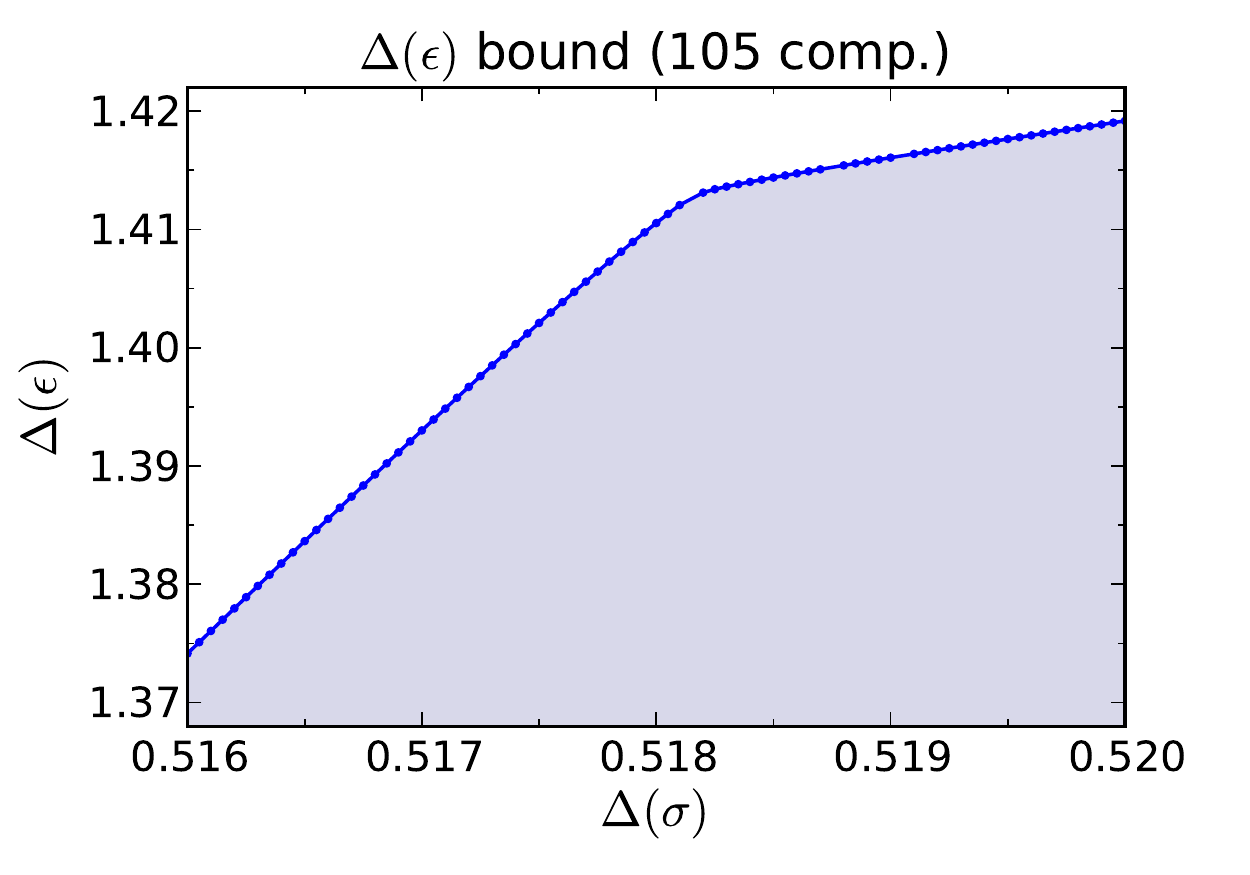}
\caption{An upper bound on the dimension of the lowest dimension scalar $\e\in \s\x\s$, as a function of $\De_\s$. The blue shaded region is allowed; the white region is disallowed. \textit{Left}: The bound at $N=78$ \cite{ElShowk:2012ht}. \textit{Right}: The bound at $N=105$, in the region near the kink. This bound is thus somewhat stronger than the previous one, and the kink is sharper.}
\label{fig-3dkinklarge}
\end{center}
\end{figure}

Consider a 3d CFT with a scalar primary operator $\s$ of dimension $\De_\s$.  In \cite{ElShowk:2012ht}, we studied the constraints of crossing symmetry and unitarity on the four-point function $\<\s\s\s\s\>$.  From these constraints, we derived universal bounds on dimensions and OPE coefficients of operators appearing in the $\s\x\s$ OPE.  Figure~\ref{fig-3dkinklarge}, for example, shows an upper bound on the dimension of the lowest-dimension scalar in $\s\x\s$ (which we call $\e$), as a function of $\De_\s$. This bound is a consequence of very general principles - conformal invariance, unitarity, and crossing symmetry - yet it has a striking ``kink" near  $(\De_\s,\De_\e)\aeq (0.518, 1.412)$, indicating that these dimensions have special significance in the space of 3d CFTs.  Indeed, they are believed to be realized in the 3d Ising CFT. 

The curves in Figure~\ref{fig-3dkinklarge} are part of a family of bounds labeled by an integer $N$ (defined in section~\ref{sec-approaching}), which get stronger as $N$ increases.   It appears likely that the 3d Ising CFT {\it saturates} the optimal bound on $\De_\e$, achieved in the limit $N\to\oo$.  Thus, in this work, we will take seriously the idea:
\begin{itemize}
\item  $\<\s\s\s\s\>$ in the 3d Ising CFT lies on the boundary of the space of unitary, crossing symmetric four-point functions.
\end{itemize}
We will further present a plausible guess for where on the boundary the 3d Ising CFT lies, and use this to formulate a precise conjecture for the spectrum of operators in the $\s\x\s$ OPE.

Although the 3d Ising CFT is certainly special, it is perhaps surprising that one might find it by studying a single four-point function.  After all, the full consistency constraints of a CFT include crossing symmetry and unitarity for {\it every} four-point function, including all possible operators in the theory.  Nevertheless, other recent work supports the idea that for some special CFTs it may be enough to consider $\<\s\s\s\s\>$.  For example, one can compute similar bounds in fractional spacetime dimension $2\leq d\leq 4$.  These bounds have similar kinks which agree with the operator dimensions present at the Wilson-Fisher fixed point near $d=4$ and the 2d Ising CFT when $d=2$ \cite{El-Showk:2013nia}.  An analogous story holds for theories with $O(n)$ global symmetry in 3d, where $O(n)$ vector models appear to saturate their associated dimension bounds \cite{Kos:2013tga}.

As a check on our conjecture, we will also apply it to the 2d Ising CFT.  We find good agreement with the known exact solution, and previous numerical explorations of the 2d bootstrap \cite{ElShowk:2012hu}.  Our study of 2d will serve as a useful guide for interpreting our results in 3d.

\subsection{Brief CFT Reminder}

Let us first recall some basic facts about CFT four-point functions that we will need in our analysis.  A four-point function of a real scalar primary $\s$ in a CFT takes the form
\be
\label{eq:generalformof4ptfunction}
\<\s(x_1)\s(x_2)\s(x_3)\s(x_4)\> = \frac{1}{x_{12}^{2\De_\s}x_{34}^{2\De_\s}} g(u,v),
\ee
where $u=\frac{x_{12}^2 x_{34}^2}{x_{13}^2 x_{24}^2}$ and $v=\frac{ x_{14}^2 x_{23}^2}{x_{13}^2 x_{24}^2}$ are conformally invariant cross-ratios.  The function $g(u,v)$ can be expanded in conformal blocks
\be
g(u,v) = \sum_{\cO\in\s\x\s} p_{\De,\ell} G_{\De,\ell}(u,v),
\label{eq:eq1}
\ee
where the sum is over primary operators $\cO$ with dimension $\De$ and spin $\ell$ appearing in the $\s\x\s$ OPE,  and the coefficients $p_{\De,\ell}=f_{\s\s\cO}^2$ are positive.  Only even spins $\ell$ can appear in $\s\x\s$, and $\De$ must obey the unitarity bounds
\be
\label{eq:unitaritybounds}
\De \geq \left\{
\begin{array}{ll}
(d-2)/2 & \textrm{if $\ell=0$ (excluding the unit operator)},\\
\ell+d-2 & \textrm{if $\ell>0$},
\end{array}
\right.
\ee
where $d$ is the spacetime dimension (we will mostly be interested in $d=3$).  We normalize $\s$ so that the OPE coefficient of the unit operator is  $f_{\s\s1}=1$ .

Finally, invariance of the four-point function (\ref{eq:generalformof4ptfunction}) under permutations of the $x_i$ implies the crossing-symmetry constraint
\be
\label{eq:crossingsymmetry}
g(u,v) = \p{\frac u v}^{\De_\s}g(v,u).
\ee
All results in this paper will be based on Eqs.~\reef{eq:eq1}, \reef{eq:crossingsymmetry} and on the information about conformal blocks reviewed in section \ref{sec:partialfractionsforblocks}.

\subsection{The Space of Unitary, Crossing-Symmetric 4-point Functions}
\label{sec:spaceofunitarycrossingsymmetric4ptfunctions}

Instead of focusing on a specific CFT, let us turn these facts around and consider all possible scalar four-point functions in any CFT.  Let $\cC_{\De_\s}$ be the space of linear combinations of conformal blocks
\be
\label{eq:crossingsymmetry1}
g(u,v)=\sum_{\De,\ell} p_{\De,\ell} G_{\De,\ell}(u,v)
\ee
such that
\begin{enumerate}
\item $(\De,\ell)$ obey the unitarity bounds (\ref{eq:unitaritybounds}),
\item $p_{0,0}=1$,\label{eq:unitopcondition}
\item $p_{\De,\ell}\geq 0$,\label{eq:unitaritycondition}
\item $g(u,v)$ is crossing-symmetric, Eq.~(\ref{eq:crossingsymmetry})\label{eq:crossingcondition}.
\end{enumerate}
We include the second condition because the unit operator should be present with $f_{\s\s1}=1$.  The third condition is because the OPE coefficients $f_{\s\s\cO}$ are real in a unitary theory, so their squares should be positive.  Note that $\cC_{\De_\s}$ depends on the parameter $\De_\s$ through the crossing constraint (\ref{eq:crossingsymmetry}).\footnote{These conditions are the obvious ones implied by conformal symmetry, unitarity, and crossing symmetry of $\<\s\s\s\s\>$.  We expect that $p_{\De,\ell}$'s in actual CFTs satisfy further conditions related to consistency of other four-point functions, and perhaps more exotic conditions like consistency of the theory on compact manifolds.  However, we do not impose these constraints in defining $\cC_{\De_\s}$.  It will be interesting and important to explore them in future work, for instance \cite{Kos:future}.}

\begin{figure}[htbp]
\begin{center}
\includegraphics[scale=0.7]{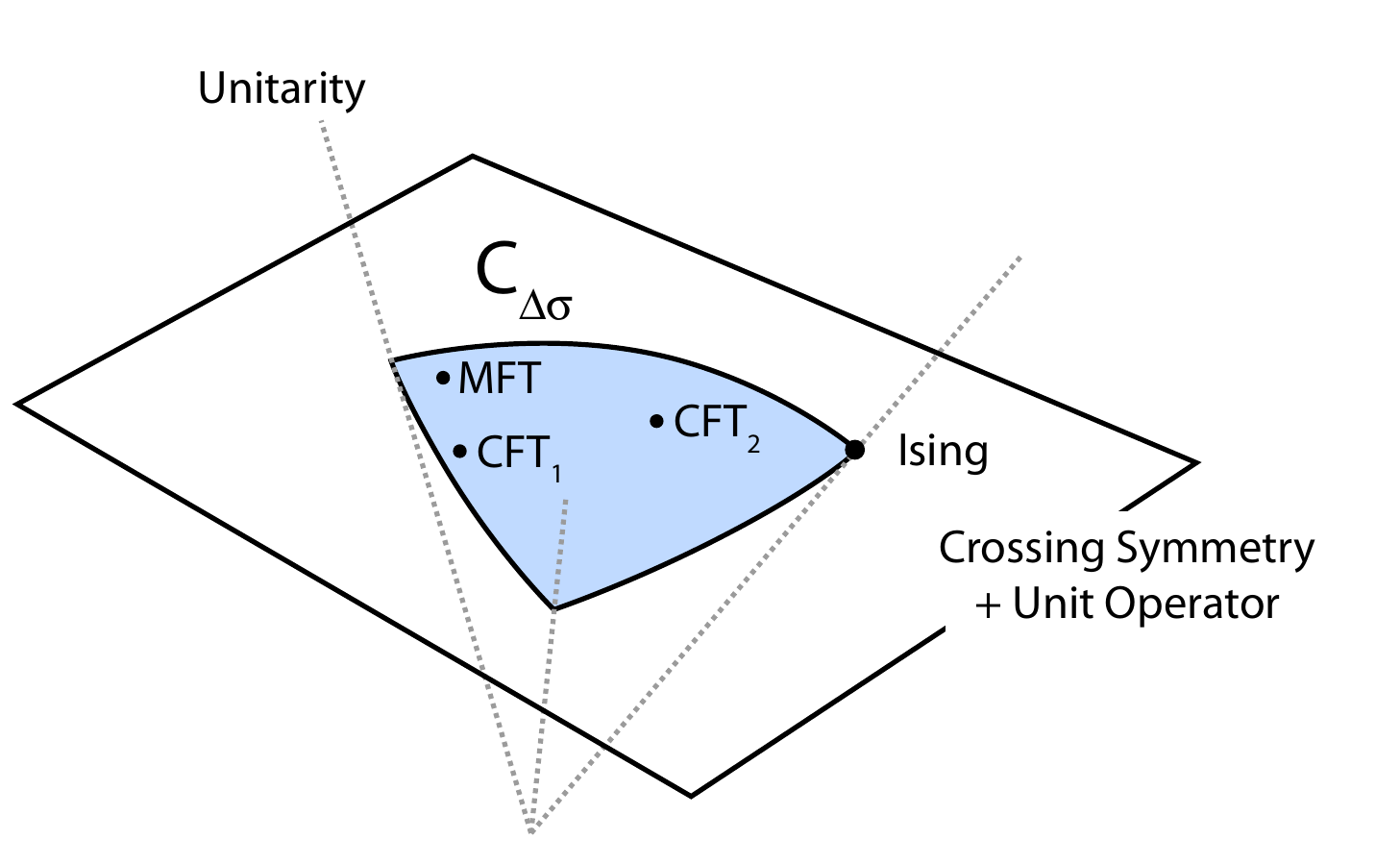}
\caption{The space $\cC_{\De_\s}$ (in blue) is the intersection of the convex cone given by the unitarity conditions $p_{\De,\ell}\geq 0$ with a hyperplane given by the affine constraints $p_{0,0}=1$ and Eq.~(\ref{eq:crossingsymmetry}).  It always contains a point corresponding to Mean Field Theory, and might contain points corresponding to other CFTs with a scalar of dimension $\De_\s$.  We conjecture that for a special value of $\De_\s$, the 3d Ising CFT lies on the boundary of $\cC_{\De_\s}$.}
\label{fig-convexspace}
\end{center}
\end{figure}

Let us make a few comments about this space.  Firstly, $\cC_{\De_\s}$ is convex.  This follows because positive linear combinations $t g_1(u,v) + (1-t) g_2(u,v)$, of four-point functions $g_1,g_2$ also satisfy the above conditions.  Geometrically, we can think of $\cC_{\De_\s}$ as a convex cone given by the unitarity condition $p_{\De,\ell}\geq 0$, subject to the affine constraints $p_{0,0}=1$ and Eq.~(\ref{eq:crossingsymmetry}).  This picture is illustrated in Figure~\ref{fig-convexspace}.

Secondly, $\cC_{\De_\s}$ is nonempty.  If a CFT with a scalar operator of dimension $\De_\s$ exists, then $\<\s\s\s\s\>$ in that theory certainly gives a point in $\cC_{\De_\s}$.  Furthermore, $\cC_{\De_\s}$ always contains a point corresponding to Mean Field Theory (MFT), where the four-point function $\<\s\s\s\s\>$ is given by a sum of products of two-point functions.\footnote{Such a scalar $\sigma$ is sometimes called a generalized free field.}  The MFT four-point function is crossing-symmetric and can be expanded as a sum of conformal blocks with positive coefficients.  The explicit coefficients $p_{\De,\ell}^\mathrm{MFT}$ can be found in \cite{Heemskerk:2009pn, Fitzpatrick:2011dm}.\footnote{MFT is not a genuine local CFT because it doesn't contain a stress-tensor.  However, it does appear as a point in $\cC_{\De_\s}$.}

What is the dimensionality of $\cC_{\De_\s}$? The answer to this question is not immediately obvious, since it is an infinite dimensional cone intersected with an infinite dimensional hyperplane. For $\De_\s$ at the unitarity bound, $\De_\s=\frac{d-2}2$, it is possible to show that $\cC_{\De_\s}$ consists of only one point: the free scalar four-point function. On the other hand, $\cC_{\De_\s}$ is likely to be an infinite-dimensional space for all $\De_\s>\frac{d-2}2$. When $\De_\s > d-2$, we can demonstrate this rigorously by exhibiting an infinite set of linearly independent crossing-symmetric, unitary four-point functions.  Namely, consider the tensor product of Mean Field Theories with scalars $\f_1$ and $\f_2$ of dimensions $\de$ and $\De_\s-\de$, and let $\s=\f_1\f_2$.  This gives a family of linearly-independent four-point functions labeled by a continuous parameter $\frac{d-2}2\leq\de\leq\De_\s/2$. While this argument does not apply for $\frac{d-2}2<\De_\s\leq d-2$, we believe that $\cC_{\De_\s}$ remains infinite-dimensional in this range. Some numerical evidence for this will be discussed in section~\ref{sec:functionals}.

\subsection{Approaching the Boundary of $\cC_{\De_\s}$}
\label{sec-approaching}

Every point on the boundary of a convex space is the maximum of some linear function over that space.  Conversely, if we have a bounded linear function on a convex space, then the maximum of that function is well-defined and generically unique.  Assuming that the 3d Ising CFT lies on the boundary of $\cC_{\De_\s}$, we should ask: {\it what does the 3d Ising CFT maximize?}. 

Before addressing this question, let us introduce some details about optimization over $\cC_{\De_\s}$.\footnote{We give a full description of our algorithm in section~\ref{sec:customsimplex}.}  To explore $\cC_{\De_\s}$ numerically, we construct it via a limiting procedure where we truncate the crossing equation (\ref{eq:crossingsymmetry}) to a finite number of constraints.\footnote{On the other hand, we will {\it not} have to restrict the dimensions $\De$ to a discrete set, as was done originally in \cite{Rattazzi:2008pe}.  For this work, we have implemented an optimization algorithm (described in section~\ref{sec:customsimplex}) that works with continuously varying $\De$.}  Specifically, define $\cC_{\De_\s}^{(N)}$ by the same conditions as $\cC_{\De_\s}$, except with the crossing-symmetry constraint replaced by
\be
\left.\ptl_u^m\ptl^n_v\p{g(u,v) - \p{\frac u v}^{\De_\s}g(v,u)}\right|_{u=v=1/4} = 0
\ee
for $N$ different (nonzero) derivatives $(m,n)$.\footnote{This equation is schematic. In section~\ref{sec:specializationofprimalsimplex} we will pass from $u,v$ to the variables $z,\bar z$, 
and take $N$ partial derivatives with respect to those variables.} The crossing-symmetric point $u=v=1/4$ is chosen as usual in the numerical bootstrap, to optimize the convergence of the conformal block expansion \reef{eq:crossingsymmetry1} \cite{Rattazzi:2008pe,Pappadopulo:2012jk}. This gives a sequence of smaller and smaller convex spaces, with $\cC_{\De_\s}$ as a limit,
\be
\cC^{(1)}_{\De_\s} \supset \cC^{(2)}_{\De_\s} \supset \dots \supset \cC_{\De_\s}.
\ee

\begin{figure}[htbp]
\begin{center}
\includegraphics[scale=0.9]{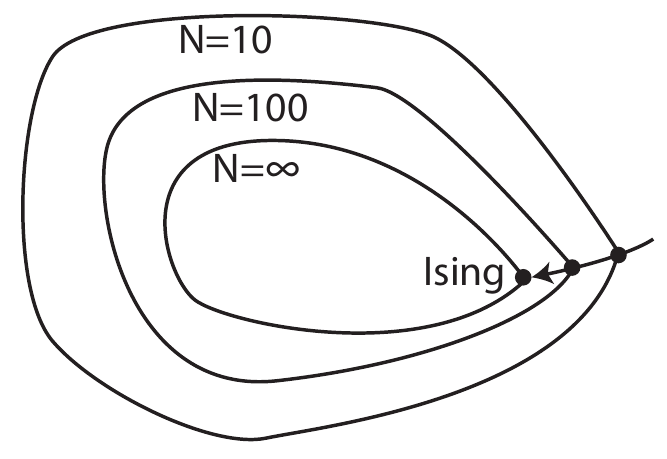}
\caption{Each number of derivatives $N$ gives an approximation to $\cC_{\De_\s}$ that shrinks as $N$ increases.  By repeatedly solving an optimization problem for each $N$, we can follow a path to the boundary of $\cC_{\De_\s}$.}
\label{fig-finite-constraints}
\end{center}
\end{figure}

Consider a function $f$ over $\cC_{\De_\s}^{(N)}$  which is maximized at a point $g_N$ on the boundary. Generically, $g_N$ will not lie inside $\cC_{\De_\s}$.  However, by following the sequence $g_N$ as $N\to\oo$, we can approach the maximum $g_*$ over $\cC_{\De_\s}$ (Figure~\ref{fig-finite-constraints}).  It's important that for functions considered below, the optimal point will be uniquely determined, i.e.~the maximum is not degenerate. This will allow us to pick out, for each $\Delta_\s$ and $N$,  the optimal spectrum and the OPE coefficients. We will see in section~\ref{sec:functionals} that the optimal spectrum at each $N$ contains order $N$ operators. As $N$ increases, we observe rapid convergence of lower dimension operators and also growth of the number of operators present. The hope is that the low lying 3d Ising CFT spectrum can be recovered in the limit.
Here we are taking inspiration and encouragement from \cite{ElShowk:2012hu}, where it was shown that in 2d a similar procedure recovers the exactly known spectrum of the 2d Ising CFT.

Note that while $g_N$ doesn't necessarily lie inside $\cC_{\De_\s}$, the value $f(g_N)$ is a rigorous upper bound on $\max_{\cC_{\De_\s}}(f)$.  Further, this bound improves as $N\to\oo$.  The bounds $f(g_N)$ are precisely those usually discussed in the conformal bootstrap literature (and shown for example in Figures~\ref{fig-3dkinklarge} and \ref{fig-ctmin-old}).  When $f$ is linear, the procedure of maximizing over $\cC_{\De_\s}$ is related by linear programming duality to the procedure of minimizing linear functionals, going back to \cite{Rattazzi:2008pe}.  From $g_N$, one can always construct an ``extremal" linear functional touching the boundary of $\cC_{\De_\s}^{(N)}$ at $g_N$, and vice versa.  In \cite{ElShowk:2012hu}, zeroes of the extremal functional were used to extract the spectrum of the optimal solution; knowing the spectrum, the squared OPE coefficients $p_{\Delta,\ell}$ could be determined by solving the crossing equation. In this work we will approach $g_N$ iteratively from inside of $\cC_{\De_\s}^{(N)}$, having a solution to crossing at each step of the maximization procedure. We will refer to the approach described here as the ``direct" or ``primal" method, as opposed to the ``dual" method of previous studies.  In principle, they are equivalent.

\subsection{Equivalence of $\De_\e$-maximization and $\CT$-minimization}
\label{sec-equivalence}
Let us return to the question: {\it what does the 3d Ising CFT maximize?}  As we have seen, there is good evidence that the 3d Ising CFT belongs to a family of solutions to crossing which, for a fixed $\Delta_\s$, maximize $\De_\eps$ (the dimension of the lowest-dimension $\bZ_2$-even scalar). However, $\De_\e$ is not a linear function on $\cC_{\De_\s}$ (the lowest dimension in a sum of four-point functions is the minimum dimension in either four-point function).  For conceptual and practical reasons, it will be useful to find an alternative characterization of the 3d Ising point in terms of a linear function on $\cC_{\De_\s}$.

In \cite{ElShowk:2012ht}, we also computed bounds on the squared OPE coefficient of {the spin 2, dimension $d$ operator. It is natural to assume that the 3d Ising CFT contains one and only one operator with such quantum numbers, namely the conserved stress tensor of the theory.}\footnote{The 3d Ising CFT can be obtained as an IR fixed point of the $\phi^4$ theory. The UV stress tensor then naturally gives rise to the IR stress tensor, but there is no reason to expect that a second operator with the same quantum numbers will emerge in the IR. This definitely does not happen in the $\eps$-expansion.}   The OPE coefficient $p_T\equiv p_{d,2}$ is related via Ward identities to the central charge $\CT$,
\be
\CT \propto \frac{\De_\s^2}{p_T},
\label{eq-pT}
\ee
with a $d$-dependent factor which depends on the normalization of $\CT$.\footnote{In our previous work \cite{ElShowk:2012ht,El-Showk:2013nia} $c$ was denoted $C_T$.} {We use the definition of the central charge $c$ as the coefficient in the two point correlation function of the stress tensor operator. This definition works for any $d$, in particular for $d=3$. In $d=2$, the central charge can also be defined as the coefficient of a central term in a Lie algebra extending the global conformal algebra (the Virasoro algebra), but an analogous interpretation for other $d$ is unknown and likely impossible.} To avoid normalization ambiguities, we will give results below for the ratio $\CT/\CT_{\text{free}}$, the latter being the central charge of the free scalar theory in the same number of dimensions. 

\begin{figure}[htbp]
\begin{center}
\raisebox{0.5cm}{\includegraphics[scale=0.85]{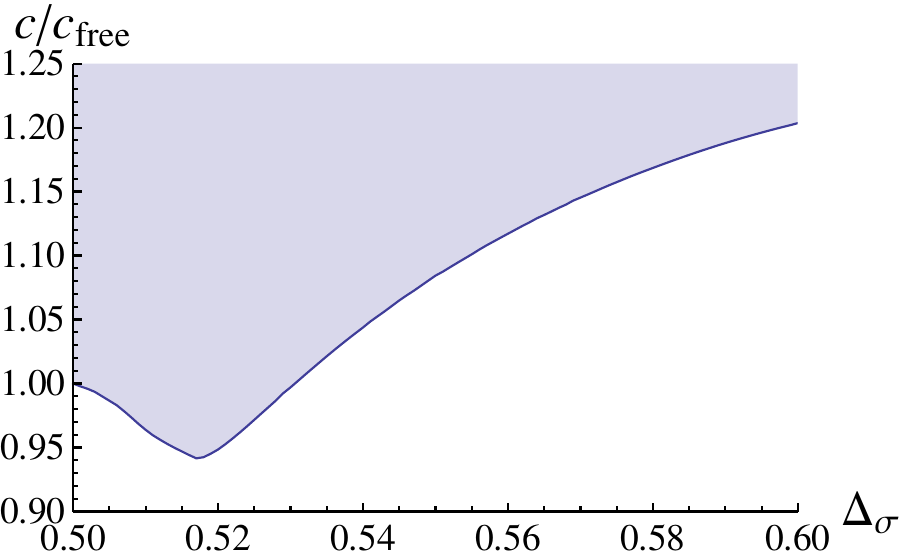}}
\includegraphics[scale=0.6]{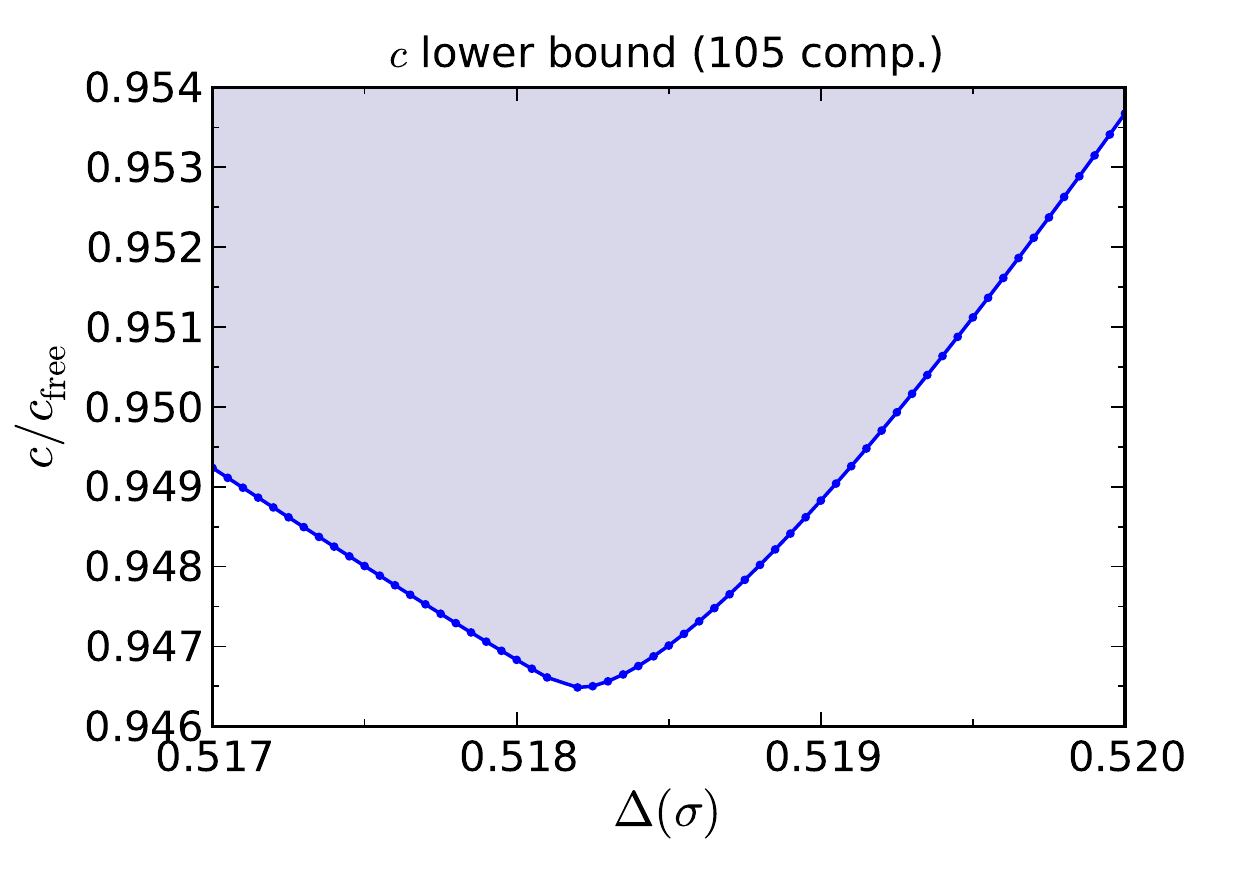}
\caption{A lower bound on $\CT$ (the coefficient of the two-point function of the stress tensor), as a function of $\De_\s$.  Left: the bound from \cite{ElShowk:2012ht} computed with $N=78$. Right: a slightly stronger bound at $N=105$ in the region near the minimum.}
\label{fig-ctmin-old}
\end{center}
\end{figure}

Figure~\ref{fig-ctmin-old} shows this lower bound on $\CT$ (equivalently, an upper bound on $p_T$) as a function of $\De_\s$. The bound displays a sharp minimum at the same value of $\De_\s$ as the kink in Figure~\ref{fig-3dkinklarge}, suggesting that the 3d Ising CFT might also minimize $\CT$.\footnote{By contrast, bounds on $c$ in 4d do not show a similar minimum~\cite{Poland:2010wg,Rattazzi:2010gj,Poland:2011ey}.}
We will call this method of localizing the 3d Ising point ``$\CT$-minimization," although it should be kept in mind that in practice we minimize $\CT$ by maximizing $p_T$, which is a linear function on $\cC_{\De_\s}$.

\begin{figure}[htbp]
\begin{center}
\includegraphics[scale=0.6]{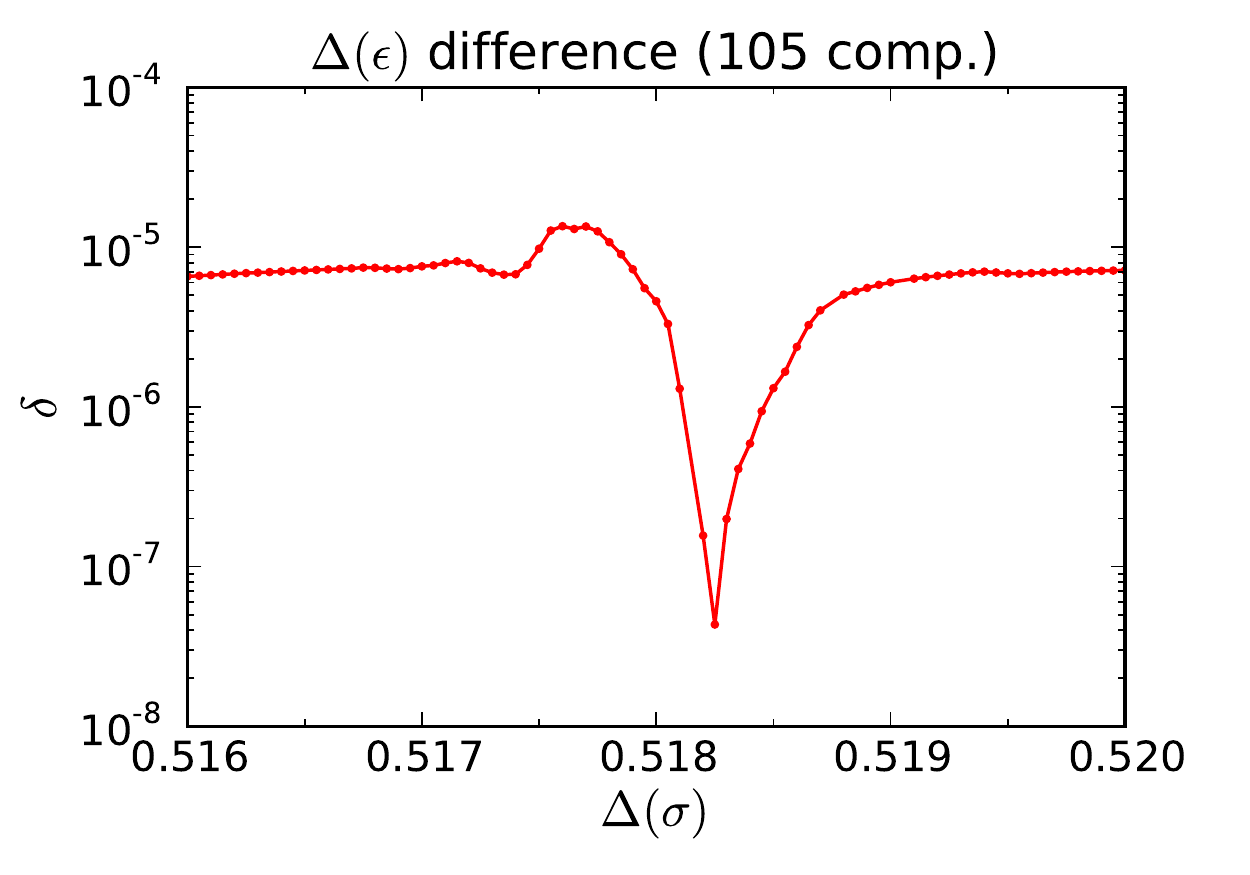}
\caption{The difference $\Delta^{\text{max}}_\eps -\Delta_\eps|_{\CT\to \text{min}}$ as a function of $\De_\s$, computed at $N=105$.}
\label{fig-difference}
\end{center}
\end{figure}

We have done numerical studies of both $\De_\e$-maximization and $\CT$-minimization, and found evidence that for all $\Delta_\sigma$ in the neighborhood of the expected 3d Ising value $\approx 0.518$ these optimization problems are solved at the same (uniquely-determined) point on the boundary of $\cC_{\De_\s}$.  In other words, they are equivalent conditions.  As an example, one can compute the maximum possible dimension $\De_\e^\mathrm{max}$, and also the value $\De_\e^*$ in the four-point function that minimizes $\CT$. As already mentioned, for each $\Delta_\sigma$ this four-point function turns out to be unique, so that $\De_\e^*$ is uniquely determined. It is then interesting to consider the difference (non-negative by definition)
\beq
\delta=\Delta^{\text{max}}_\eps -\Delta_\eps^*,\quad \Delta_\eps^*\equiv \Delta_\eps|_{\CT\to \text{min}}.
\eeq
We plot this difference in Figure~\ref{fig-difference}, where we have chosen $N=105$.  The difference is tiny for all values of $\De_\s$, and near one point it drops sharply towards zero --- as we will see below, that's where the 3d Ising CFT sits.  We expect that as one increases the precision by taking $N\to \oo$, the difference $\de$ should go exactly to zero at the 3d Ising point.

\begin{figure}[htbp]
\begin{center}
\includegraphics[scale=0.35]{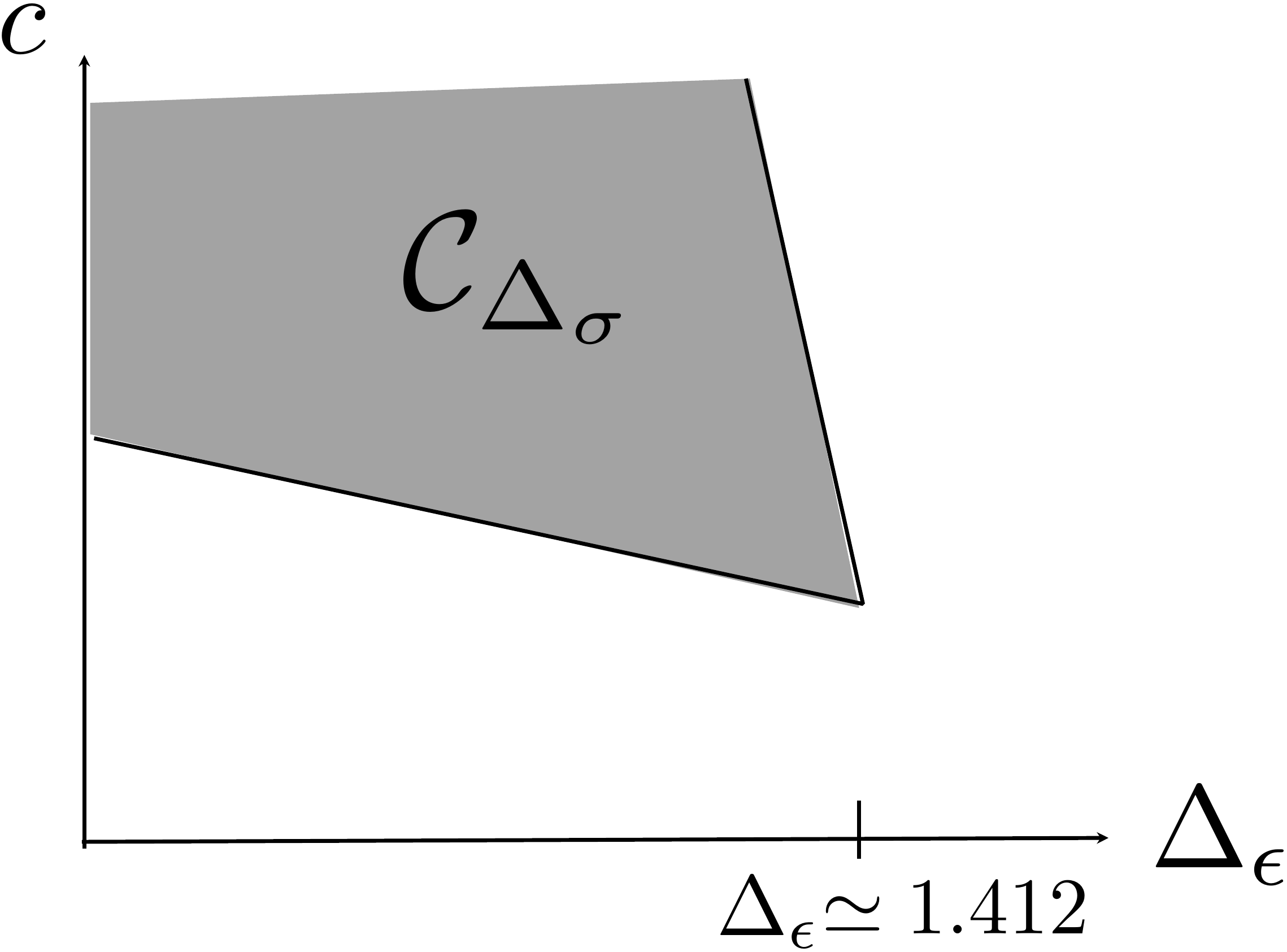}
\caption{Presumed geometry of $\cC_{\De_\s}$ near the 3d Ising CFT. Its boundary must have a corner point which can be reached by either maximizing $\Delta_\eps$ or minimizing $\CT$.} 
\label{deCT}
\end{center}
\end{figure}

A natural explanation of these observations is that, for a fixed $\Delta_\sigma$ there exists a corner point on the boundary of $\cC_{\De_\s}$, which can be reached by maximizing any of several different functions --- in particular, both $\De_\e$ and $p_T$ are suitable (Figure~\ref{deCT}).  Although this corner is roughly present in the approximations $\cC_{\De_\s}^{(N)}$, it should emerge sharply as $N\to\oo$. 

In what follows, we will mainly be using $\CT$-minimization rather than $\Delta_\eps$-maximization to localize the 3d Ising CFT.
One reason for this is practical: since $p_T$ is a linear function on $\cC_{\De_\s}$, maximizing $p_T$ turns out to be much easier than maximizing $\Delta_\eps$ in our numerical approach (see section \ref{sec:deltaemaximization} for an explanation). 

In the above discussion we varied $\Delta_\s$ in an interval around 0.518. The true 3d Ising value of $\Delta_\s$ should correspond to the position of kinks in the given $\Delta_\eps$ and $c$ bounds. Clearly, \textit{overall} $\Delta_\eps$-maximization is not the right criterion to pick out the kink in the $(\Delta_\s,\Delta_\eps)$ plane (Figure \ref{fig-3dkinklarge}). We could of course cook up an ad hoc linear combination of $\Delta_\s$ and $\Delta_\eps$ whose maximum would pick out the kink. An especially neat feature of the $\CT$-minimization is that no such ad hoc procedure is necessary: the kink is at the same time the overall minimum of $\CT$ when $\Delta_\s$ is allowed to vary. It is worth emphasizing that the factor $\Delta_\s^2$ in \reef{eq-pT} is important for achieving this property, i.e.~without this factor $1/p_T$ would have a kink but not an overall minimum at the 3d Ising point.  Thus it is really the quantity $\CT$, rather than the OPE coefficient of $T_{\mu\nu}$, which seems to enjoy a special and non-trivial role in determining $\Delta_\s$ for the 3d Ising model.

This last point allows us to conjecture that the 3d Ising CFT minimizes the central charge $\CT$ over the space of unitary, crossing symmetric four-point functions.  More succinctly:
\be
\De_\s, g(u,v)\textrm{ in 3d Ising}
=
\mathop{\mathrm{argmin}}\limits_{\De_\s,\,g(u,v)\in \cC_{\De_\s}} \CT,
\label{eq-conj}
\ee
up to two qualifications mentioned below.

There are clear conceptual advantages to phrasing our conjecture in terms of the $\CT$ minimum instead of the $\De_\e$ kink.  The stress tensor $T_{\mu\nu}$ is arguably more special than the scalar operator $\e$ --- it is a conserved current present in every local CFT.  Further, the central charge $\CT$ can be interpreted loosely as a measure of how ``simple" a CFT is; our conjecture then implies that the 3d Ising CFT is the ``simplest" 3d CFT.\footnote{We are aware of the fact that $\CT$ is not always monotonic under RG-flow, e.g. \cite{Nishioka:2013gza}.}  In particular, the 3d Ising CFT is as far as possible from Mean Field Theory, which has $\CT=\oo$.

The two qualifications concerning \reef{eq-conj} are as follows. Firstly, the minimum over $\Delta_\s$ is local; it is supposed to be taken in the region
sufficiently close to the unitarity bound, $0.5\le \Delta_\s\lesssim 1$. For larger $\Delta_\s$ the bound in Figure \ref{fig-ctmin-old} curves down and approaches zero. Although it seems plausible that the 3d Ising CFT has \textit{globally} minimal $\CT$ among all unitary 3d CFTs, this conclusion cannot be reached by studying just a single four-point function of the theory.

Secondly, the minimum over $\cC_{\De_\s}$ in Eq.~\reef{eq-conj} must be computed assuming an additional lower cutoff on $\Delta_\eps$. The point is that in the region $\Delta_\eps<1$ there exist $\CT$ minima with $\CT<\CT_{\text{3d Ising}}$. These clearly have nothing to do with the 3d Ising CFT, which is known to have $\Delta_\eps\approx 1.412$. In fact, we know of no examples of 3d unitary CFTs with $\Delta_\eps<1$, and we suspect that these extra minima are altogether unphysical. In this work we eliminate them by imposing a cutoff $\Delta_\eps\ge \Delta_\eps^{\text{cutoff}}\approx  1$ (this was already done in producing Fig.~\ref{fig-ctmin-old}). The end results are not sensitive to the precise value of this cutoff.
In particular, the final spectra do not contain an operator at the cutoff.

From a practical standpoint, $\CT$-minimization or $p_T$-maximization over $\cC_{\De_\s}$ is a linear programming problem which can be attacked numerically on a computer.\footnote{Because $\De$ can vary continuously, $p_T$-maximization should more properly be called a {\it semi-infinite} program, although we will not be careful about this distinction.}  For this purpose, we have implemented a customized version of Dantzig's simplex algorithm, capable of dealing with the continuous infinity of possible dimensions $\De$, and exploiting special structure in the conformal blocks $G_{\De,\ell}$.  Our algorithm is described in detail in section~\ref{sec:customsimplex}.  In the next section, we will apply it to $p_T$-maximization and study the spectrum of the 3d Ising CFT.

\section{The 3d Ising Spectrum from $\CT$-minimization}
\label{sec:3dspec}

In the previous section, we conjectured that the 3d Ising CFT lives at the overall $\CT$ minimum in Figure \ref{fig-ctmin-old}. In this section, we will focus on the interval of $\De_\s\in[0.5179,0.5185]$ containing this minimum. We will perform a high-precision exploration of this interval, including as many as $N=231$ derivative constraints. This is the highest number of constraints ever used in the analysis of a single bootstrap equation, and is a threefold increase compared to $N=78$ used in our previous work \cite{ElShowk:2012ht}.\footnote{In the analysis of CFTs with global symmetry \cite{Poland:2011ey}, up to 66 constraints were used per each of the three (for $SO(N)$) or six (for $SU(N)$) bootstrap equations present in that case. In \cite{Kos:2013tga}, up to 55 constraints were used per bootstrap equation.  See Appendix~\ref{app:semidefinitecomparison} for a discussion of the methods used in these works.} 

We will obtain a number of predictions for the operator dimensions and the OPE coefficients in the 3d Ising CFT. Comparison with previously known results by other techniques is postponed to section \ref{sec:comp}.

\subsection{$\Delta_\s$ and $\CT$}
\label{sec:ds}

\begin{figure}[htbp]
\begin{center}
\includegraphics[scale=0.8]{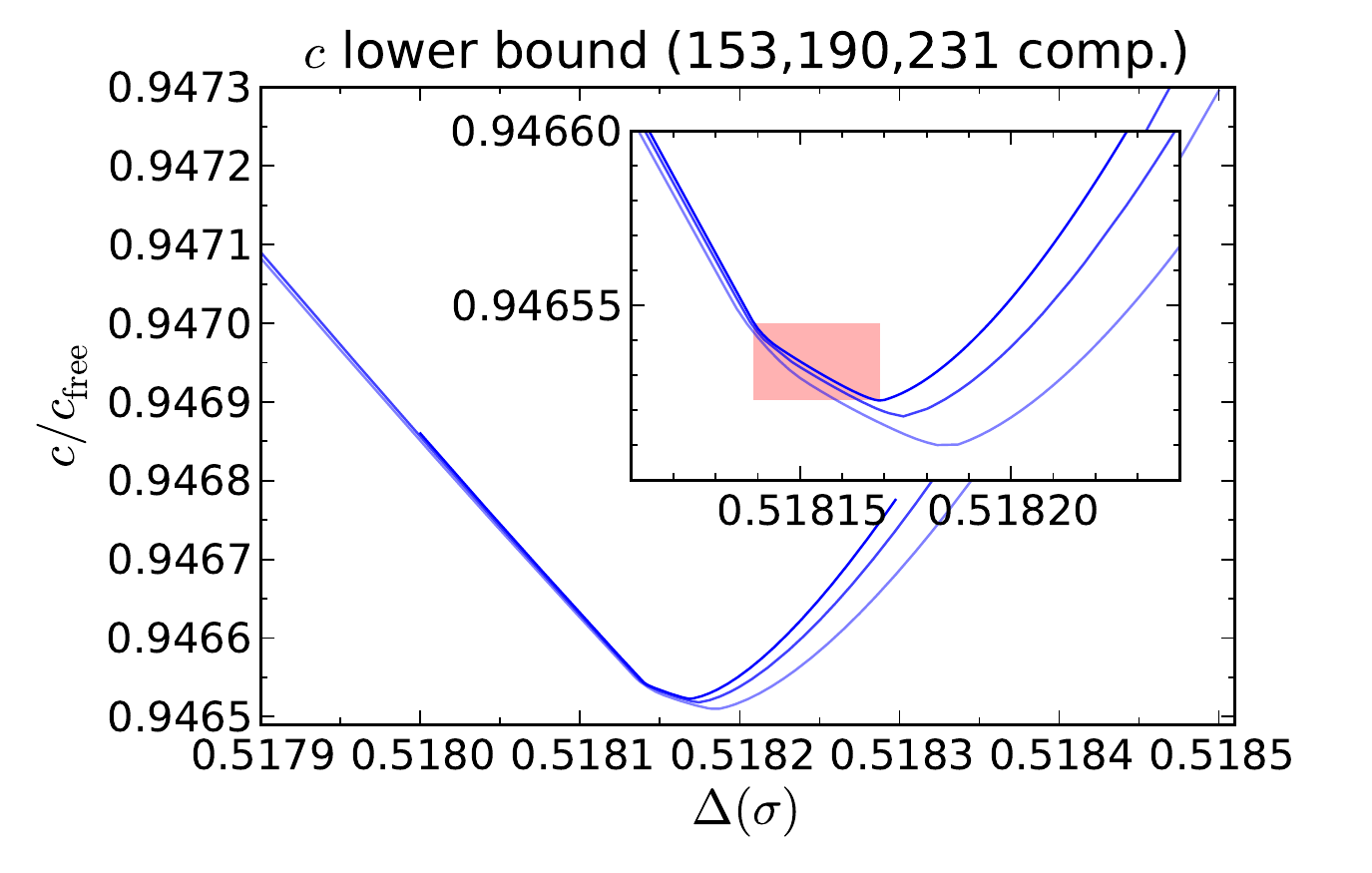} 
\caption{A lower bound on $\CT$ in the region of $\Delta_\s$ close to the minimum in Figure \ref{fig-ctmin-old}. We plot three bounds, computed with $N=231$ (dark blue), 190 (lighter blue), 153 (lightest blue). The pink rectangle in the zoomed inlay corresponds to our conservative upper and lower bounds \reef{eq-ds-ct} for the location of the minimum for $N\to\infty$, which according to our conjecture should correspond to the 3d Ising CFT.}
\label{sev_ct}
\end{center}
\end{figure}

In Figure \ref{sev_ct}, we show the $\CT$ minimization bound computed as a function of $\Delta_\s$ in the region near the overall $\CT$ minimum, for $N=153,190,231$.\footnote{Here and in subsequent plots $N=231$ data cover a smaller subinterval $\De_\s\in[0.5180,0.5183]$.}
The shape of these bounds can be schematically described as follows: on the left an almost-linearly decreasing part, then a ``kink'' and a short middle part leading to the overall minimum, and finally a monotonically increasing part on the right. When $N$ is increased, the bound is getting visibly stronger on the right, but is essentially stable on the left. As a result the minimum is shifting somewhat upward and to the left, while the kink is barely moving. 

In these plots, the middle part of the bounds between the kink and the minimum is shrinking monotonically when $N$ is increased. It looks certain that the minimum is set to merge with the kink in the $N\to\infty$ limit. According to our conjecture, this merger point should correspond to the 3d Ising CFT. Therefore, the positions of the kink and the minimum at $N=231$ give upper and lower bounds on $\Delta_\s$ and $\CT$ in the 3d Ising CFT (pink rectangle in the zoomed inlay in Figure \ref{sev_ct}):
\beq
\Delta_\sigma=0.518154(15),\quad \CT/\CT_{\text{free}}=0.946534(11)\,.
\label{eq-ds-ct}
\eeq
For simplicity, we give symmetric error bars around the center of the middle part of the $N=231$ bound. However, from the way the middle part of the bounds is shrinking mostly from the right, we can foresee that the true 3d Ising CFT values lie probably closer to the left kink.

\subsection{Extracting the Spectrum and OPE Coefficients in $\sigma\times\sigma$}

The $\CT$-minimization bounds in Figure \ref{sev_ct} actually contain much more information than plotted there. For each point saturating those bounds, there is a \textit{unique} unitary four-point function $\langle\sigma\sigma\sigma\sigma\rangle$ which solves the corresponding $N$ crossing constraints. It is very interesting to study how the spectrum of operators appearing in the conformal block decomposition of this four-point function, and their OPE coefficients, depend on $\Delta_\s$. This will be done in the next section for the leading scalar, and in the following sections for the higher states.

We should stress that no additional computation is needed to extract the solution to crossing corresponding to the minimal $\CT$. The minimization algorithm starts somewhere inside the allowed region (i.e.~above the bound) and performs a series of steps. Each step replaces a solution to crossing by another solution with a strictly smaller $\CT$, thus moving towards the boundary. After many steps (tens of thousands), $\CT$ stops varying appreciably, and the algorithm terminates. We thus obtain both the minimal $\CT$ value and the corresponding solution to crossing. Empirically, we observe that the spectrum and the OPE coefficients change a lot from one step to the other in the initial phases of the algorithm, while in the end they stabilize to some limiting values, which depend just on $\Delta_\s$ and the value of $N$ we are working at. These limiting values depend smoothly on $\Delta_\s$, except for some interesting rearrangements in the higher states happening near the 3d Ising point, which will be discussed below.
The smoothness of these limits is by itself also evidence for their uniqueness. As mentioned above, to reach the boundary, the simplex method must perform tens of thousands of small steps, different for even nearby values of $\Delta_\s$. The absence of uniqueness would be detectable by jittering in the plots below, but we observe no such jittering. 

After these general preliminary remarks, let us explore the spectrum and the OPE coefficients corresponding to the $\CT$ bounds in Figure \ref{sev_ct}. 

\subsection{The Leading $\bZ_2$-Even Scalar $\eps$}
\label{sec:de}

\begin{figure}[htbp]
\begin{center}
\includegraphics[scale=0.8]{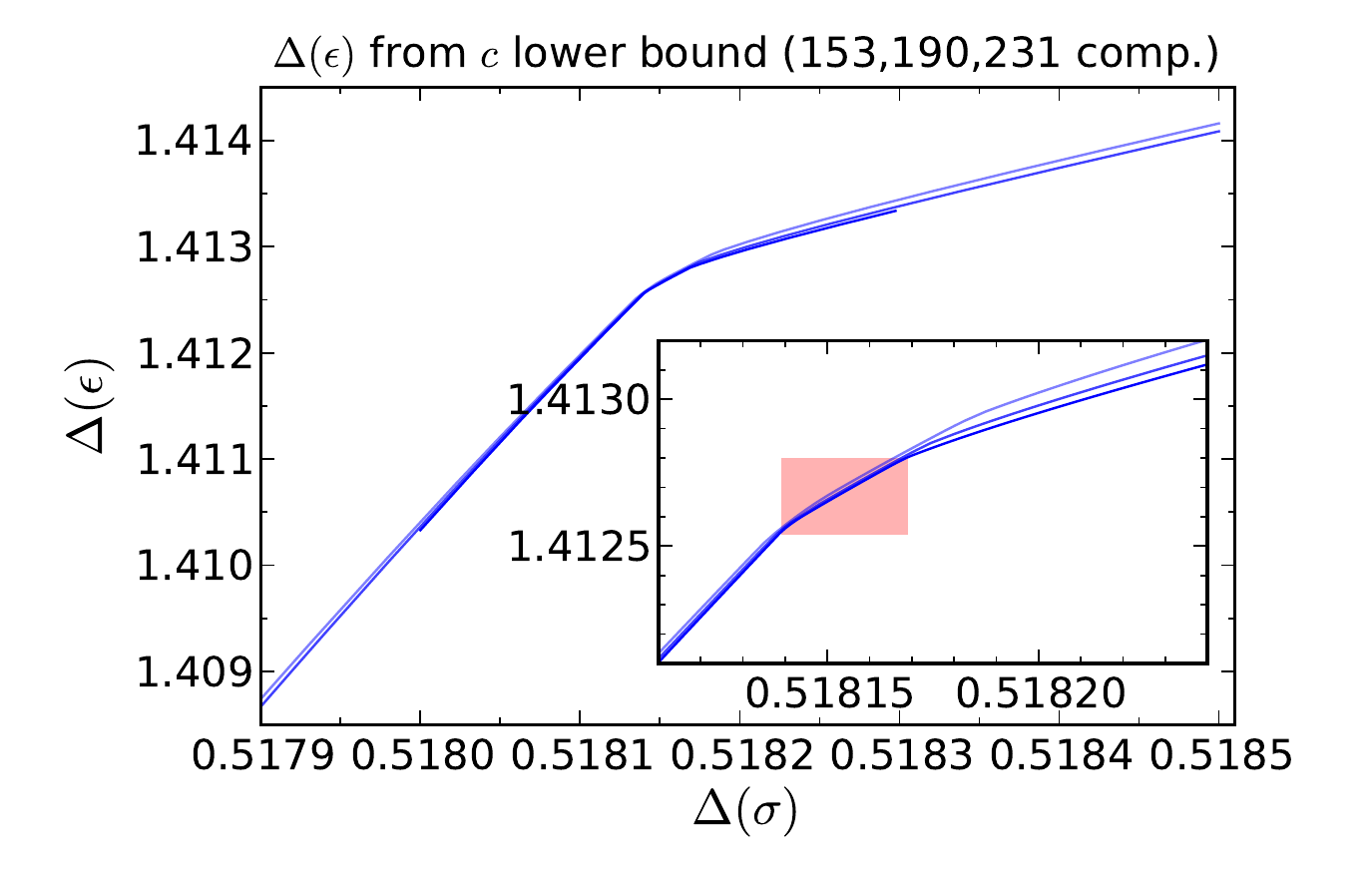}
\caption{The dimension of the leading scalar $\eps\in\sigma\times\sigma$, from the four-point functions realizing the minimal $\CT$ bounds in Figure \ref{sev_ct}, with the same line color assignment. The pink rectangle has the same horizontal extension as in Figure \ref{sev_ct}; its vertical extension gives our prediction \reef{eq-de} for $\Delta_\eps$ in the 3d Ising CFT.}
\label{sev_de}
\end{center}
\end{figure}

In this section, we are interested in the leading scalar operator $\eps\in\sigma\times\sigma$. In Figure \ref{sev_de} we plot its dimension $\Delta_\eps$ as a function of $\Delta_\s$. 

The curves in Figure \ref{sev_de} look qualitatively similar to the $\Delta_\eps$ bound in Figure \ref{fig-3dkinklarge}, although it should be kept in mind that they have been computed by a different method. In Figure \ref{fig-3dkinklarge}, we were maximizing $\Delta_\eps$, while in Figure \ref{sev_de} we minimized $\CT$ and extracted $\Delta_\eps$ corresponding to this minimum. Nevertheless, as discussed in section \ref{sec-equivalence}, the two methods give very close results near the 3d Ising point, so here we are using the $\CT$-minimization method.

The plots in Figure \ref{sev_de} have a narrowly localized kink region, which keeps shrinking as $N$ is increased. Just as we used the $\CT$ bounds to give predictions for $\Delta_\s$ and $\CT$, we can now use Figure \ref{sev_de} to extract a prediction for $\Delta_\eps$. The upper and lower bounds are given by the pink rectangle in the zoomed inlay. Notice that the horizontal extension of this rectangle is exactly the same as in Figure \ref{sev_ct}---the changes in slope happen at precisely the same $\Delta_\s$'s as for the $\CT$ bounds. This is not surprising since $\CT$ and $\Delta_\eps$ enter as variables in the same bootstrap equation, and any non-analyticity in one variable should generically have some reaction on the others. On the other hand, the vertical extension of the rectangle gives our confidence interval for $\Delta_\eps$: 
\beq
\Delta_\eps = 1.41267(13)\,.
\label{eq-de}
\eeq

\begin{figure}[htbp]
\begin{center}
\includegraphics[scale=0.8]{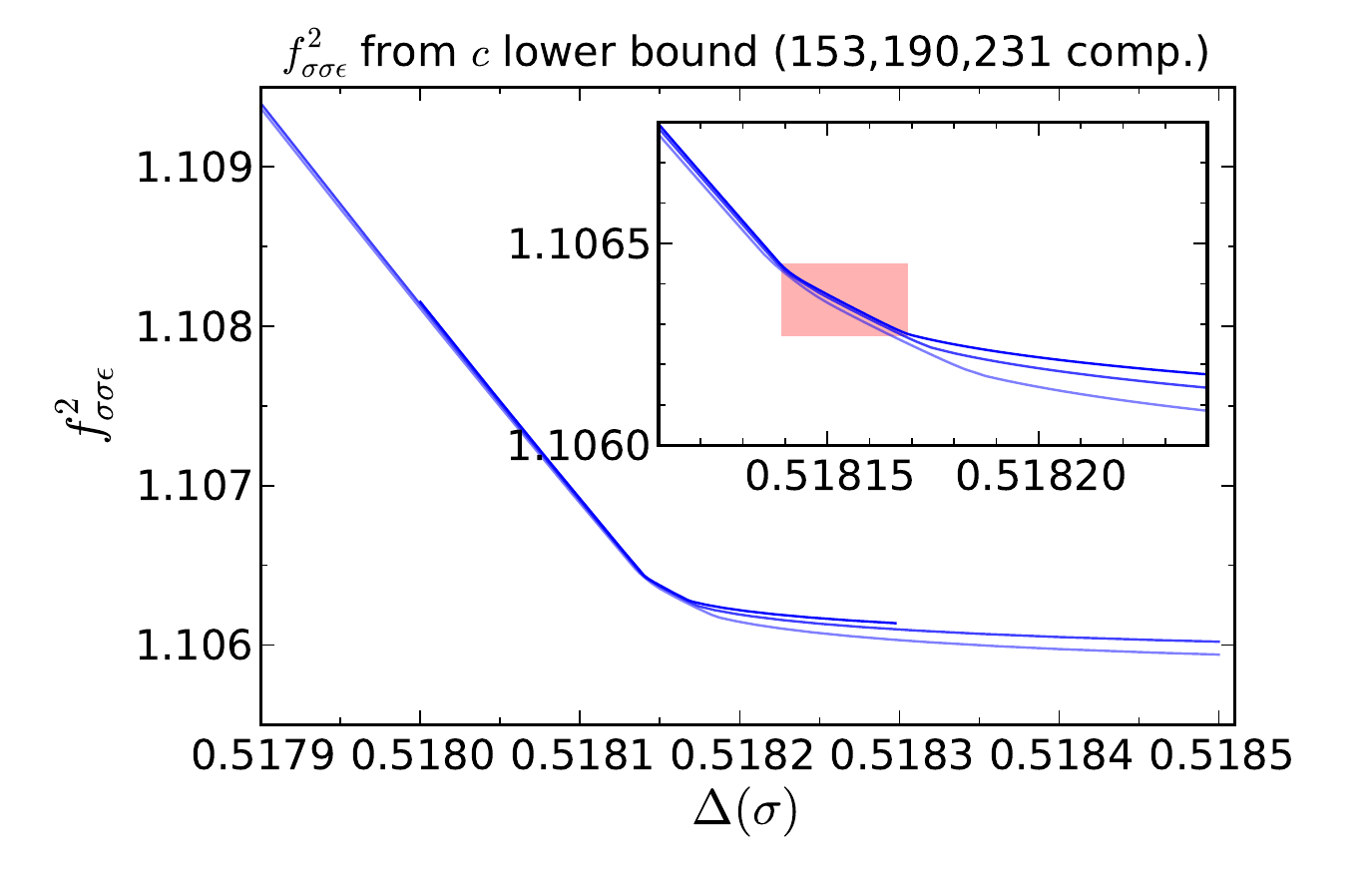}
\caption{Same as the previous figure, for the squared OPE coefficient $f_{\sigma\sigma\eps}^2$.}
\label{sev_de-ope}
\end{center}
\end{figure}

In Figure \ref{sev_de-ope} we repeat the same exercise for the $\eps$'s OPE coefficient. Again, the horizontal extension of the pink rectangle is the same as in Figure \ref{sev_ct}, while its vertical extension gives our prediction for the OPE coefficient:
\beq
f_{\sigma\sigma\eps}^2=1.10636(9)\,.
\eeq

In this paper we normalize the OPE coefficients and conformal blocks in the same way as in \cite{ElShowk:2012ht}. For scalars in 3d this normalization coincides with the most commonly used normalization of OPE coefficients through the three-point function:
\beq
\langle \cO_1(x_1) \cO_2(x_2)\cO_3(x_3) \rangle = \frac {f_{123}}
{x_{12}^{\Delta_1+\De_2-\De_3} x_{13}^{\De_1+\De_3-\De_2} x_{23}^{\De_2+\De_3-\De_1} }\,,
\eeq
where all scalars are assumed to have a unit-normalized two-point function. For example, in Mean Field Theory we have $f_{\phi\phi \cO}=\sqrt{2}$ where $\phi$ and $\cO=\phi^2/\sqrt{2}$ are unit-normalized.

\subsection{Higher Scalars: General Features}
\label{sec:spin0}

\begin{figure}[htbp]
\begin{center}
\includegraphics[scale=0.6]{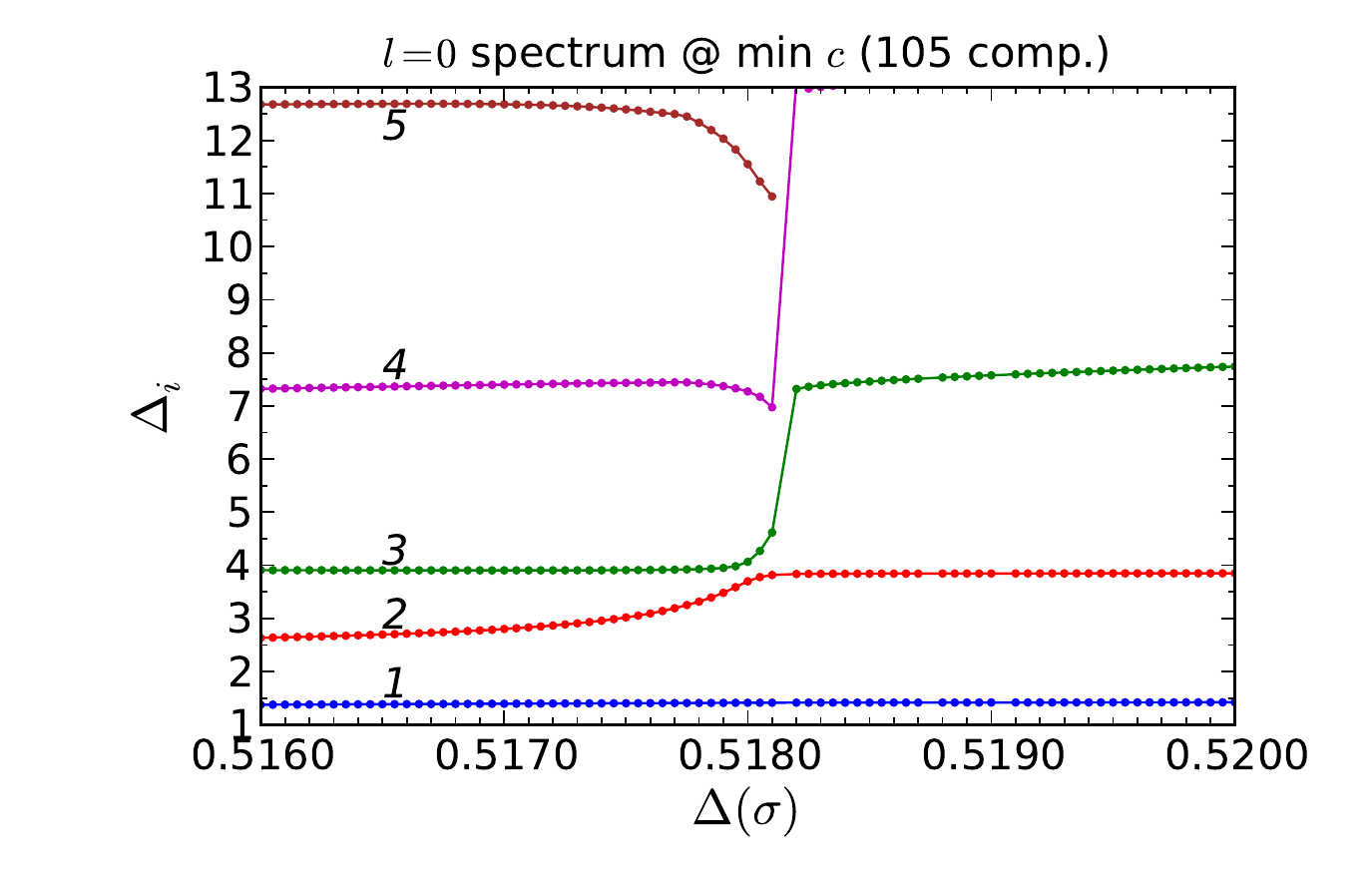}
\hspace{-20pt}
\includegraphics[scale=0.6]{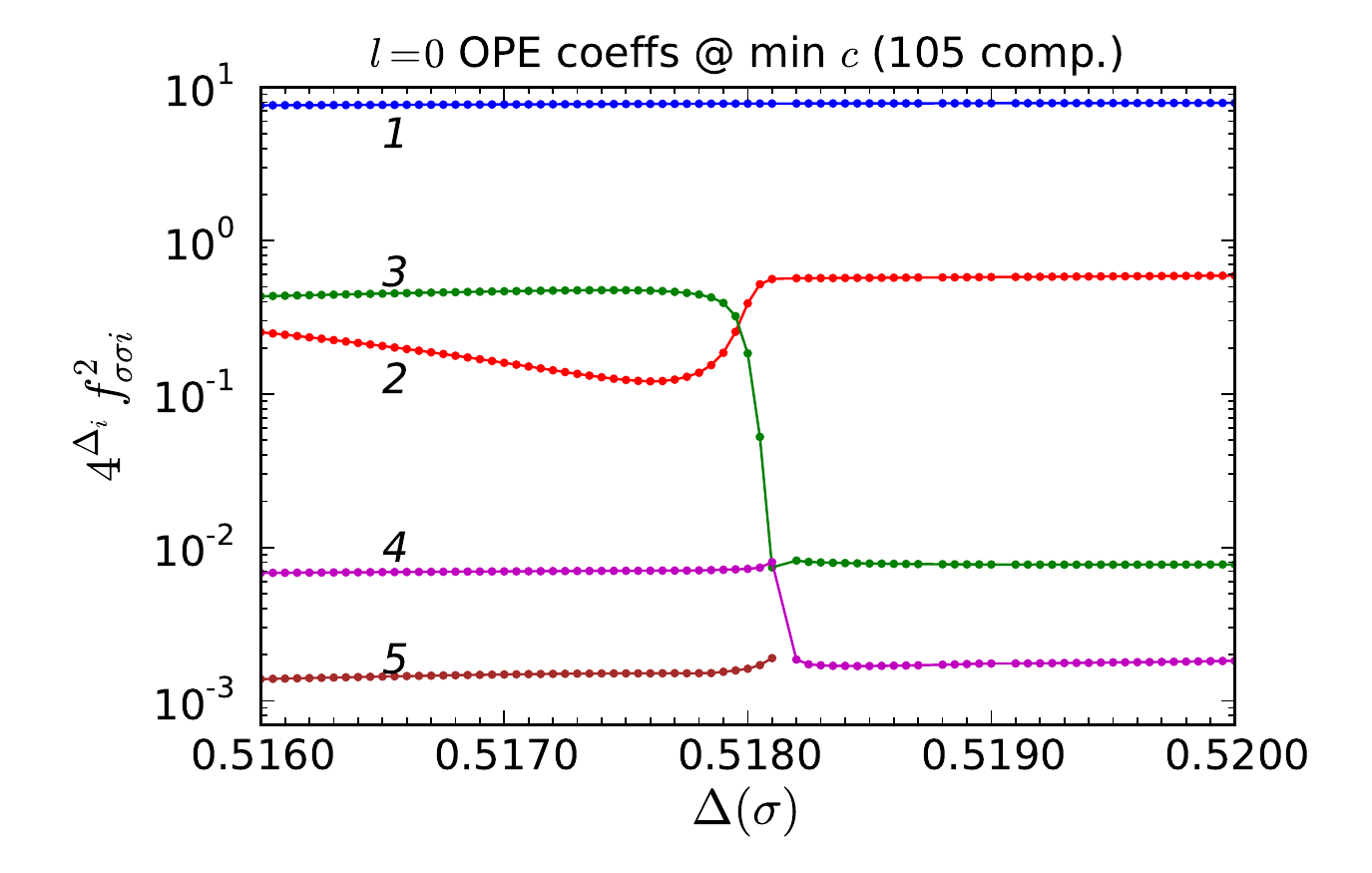}
\caption{Scalar operators with dimensions with $\Delta\le 13$ and their squared OPE coefficients (matching color and numbering). Here we are using the same $N=105$ data as in Figure \ref{fig-difference} (right). The squared OPEs are plotted multiplied by $4^{\Delta_i}$ to somewhat correct for the disparity in size; this is also the natural normalization from the point of view of the OPE convergence estimates \cite{Pappadopulo:2012jk}. 
}
\label{fig-nmax13-spin0}
\end{center}
\end{figure}

\begin{figure}[htbp]
\begin{center}
\includegraphics[scale=0.6]{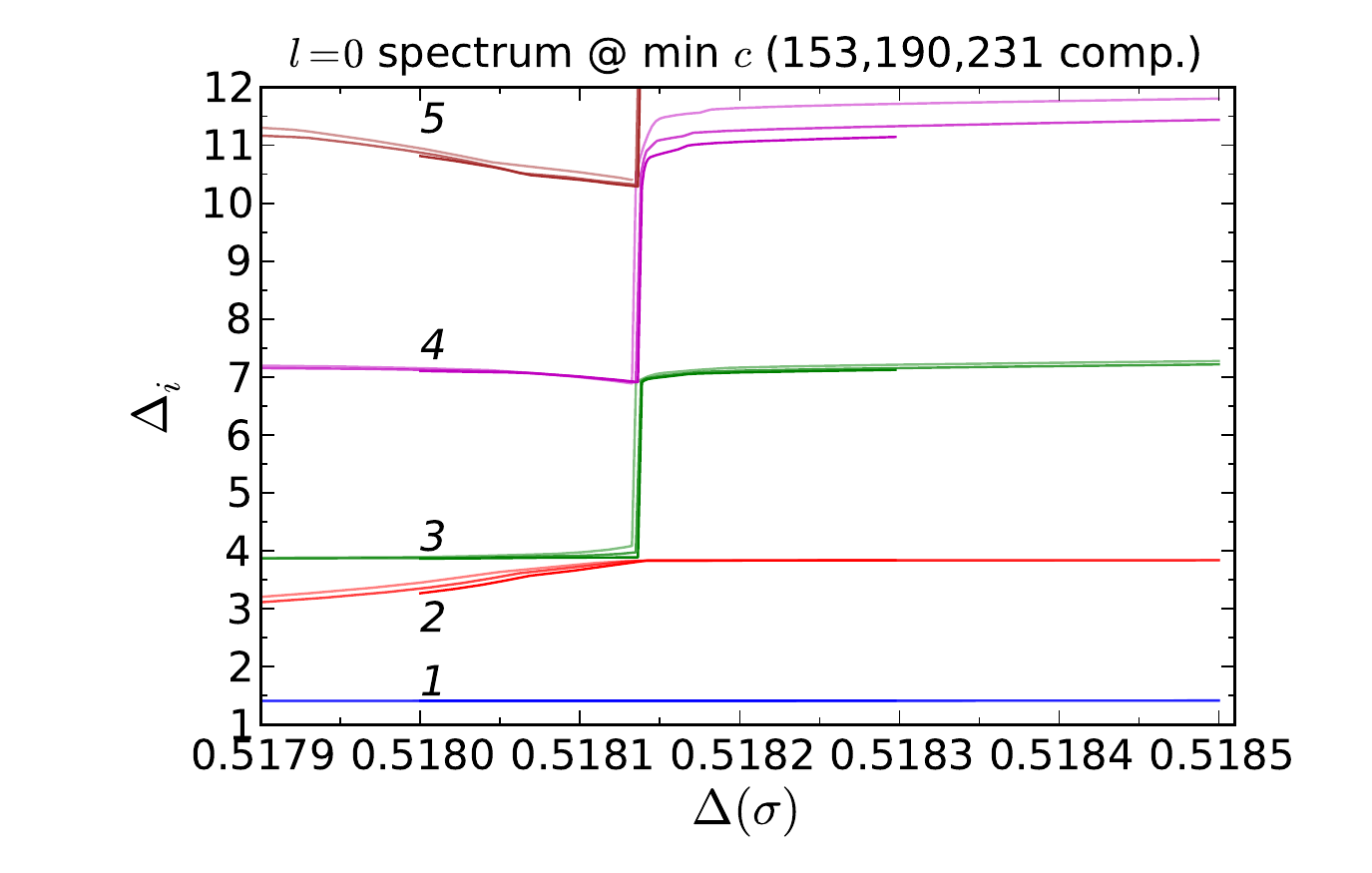}
\hspace{-20pt}
\includegraphics[scale=0.6]{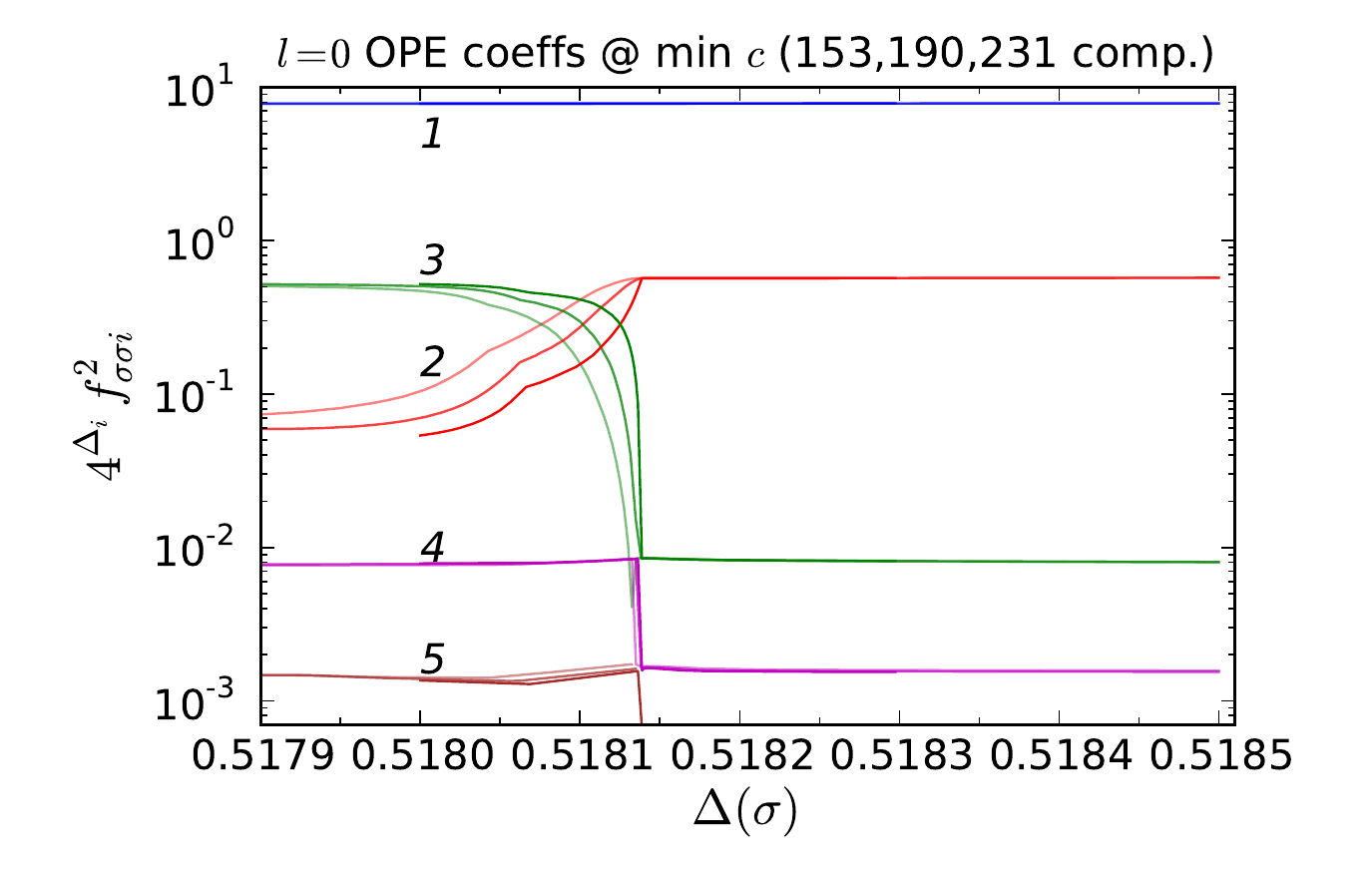}
\caption{Scalar operators with dimensions with $\Delta\le 13$ and their squared OPE coefficients, from the four-point functions realizing the $\CT$ bounds in Figure \ref{sev_ct}: $N=231$ (darker shades), 190 (lighter shades), 153 (lightest shades). }
\label{fig-ct-spin0}
\end{center}
\end{figure}

Let us now take a look at the higher scalars in the four-point function minimizing $\CT$. In Figure \ref{fig-nmax13-spin0}, we show how the scalar dimensions and OPE coefficients vary in the interval $\Delta_\s\in[0.516,0.52]$. These plots correspond to the $\CT$ bound at $N=105$ from Figure \ref{fig-difference}. In Figure \ref{fig-ct-spin0}, we zoom in on the interval close to the 3d Ising point. Here we are using higher $N$ data corresponding to the $\CT$ bounds in Figure \ref{sev_ct}.

\begin{figure}[htbp]
\begin{center}
\includegraphics[scale=1]{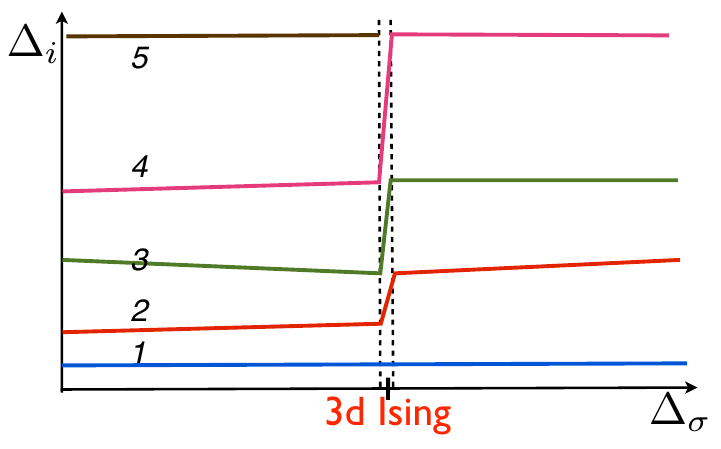}
\includegraphics[scale=1]{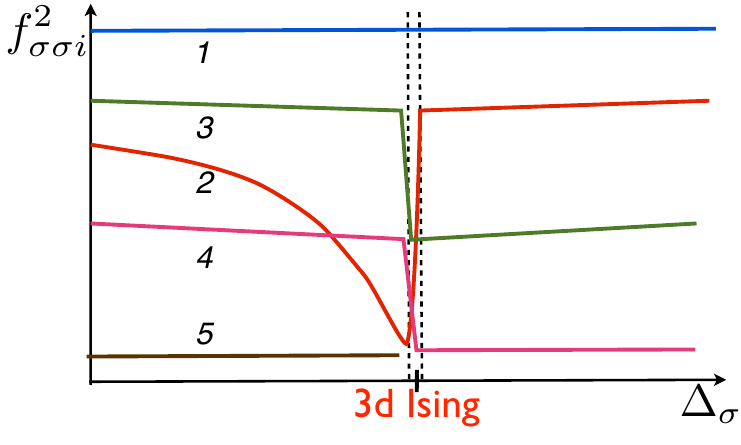}
\caption{
An idealized rendering of the likely scalar operator behavior at a much larger $N$. Between the dashed lines is the transition region. We believe that all operator dimensions should go to some slowly varying limits outside of this ever-shrinking region. In particular the red (no.2) operator which is currently showing larger variation than say the green (no.3) operator should also become slowly varying. All OPE coefficients should also become slowly varying except the red (no.2) OPE coefficient going to zero when approaching the 3d Ising point from below. Apart from the red (no.2) operator disappearing, there is one-to-one correspondence between the spectrum and OPE coefficients above and below the transition region. In the $N\to\infty$ limit, the spectrum and OPE coefficients should vary continuously (although not necessarily smoothly) across the 3d Ising point, with one operator disappearing.}
\label{fig-decoup}
\end{center}
\end{figure}

In these plots we are showing all scalars with $\Delta\le 13$. The blue (no.1) curves correspond to the leading scalar $\eps$. As we have seen in the previous section, its dimension and OPE coefficient vary slowly; at the scale of this plot they look essentially constant. It is then shocking to see how much the higher scalar curves are varying on the same scale. The most salient properties of these curves can be summarized as follows:
\begin{enumerate}
\item The higher scalar operator dimensions and OPE coefficients vary mildly, except at the 3d Ising point, where they experience rapid, near-discontinuous changes. The transition region where these changes happen is shrinking as $N$ is increased (this is especially noticeable in the OPE coefficient plot in Figure \ref{fig-ct-spin0}).

\item The effect of the above changes is just to shift the higher scalar spectrum by one operator dimension up, \emph{\`a la} Hilbert's infinite hotel. 
The operator marked no.2 (red) of dimension $\approx3$ below the 3d Ising point disappears. As a result the higher scalar spectrum and the OPE coefficients above the 3d Ising point (and past the transition region) are the same as below, minus the disappearing operator.

\item The OPE coefficient of the disappearing operator no.2 tends to zero approaching the 3d Ising point. This property may not be obvious from the shown plots, as it's obscured by the presence of the transition region, but we believe that it should become exact in the limit $N\to\infty$. What we mean is that the red (no.2) OPE coefficient curve in Figure \ref{fig-nmax13-spin0} is heading towards zero before entering the transition region, at which point it's shooting up to reconnect with the green (no.3) curve. From Figure \ref{fig-ct-spin0} we can see that the minimum value this curve reaches before shooting up becomes smaller and smaller as $N$ becomes larger. In Figure \ref{fig-decoup} we give an idealized sketch showing what the above plots should look like at a much larger $N$ than considered here.
\end{enumerate}

The first hints of property 1 were noticed in our previous work \cite{ElShowk:2012ht},\footnote{And even earlier in the two-dimensional case in \cite{Rychkov:2011et}.} where we presented an \emph{upper bound} on the dimension of the subleading scalar $\Delta_{\eps'}$, fixing $\Delta_\eps$ to its maximal allowed value, i.e.~the boundary in Figure \ref{fig-3dkinklarge}. That upper bound\footnote{As well as an analogous bound on the dimension of the subleading spin 2 operator.} showed strong variation at the 3d Ising point, just as curve no.2 does. As we now realize, that upper bound was in fact more than simply a bound---it was showing the unique value of $\Delta_{\eps'}$ allowed at the boundary. 

Although Properties 2 and 3 are exhibited here for the first time, in retrospect they are in fact very natural and connected to the very existence of the 3d Ising kink. Imagine approaching the 3d Ising point along the boundary of the $\CT$ lower bound in Figure \ref{sev_ct}. All along the bound we have a solution to crossing. If everything in this solution varies smoothly, we can analytically continue the solution beyond the 3d Ising point. Yet we know that this is impossible, since the bound shows a kink there. So something must happen which invalidates the analytically continued solution. The simplest obstruction is if some $p_{\Delta,\ell}$ hits zero at the 3d Ising point, so that the analytic continuation has negative $p_{\Delta,\ell}$ and violates unitarity. Property 3 means that such an obstruction is indeed encountered when approaching 3d Ising from below, as one $p_{\Delta,\ell=0}$ hits zero. We will see in section \ref{sec:spin2} that an obstruction of the same type occurs when approaching the 3d Ising CFT from above, except that in this case the operator whose OPE coefficient hits zero has $\ell=2$.

Property 2 implies a practical way to extract the 3d Ising CFT spectrum---it is given by the operators dimensions which are present on both sides of the transition region. The operator no.2 (red curve) on the left of the transition region is thus excluded, since it decouples at the 3d Ising point. Looking at Figure \ref{fig-ct-spin0} and applying this rule, we expect that the second $\bZ_2$-even scalar after $\eps$ has dimension $\approx4$, since this dimension is present both below (green no.3) and above (red no.2) the transition region. In the next section we will be able to determine its dimension and the OPE coefficient rather precisely. A third $\bZ_2$-even scalar $\eps''$ appears at dimension $\approx 7$, and a fourth $\eps'''$ at $\approx 10.5$. These estimates (especially the one for $\eps'''$) are preliminary, since the curves corresponding to these operators show non-negligibile variation for values of $N$ in Figure \ref{fig-ct-spin0}. In particular, we prefer not to assign error bars to them.

Although we do not show scalars of $\Delta>13$ in Figure \ref{fig-ct-spin0}, we checked that the same qualitative behavior continues. In particular, the operator dimensions continue shifting by one operator across the transition region, and no additional operator decoupling is observed, so that the red no.2 line below the 3d Ising point is the only one which decouples. The higher in dimension one goes, the stronger is variation with $N$. This loss of sensitivity is expected in view of the exponential decoupling of high dimension operators in a four-point function of low-dimension operators \cite{Pappadopulo:2012jk}. A natural way to boost sensitivity to high exchanged dimensions might be to raise the dimension of the external operators, by considering e.g.~the four-point function $\langle\eps\eps\eps\eps\rangle$. It would be important to try this in the future. 

\subsection{The Second $\bZ_2$-Even Scalar $\eps'$}
\label{sec:eps'}

\begin{figure}[htbp]
\begin{center}
\includegraphics[scale=0.6]{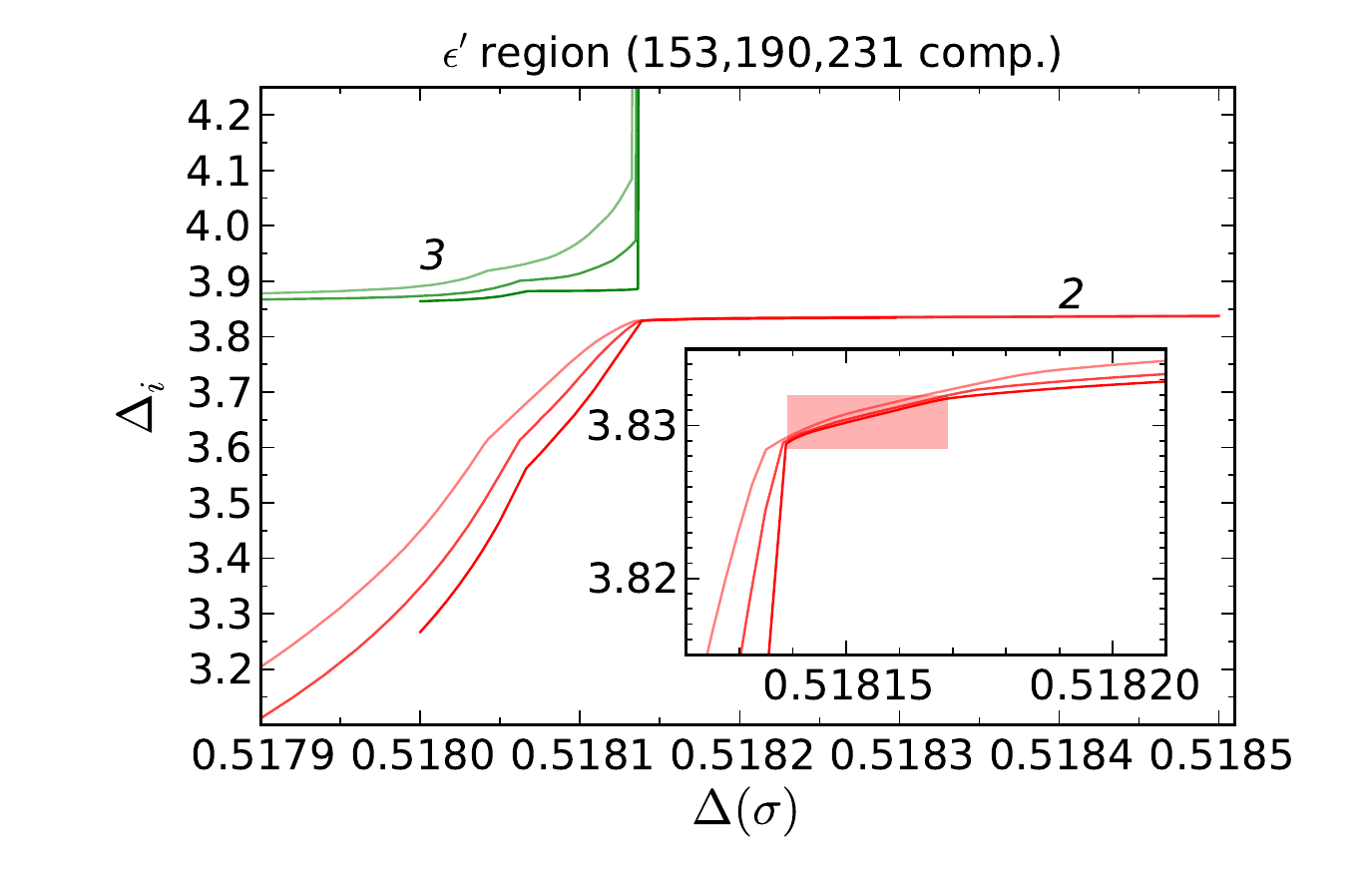}
\hspace{-20pt}
\includegraphics[scale=0.6]{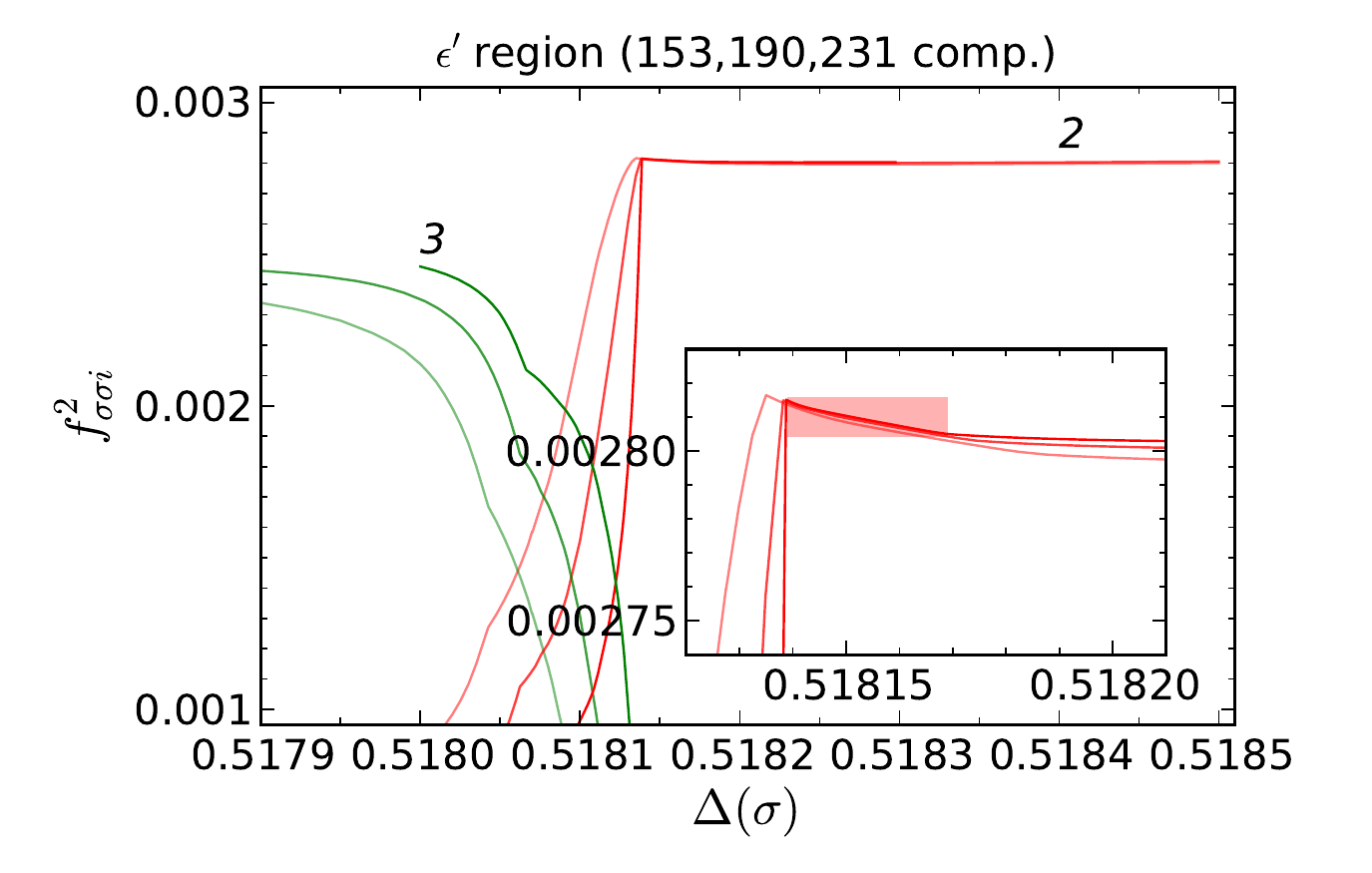}
\caption{The zoom on the part of the scalar spectrum relevant for the extraction of the $\eps'$ parameters. The operator colors and numbering are the same as in Figure \ref{fig-ct-spin0}. Note that compared to Figure \ref{fig-ct-spin0}, the OPE coefficients are here plotted without the $4^\Delta$ factor. }
\label{fig-ct-eps1}
\end{center}
\end{figure}

Having discussed the general properties of the 3d Ising scalar spectrum in the previous section, here we will focus on the second $\bZ_2$-even scalar $\eps'$, which is the operator of dimension $\approx4$ present both below (green no.3) and above (red no.2) the transition region in Figure \ref{fig-ct-spin0}. In Figure \ref{fig-ct-eps1} we give a vertical zoom of the part of Figure \ref{fig-ct-spin0} where the green no.3 and the red no.2 lines approach each other. 

We fixed the horizontal extension of the pink rectangles in these plots equal to the 3d Ising $\Delta_\sigma$ range previously determined in sections \ref{sec:ds} and \ref{sec:de}. We see that this range falls on the part of the red no.2 plateau which is quite converged. The vertical extension of these rectangles then gives our best estimates of the $\eps'$ dimension and the OPE coefficient:
\beq
\Delta_{\eps'}=3.8303(18),\qquad f^2_{\sigma\sigma\eps'}=0.002810(6).
\eeq

In contrast to the red no.2 plateau, the green no.3 curves to the left of the 3d Ising point are still changing significantly with $N$. As explained in the previous section, we expect that in the $N\to\infty$ limit the green no.2 curves will reach a limiting plateau continuously connecting to the red no.2 plateau. Although this has not happened yet on the scale of Figure \ref{fig-ct-eps1}, the tendency is clearly there.

\subsection{Spin 2 Operators}
\label{sec:spin2}

\begin{figure}[htbp]
\begin{center}
\includegraphics[scale=0.6]{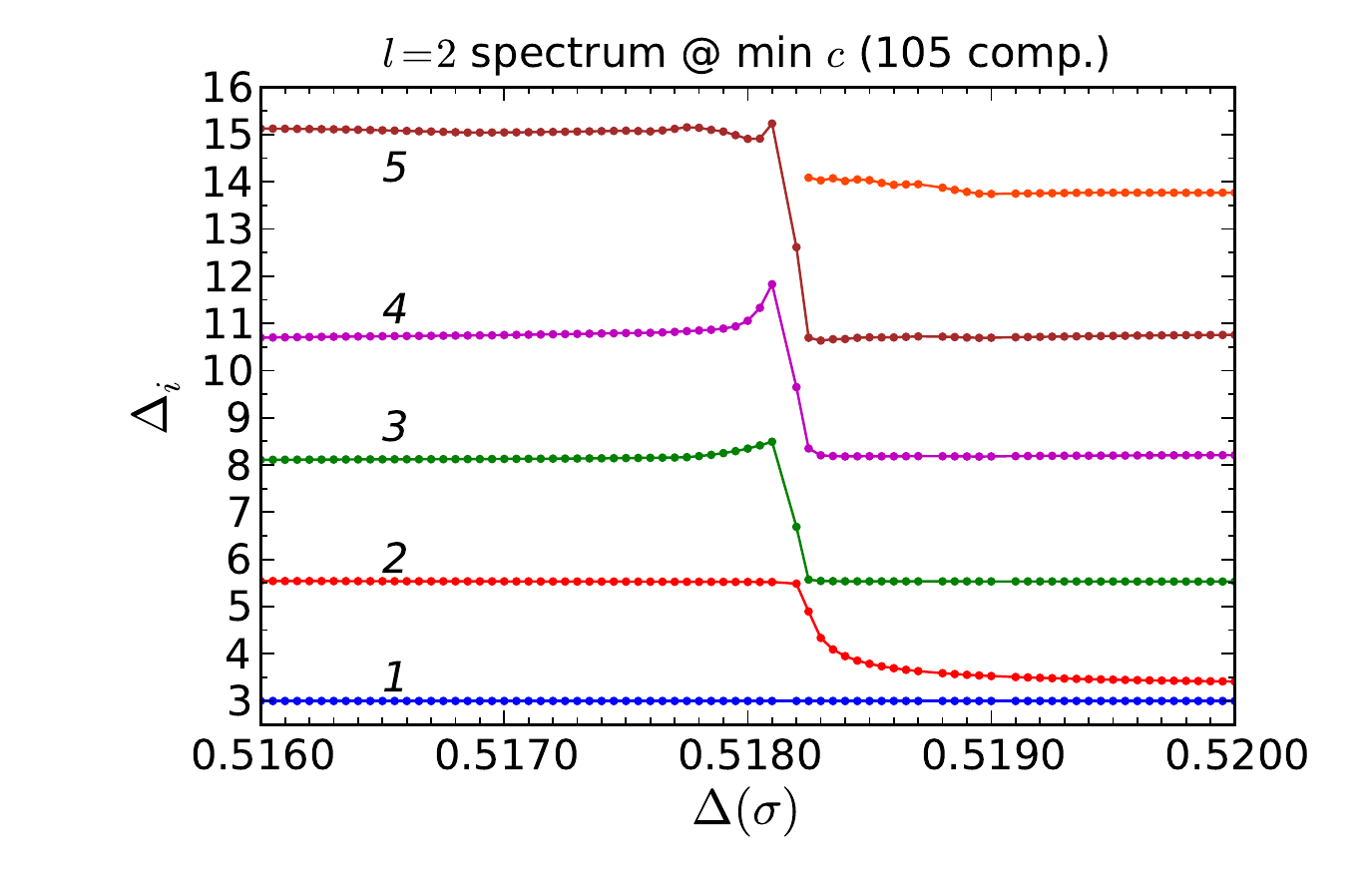}
\hspace{-20pt}
\includegraphics[scale=0.6]{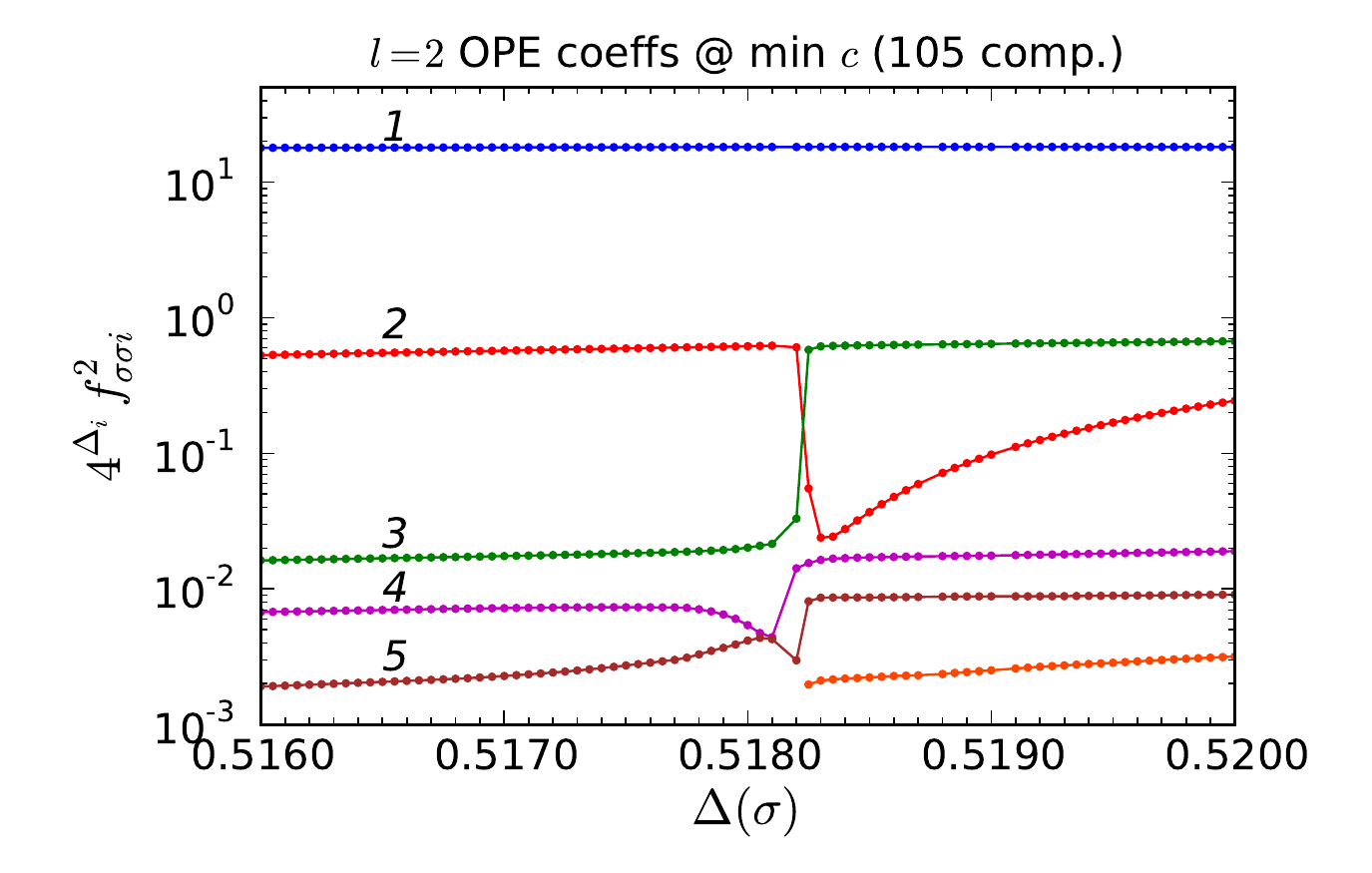}
\caption{Spin 2 operators of dimension $\Delta<16$ and their OPE coefficients, at $N=105$.}
\label{fig-nmax13-spin2}
\end{center}
\end{figure}

\begin{figure}[htbp]
\begin{center}
\includegraphics[scale=0.6]{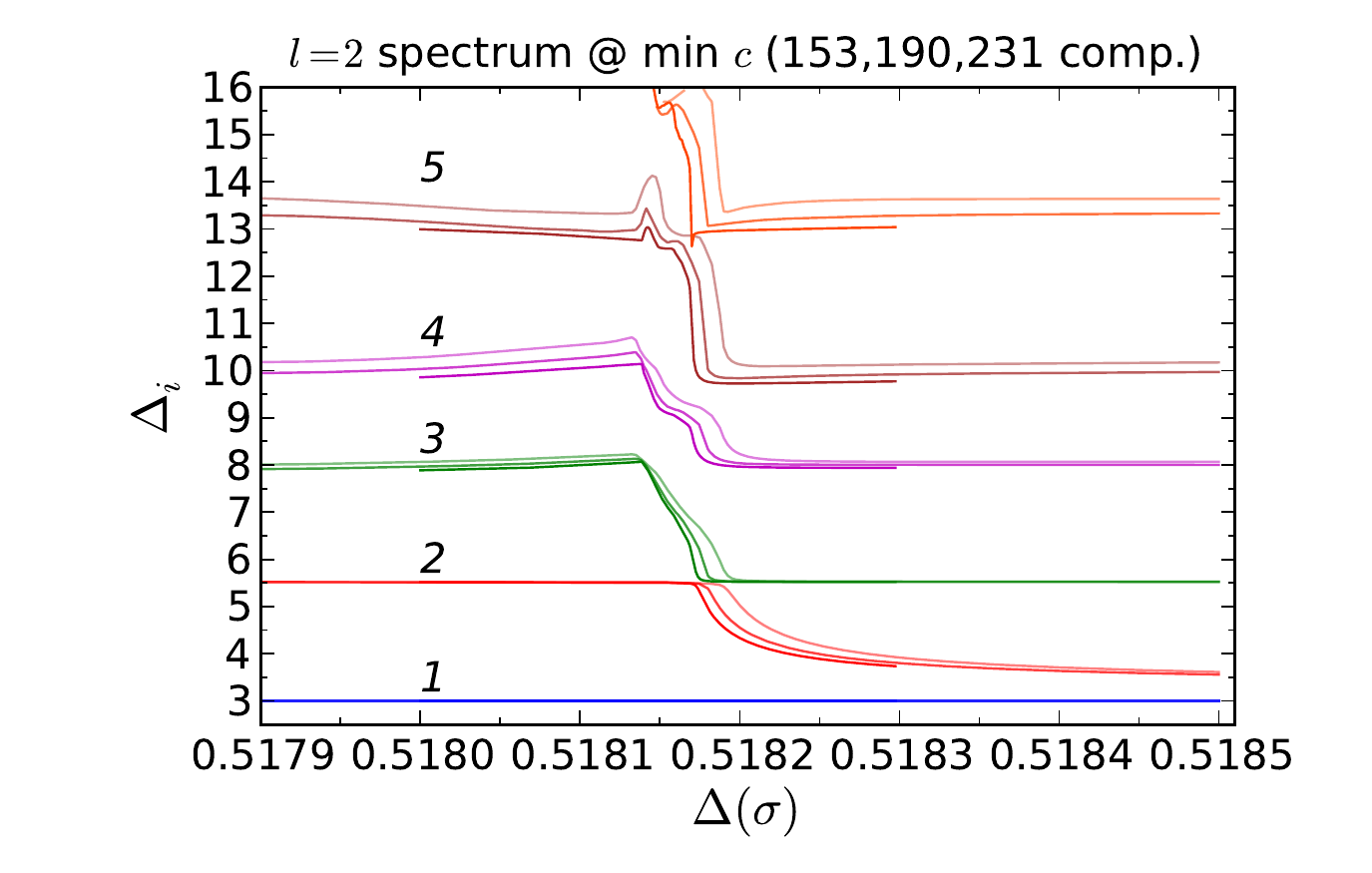}
\hspace{-20pt}
\includegraphics[scale=0.6]{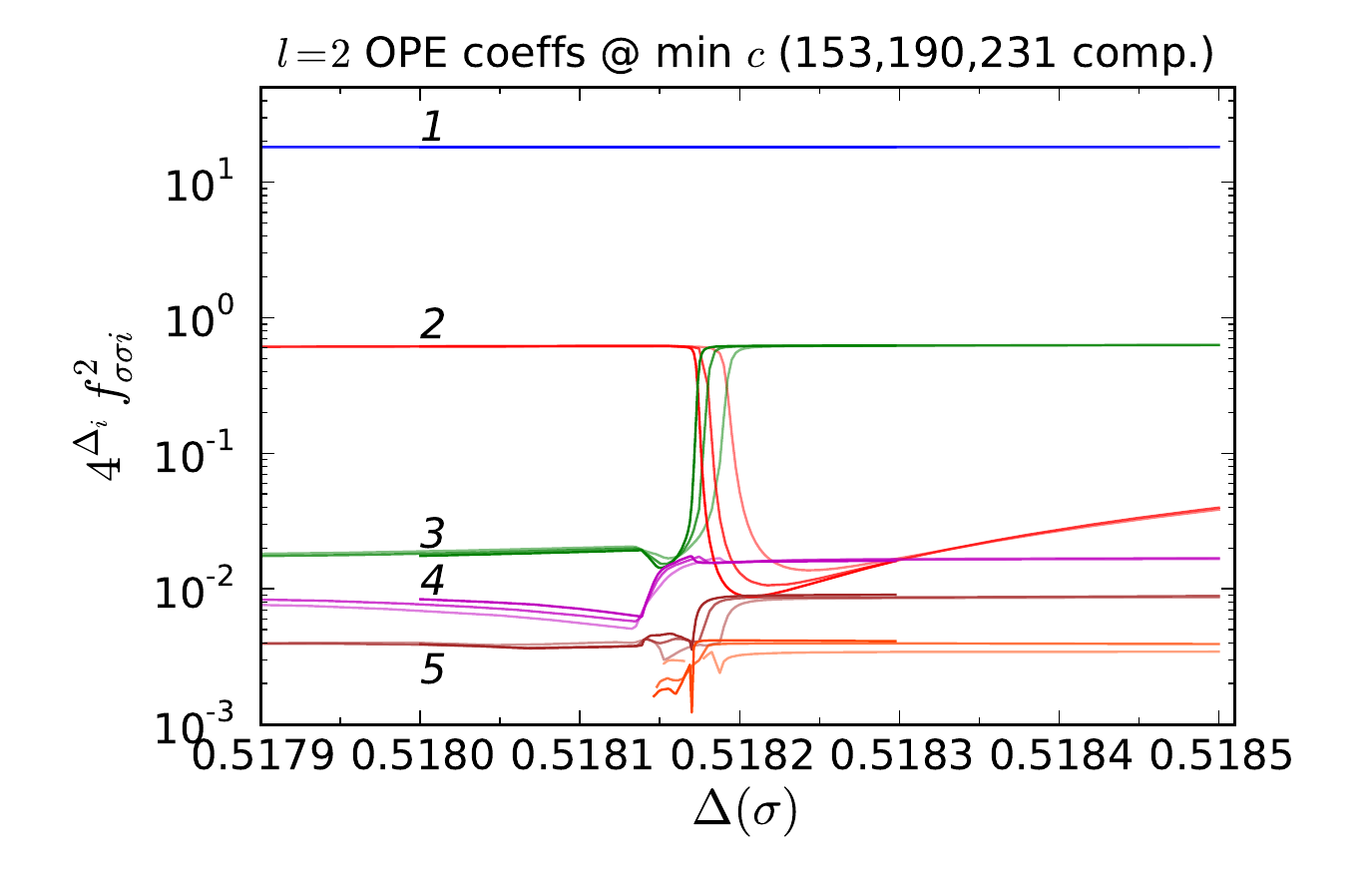}
\caption{Spin 2 operators of dimension $\Delta<16$ and their OPE coefficients, at $N=231,190,153$.}
\label{fig-ct-spin2}
\end{center}
\end{figure}

In this section we analogously consider the $\ell=2$ operators in the $\sigma\times\sigma$ OPE. In Figure \ref{fig-nmax13-spin2} we show the $\ell=2$ spectrum and OPE coefficients at $N=105$ in a wider range, and in Figure \ref{fig-ct-spin2} we zoom in at the transition region at higher $N$.

We see from these plots that there are many similarities in the general features of $\ell=0$ and $\ell=2$ spectra. The lowest operator in the spectrum is now the stress tensor, its dimension is 3 and independent of $\Delta_\sigma$, Its OPE coefficient varies slightly, according to the $\CT$-minimization bounds shown above, but on this scale this variation is not noticeable. 

For the higher spectrum, the analogues of Properties 1,2,3 from section \ref{sec:spin0} are true, with one difference: 
the red no.2 operator now decouples approaching the 3d Ising point from the right instead of from the left. The fact that its OPE coefficient tends to zero in this limit (and for $N\to\infty$) is even more evident here than it was for its cousin in section \ref{sec:spin0}. As promised, the existence of this decoupling operator provides an obstruction for the analytic continuation across the kink, approaching it from above.

Leaving out the decoupling operator and interpolating the plateaux above and below the transition region in Figure \ref{fig-ct-spin2}, we get an approximate $\ell=2$ spectrum in the 3d Ising CFT. Apart from the stress tensor, it should contain operators of dimensions $\approx 5.5,8,10,13,\ldots$

\begin{figure}[htbp]
\begin{center}
\includegraphics[scale=0.6]{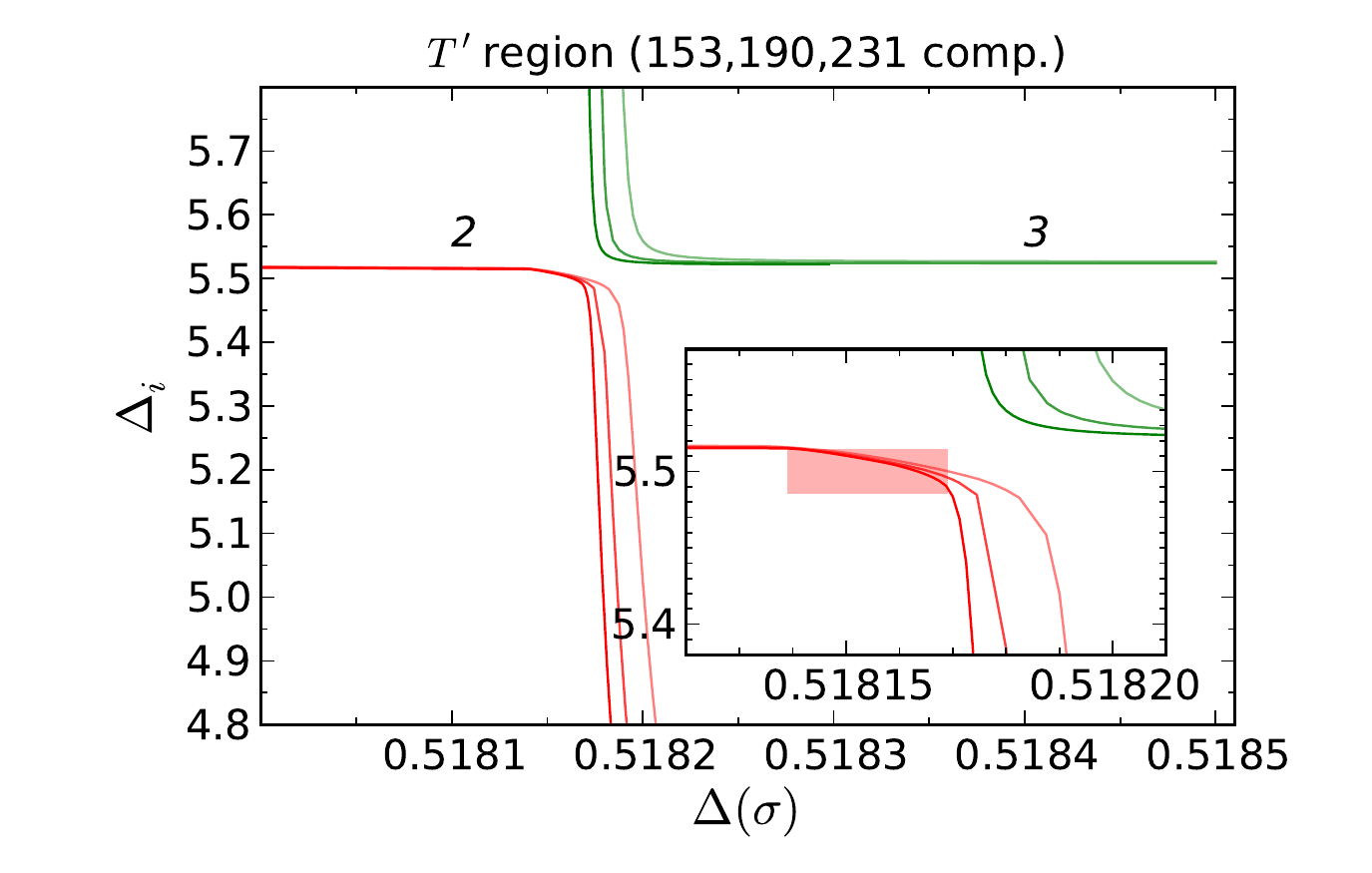}
\hspace{-20pt}
\includegraphics[scale=0.6]{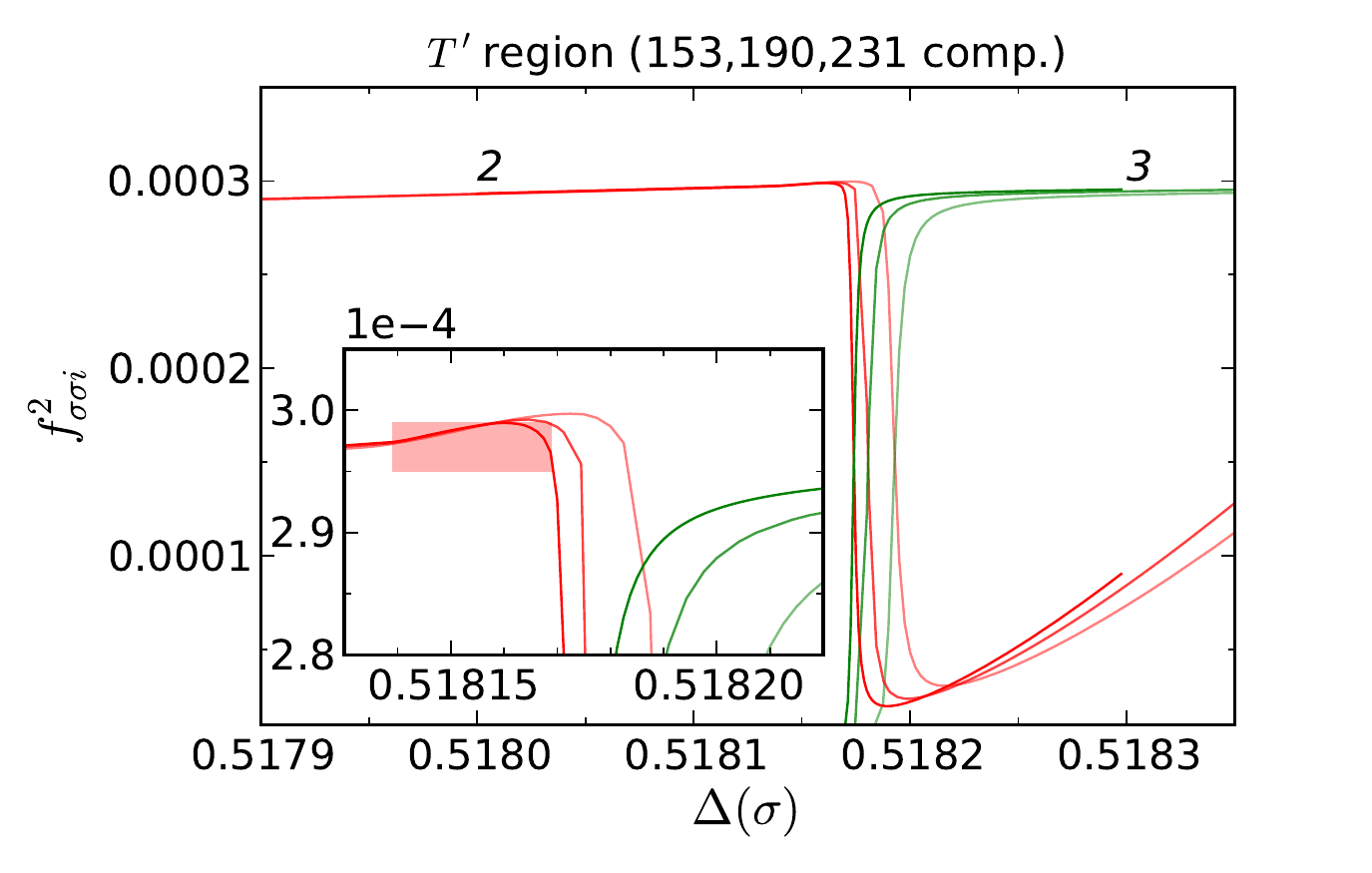}
\caption{The zoom on the region relevant for the extraction of the $T'$ operator parameters. The operator colors and numbering are the same as in Figure \ref{fig-ct-spin2}. Note that the OPE coefficients are plotted without the $4^\Delta$ factor. }
\label{fig-ct-T1}
\end{center}
\end{figure}

We will now determine the parameters of the first of these subleading operators, call it $T'$, similarly to how we fixed $\eps'$ in section \ref{sec:eps'}. In Figure \ref{fig-ct-T1} we zoom in on the region of near level-crossing between the red no.2 and the green no.3 curves. The horizontal extent of the pink rectangle coincides with the 3d Ising $\Delta_\sigma$ confidence interval. We are again lucky in that this interval falls on the part of the red plateau which looks reasonably converged. So we determine the vertical extension of the rectangles by fitting in the $N=231$ curves, and obtain the following estimates:
\beq
\Delta_{T'}=5.500(15),\qquad f^2_{\sigma\sigma T'}=2.97(2)\times 10^{-4}\,.
\eeq

From Figures \ref{fig-ct-eps1} and \ref{fig-ct-T1} it can be seen clearly that the left (right) end of the $\Delta_\sigma$ confidence interval coincides with the location of the operator jumps in the $\ell=0$ ($\ell=2$) sector. Thus looking at these jumps gives an equivalent method to localize the 3d Ising point.

\subsection{Higher spins}
\label{sec:higher}
The $\sigma\times\sigma$ OPE is also expected to contain infinitely many primary operators of spin $\ell=4,6,8,\ldots$ It would be of course interesting to learn something about their dimensions from the bootstrap. Particularly interesting are the operators of the smallest dimension for each spin, let's call them $C_\ell$. Being an interacting CFT, the critical Ising model cannot contain conserved higher spin currents \cite{Maldacena:2011jn}. Thus operators $C_\ell$ should have a positive anomalous dimension:
\beq
\Delta(C_\ell)=\ell+d-2+\gamma_\ell,\qquad \gamma_\ell>0\,.
\eeq
The dimension of $C_4$ is known with some precision \cite{omega_NR}:
\beq
\label{eq:C4}
\Delta(C_4)=5.0208(12)\,.
\eeq
Dimensions of higher $C_\ell$ are not accurately known, although at the Wilson-Fisher fixed point in $d=4-\eps$ \cite{Wilson:1973jj}
\beq
\gamma_\ell=\frac{\eps^2}{54}\left(1-\frac6{\ell(\ell+1)}\right)+O(\eps^3)\,.
\eeq

There are also two general results about the sequence of $\gamma_\ell$, which are supposed to be valid in any CFT. Firstly, the large $\ell$ limit of $\gamma_\ell$ has to be equal twice the anomalous dimension of $\sigma$ \cite{Fitzpatrick:2012yx,Komargodski:2012ek}:\footnote{This is sometimes called the Callan-Gross relation since it was first noticed in perturbation theory in \cite{Callan:1973pu}.}
\beq
\label{eq:CG}
\lim_{\ell\to\infty}\gamma_\ell= 2\gamma_\sigma \equiv 2\Delta_\sigma-1\,.
\eeq
The asymptotic rate of approach to the limit is also known, see \cite{Fitzpatrick:2012yx,Komargodski:2012ek}.

Secondly, we have ``Nachtmann's theorem" \cite{Nachtmann:1973mr}, which says that the sequence $\gamma_\ell$ is monotonically increasing and upward convex. This remarkable result is on a somewhat less solid footing than \reef{eq:CG}. As was emphasized in its recent discussion in \cite{Komargodski:2012ek}, its derivation uses the polynomial boundedness of a certain scattering amplitude, which is still conjectural (although plausible) at present. Depending on the degree of the polynomial boundedness, Nachtmann's theorem may hold not for all $\ell$ but in the range $\ell\ge \ell_0$.

It would be therefore very interesting to determine $\gamma_\ell$ in the 3d Ising CFT using the bootstrap, and see if they satisfy the above two general properties. 

In principle, the determination of $\gamma_\ell$ for $\ell\ge4$ using the $c$-minimization is as straightforward as the determinations of $\ell=0,2$ operator dimensions discussed in the previous sections. The extremal $c$-minimization spectra that we obtain do contain higher spin operators, and we can easily identify the lowest operator for each spin.

In practice, however, this procedure for $\ell\ge4$ turns out to be numerically somewhat less stable than for $\ell=0,2$. This must be somehow related to the fact that the anomalous dimensions $\gamma_\ell$ are all expected to be small yet nonzero. Since conformal blocks of spin $\ell\ne0$ operators are continuous in the limit of $\Delta$ approaching the unitarity bound, it's not easy for our algorithm to distinguish an exactly conserved operator from one with a small anomalous dimension. Moreover, in this case, $d=2$ does not provide any guidance as there the higher spin operators \textit{are} conserved and our algorithm has no difficulty in reconstructing the $\ell > 2$ spectrum (see section \ref{sec:2dspec} below).

In spite of these numerical problems, our calculations do show that the higher spin currents acquire positive anomalous dimensions. We managed to extract anomalous dimensions up to spin $\ell\simeq40$. Although precision needs to be improved, the extracted values are largely consistent with the Callan-Gross relation \reef{eq:CG}. Nachtmann's theorem also seems to hold, in the full range of $\ell\ge4$. Our value of $\Delta(C_4)$ is roughly consistent with \reef{eq:C4}.

These preliminary results are encouraging, and we give them here since they may stimulate further work on Nachtmann's theorem. Nevertheless, we prefer to postpone detailed plots and precision determinations until we have the higher spin sector under better control. 

\section{Comparison to Results by Other Techniques}
\label{sec:comp}

In the previous section, we used the conformal bootstrap method together with our $\CT$-minimization conjecture to determine several operator dimensions and OPE coefficients of the 3d Ising CFT. We will now compare our results to prior studies of the 3d Ising model at criticality, by other techniques. 

\noindent\textit{Operator dimensions and critical exponents}

One well studied class of universal quantities characterizing the critical point are the critical exponents.\footnote{Another such class are the amplitude ratios. These are related to IR-dominated properties of RG flows produced when the theory is perturbed away from the critical point. Unlike the critical exponents, the amplitude ratios cannot be easily computed in terms of the CFT parameters.} They have simple expressions in terms of the CFT operator dimensions. In particular, the well-known critical exponents $\eta$, $\nu$, $\omega$ can be expressed via 
the dimensions of $\sigma$, $\eps$, and $\eps'$ from the formulas:
\beq
\Delta_\sigma=1/2+\eta/2,\qquad \Delta_\eps=3-1/\nu, \qquad \Delta_{\eps'}=3+\omega\,.
\label{eq:dim-exp}
\eeq

In Table \ref{tab-prior} we quote some notable prior determinations of these exponents. The first two determinations from \cite{Guida:1998bx} are by field theoretic (FT) techniques (the $\eps$-expansion and fixed-dimension expansion). For $\eta$ and $\nu$, these are an order of magnitude less precise than the best available determination from the lattice: the Monte Carlo simulations (MC) and the high-temperature expansion (HT). It is again the MC which provides the best estimate of $\omega$, followed by FT, and HT.
\begin{table}[htbp]
\centering
\begin{tabular}{cccccc}
ref & year & Method & $\nu$ & $\eta$ & $\omega$\\
\hline
\cite{Guida:1998bx} & 1998 & $\eps$-exp & 0.63050(250) & 0.03650(500) & 0.814(18)\\
\cite{Guida:1998bx} & 1998 & 3D exp & 0.63040(130) & 0.03350(250) & 0.799(11)\\
\cite{Campostrini} & 2002 & HT & 0.63012(16) & 0.03639(15) & 0.825(50)\\
\cite{Deng} & 2003 & MC & 0.63020(12) & 0.03680(20) & 0.821(5)\\
\cite{Hasenbusch2010} & 2010 & MC & 0.63002(10)  & 0.03627(10)   & 0.832(6)\\
\hline
&this work && 0.62999(5) & 0.03631(3) & 0.8303(18)
\end{tabular}
\caption{The most precise prior determinations of the leading critical exponents, compared to our results.}
\label{tab-prior}
\end{table}

In the same table we give the values of $\eta$, $\nu$, $\omega$ obtained via \reef{eq:dim-exp} using the values of operator dimensions from the previous section. Comparing our values with the recent precise MC determinations by Hasenbusch \cite{Hasenbusch2010}, we see that the agreement is very good, our results being factor 2 - 3 more precise. The agreement with the older MC results \cite{Deng} is not as good, with about 2$\sigma$ tension in every exponent. Our results clearly favor \cite{Hasenbusch2010} over \cite{Deng}.

\noindent\textit{Subleading correction-to-scaling exponents}

Another method for computing the critical exponents is the functional renormalization group (FRG), in its various incarnations. For the leading exponents $\eta$, $\nu$, $\omega$, it is not currently competitive with the methods listed in Table \ref{tab-prior} (see \cite{Canet:2003qd,Litim:2010tt} for state-of-the-art studies). On the other hand, an advantage of this method is that it can compute the subleading correction-to-scaling exponents. Here we will be particularly interested in the subleading exponents $\omega_{2,3}$ related to the dimensions of the higher $\bZ_2$-even scalar operators $\eps''$ and $\eps'''$:
\beq
\Delta_{\eps''}=3+\omega_2,\qquad \Delta_{\eps'''}=3+\omega_3.
\eeq 

In Table \ref{tab-frg} we collect several available FRG determinations of $\omega_{2,3}$ and compare them with our results. 
The oldest calculation is by the ``scaling field" method \cite{scaling-field}, which performs an RG flow in a local operator basis truncated to finitely many operators and optimized to minimize the truncation effects. The largest bases had relatively few ($\le 13$) operators, but unlike the calculations discussed below they included some operators with two and four derivatives. This method gives the leading critical exponents in agreement, within cited errors, with the more precise determinations from Table \ref{tab-prior}. On the other hand, their value of $\omega_2$ is much smaller than our estimate $\omega_2\approx 4$.

Then there is a group of more recent calculations which truncate the flowing Lagrangian to the standard kinetic term plus terms with an arbitrary number of $\phi$'s but no derivatives; these are marked as $\del^0$ in the table. By the nature of such a truncation, all these calculations give $\eta=0$. The other exponents vary significantly depending on the type of cutoff used to regularize the RG flow. The background field cutoff gives $\nu$ and $\omega$ closest to our determinations; it also gives the values of $\omega_{2,3}$ closest to our estimates.
\begin{table}
\centering
\begin{tabular}{cclll|ccc}
ref & year & FRG variant & $\omega_2$ & $\omega_3$ & $\nu$ & $\eta$ & $\omega$\\
\hline
\cite{scaling-field} & 1984 & scaling field (13 ops) & 1.67(11) & --- & 0.626(9) & 0.040(7) & 0.855(70)\\
\cite{Comellas:1997tf} & 1997 & $\del^0$, sharp cutoff & 2.8384 & 5.1842 & 0.6895 & 0 & 0.5952\\
\cite{Litim:2002cf} & 2002 &$\del^0$, optimized cutoff & 3.180 & 5.912 & 0.6496 & 0 & 0.6557 \\
\cite{Litim:2003kf} & 2003 & $\del^0$, quartic cutoff & 3.048 & 5.63  &0.6604 & 0  &0.6285\\
 & & $\del^0$, background field                                & 3.6845 & 7.038 & 0.6260 & 0 & 0.7622 \\
\hline
&this work && $\approx 4$ & $\approx 7.5$ & 0.62999(5) & 0.03631(3) & 0.8303(18)\end{tabular}
\caption{Functional renormalization group determinations of the subleading correction-to-scaling exponents, compared to our results. For completeness and ease of comparison, we cite also the results for the leading exponents obtained in the same studies.}
\label{tab-frg}
\end{table}

In view of the remaining discrepancies, it would be interesting to upgrade the recent FRG calculations of the subleading exponents by increasing the derivative truncation order to $\del^2$ and $\del^4$. While such calculations have been performed for the leading exponents and noticed to improve their estimates \cite{Canet:2003qd,Litim:2010tt}, we are not aware of any results for $\omega_{2,3}$.

\noindent\textit{OPE coefficient $f_{\s\s\eps}$ }

Hardly anything is known about the OPE coefficients in the 3d Ising CFT. Some universal couplings were studied by Henkel \cite{Henkel:1986ia}, but as we will now explain they are not equal to the flat space OPE coefficients. He studied a 2d Hamiltonian with a quantum critical point known to belong to the 3d Ising universality class. The Hamiltonian was diagonalized on a family of square $N\times N$ lattices with periodic boundary conditions ($N\le5$). Continuum limit quantities were extracted by applying finite-size scaling analysis to matrix elements of lattice versions of $\eps$ and $\sigma$ operators sandwiched between the vacuum and the lowest nontrivial $\bZ_2$-even and $\bZ_2$-odd eigenstates:\footnote{Henkel refers to $|-\rangle$ and $|+\rangle$ as $|\sigma\rangle$ and $|\eps\rangle$, but as will see below this notation may lead to a confusion in his geometry.}
\beq
\langle 0|\eps|+\rangle\quad\text{and}\quad \langle 0|\sigma|-\rangle\,.
\eeq
This gave reasonable, although not very precise, estimates for the continuum operator dimensions: $\Delta_\s=0.515(9)$ and $\Delta_\e=1.42(2)$. 
He then considered universal quantities related to the properly normalized matrix elements:
\beq
C_{\sigma\sigma\eps}\propto \langle -|\sigma|+\rangle,\qquad C_{\sigma\eps\sigma}\propto \langle -|\eps|-\rangle\,.
\eeq
Are these in any way related to $f_{\sigma\sigma\eps}$? In 2d, the answer would be yes, but in 3d it's no, for the following reason. Using periodic square lattices means that the critical 3d theory is put on $T^2\times \bbR$. Since this geometry is not Weyl-equivalent to flat 3d space, there is no way to relate the $C$'s to the flat space OPE coefficients. A priori, the two $C$'s do not even have to be equal; and in fact \cite{Henkel:1986ia} finds unequal values:
\beq
C_{\sigma\sigma\eps}\approx 0.85,\qquad C_{\sigma\eps\sigma}\approx 1.10\,.
\eeq 

In order to measure $f_{\sigma\sigma\eps}$, one would have to put the theory on $S^2\times \bbR$. Via the radial quantization state-operator correspondence, the states on $S^2$ are obtained by the flat-space operators inserted at the origin and at infinity.
The lowest states $|-\rangle$ and $|+\rangle$ can then be identified with $|\sigma\rangle$ and $|\eps\rangle$, and we would have 
\beq
C_{\sigma\sigma\eps}=C_{\sigma\eps\sigma}=f_{\sigma\sigma\eps}\quad(\text{on $S^2\times \bbR$})\,.
\eeq

In connection with this we would like to point out a recent paper \cite{Brower:2012vg} which measures $\Delta_\s$ via a MC simulation of the statistical mechanics 3d Ising model in the $S^2\times \bbR$ geometry, approximating $S^2$ by a discretized icosahedron.\footnote{An earlier reference \cite{Weigel2000721} measured $\Delta_\sigma$ and $\Delta_\eps$ approximating $S^2$ by a discretized cube.} It would be interesting to measure $f_{\sigma\sigma\eps}$ by this or another technique, so that we would have something to compare to.

\noindent\textit{Central charge $\CT$ }

The prior knowledge of $\CT$ can be summarized by the $\eps$-expansion formula \cite{Hathrell:1981zb,Jack:1983sk,Cappelli:1990yc,Petkou:1994ad}:
\beq
\CT/\CT_{\text{free}}=1-\frac{5}{324}\eps^2+\ldots\,.
\eeq
Preliminary bootstrap determinations of $\CT$ in 3d were given in \cite{ElShowk:2012ht,El-Showk:2013nia}. Moreover, Ref.~\cite{El-Showk:2013nia} checked the above formula via the conformal bootstrap analysis in a fractional number of spacetime dimensions $d=4-\eps$. Good agreement was observed for $\eps\lesssim 0.3$, while in 3d ($\eps=1$) corrections due to the unknown $O(\eps^3)$ and higher terms must be significant. 

It would be interesting if the $10^{-5}$ precision of our $\CT$ determination in section \ref{sec:ds} could be approached or matched by any other technique.

\section{2d Checks}
\label{sec:2dchecks}

So far in this paper we were applying the bootstrap methods and the $\CT$-minimization conjecture to extract information about the 3d Ising CFT. 
Since it is known that the 3d Ising CFT is a member of the family of the Wilson-Fisher CFTs interpolating between 2 and 4 dimensions, it would be interesting to apply the same techniques in $2\le d<4$ to learn more about this family, which so far has been studied mostly using the $\eps$-expansion
\cite{LeGuillou:1987ph}.\footnote{See also \cite{Codello:2012sc} for a study based on the functional renormalization group.} First steps in this direction were made in our paper \cite{El-Showk:2013nia}, where we used the position of the $\Delta_\eps$-maximization kink as a function of $2\le d<4$ to extract $\Delta_\sigma$, $\Delta_\eps$, and (from the extremal solution) $\CT$.

Postponing a more detailed study of fractional dimensions to the future, in this section we will discuss in some detail the other integer case, $d=2$. This case has already played a very stimulating role in the development of our techniques. It was first observed in \cite{Rychkov:2009ij} that the 2d $\Delta_\eps$-maximization plot has a kink whose position could be identified with the exactly known operator dimensions of the 2d Ising CFT $\Delta_\s=1/8$, $\Delta_\eps=1$. The sharp variation of the subleading scalar dimension near the kink was also first noticed in 2d \cite{Rychkov:2011et}. Both these facts turned out to have close analogues in 3d.


In this section, we will go beyond the previous 2d studies and apply our method to extract the low-lying $\sigma\times\sigma$ spectrum and OPE coefficients in the neighbourhood of the 2d Ising CFT. At the 2d Ising point we will be able to improve on the accuracy of previous results \cite{ElShowk:2012hu} and determine $\Delta_\sigma$ itself from the position of the kink. In a region to the right of the kink our method closely reconstructs a certain family of solutions to crossing symmetry related to the 2d minimal models. In particular, the spectrum of quasiprimaries will respect the Virasoro symmetry, even though our bootstrap equations have only global $SL(2,\mathbb{C})$ symmetry built in from the start.

Apart from checking the $\CT$-minimization method against the exact solution, the 2d study will also shed some light on the operator decoupling phenomenon observed in 3d when approaching the Ising point. As we will see, an analogous decoupling happens in 2d. While in 3d the reasons behind the decoupling remain unknown, in 2d it can be given a natural explanation, in terms of null states present in the 2d Ising CFT.

\subsection{Spectrum Extraction at Fixed $\Delta_\sigma=1/8$}
\label{sec:2dspec}

In section \ref{sec:3dspec}, spectrum determination in the 3d Ising CFT consisted of two steps. First, a range for $\Delta_\s^{\text{3d Ising}}$ was identified near the minimum of the $\CT(\Delta_\s)$ curve. Second, spectra corresponding to the solutions in this range were looked at and interpreted under the assumptions of convergence, operator rearrangement and decoupling. 

In 2d,  $\Delta_\s$ is known exactly, so it is possible to eliminate the first step and study the second step in isolation \cite{ElShowk:2012hu}. We fix $\Delta_\s=1/8$ and maximize $\Delta_\eps$, which gives a value very close to the exact value $\Delta_\eps=1$.\footnote{$\Delta^{\text{max}}_\eps=1.000003$ for $N=60$ \cite{ElShowk:2012hu}.} We then extract the low-lying spectrum and the OPE coefficients corresponding to the maximal $\Delta_\eps$. 
\begin{figure}[htbp]
\begin{center}
\includegraphics[scale=0.6]{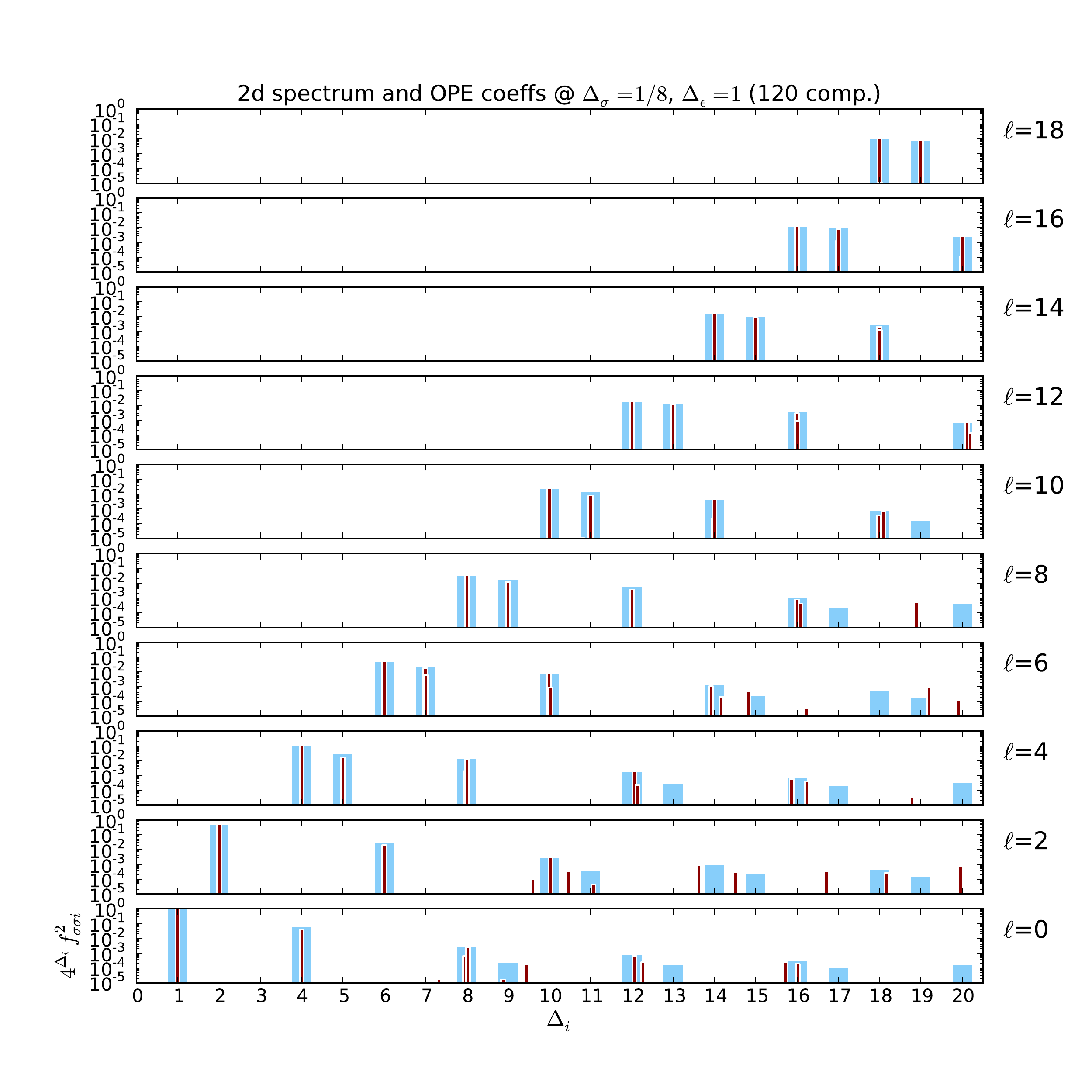}
\caption{The extremal 2d spectrum extracted at $\Delta_\s=1/8$. This plot was produced using our code at $N=120$. Such results were first obtained in \cite{ElShowk:2012hu} at $N=60$, using the dual method.}
\label{fig-fixed-ds-de}
\end{center}
\end{figure}

The results (thin red bars) are shown and compared to the exact 2d Ising data (wide blue bars) in Figure \ref{fig-fixed-ds-de}, for spins $\ell\le 18$ and dimensions $\Delta\le 20$. Let us first discuss the exact data. The blue bars are centered at the positions of the $\bZ_2$-even quasiprimaries $\mathcal{O}_i\in \sigma\times\sigma$ in the 2d Ising CFT, the height of the bars given by the squared OPE coefficients.\footnote{The OPE coefficients were obtained by expanding the exactly known four-point function $\langle\sigma\sigma\sigma\sigma\rangle$ into $SL(2,\mathbb{C})$ conformal blocks. When there are several quasiprimaries with the same dimension and spin, they are lumped together by summing their OPE coefficients squared.} All these operators are obtained from $\mathds{1}$ and $\eps$ by acting with higher Virasoro generators $L_{-n}$, $n\ge 2$, and keeping track of the null state conditions specific to the 2d Ising CFT. 
For example, the leading twist trajectory $\Delta=\ell$ (the highest diagonal) are the Virasoro descendants of the identity. Four and eight units above it lie the operators obtained by acting with $L_{-2}\bar L_{-2}$ and $L_{-4}\bar L_{-4}$. Twist one operators are the $\eps$ and its Virasoro descendants; this ``$\eps$-trajectory" has a gap for $\ell=2$ because the $\eps=\phi_{1,2}$ is degenerate on level 2, and the state in question is null. For the same reason, the $\eps$-trajectory has daughters
at $8,12,16\ldots$ but not 4 units higher.

Now consider the red bars, whose position and height show the extremal spectrum and the OPE coefficients extracted by our algorithm at $N=120$.
The agreement for the first three operator trajectories is almost too good. The fourth trajectory can also be divined, although not as cleanly, at the positions where the numerical spectrum has several nearby operators. For even higher twists the sensitivity of our algorithm becomes insufficient, and the numerical spectrum no longer correlates with the exact one. (When $N$ is increased, the sensitivity loss threshold is pushed to larger and larger twists.)

One important lesson from this plot is that for an operator to be numerically extractable, it better have a sizable OPE coefficient: the four extracted trajectories consist of the operators with the largest coefficients. To reach this conclusion it's important to normalize OPE coefficients using the natural normalization of \cite{Pappadopulo:2012jk}.

\subsection{$\CT$-Minimization}

In the previous section we set $\Delta_\s$ to the precisely known 2d Ising CFT value $\frac 18$. In this section we will instead mimic more closely our 3d study, performing a $\CT$-minimization scan in the neighborhood of the 2d Ising point. 

\begin{figure}[htbp]
\begin{center}
\includegraphics[scale=0.6]{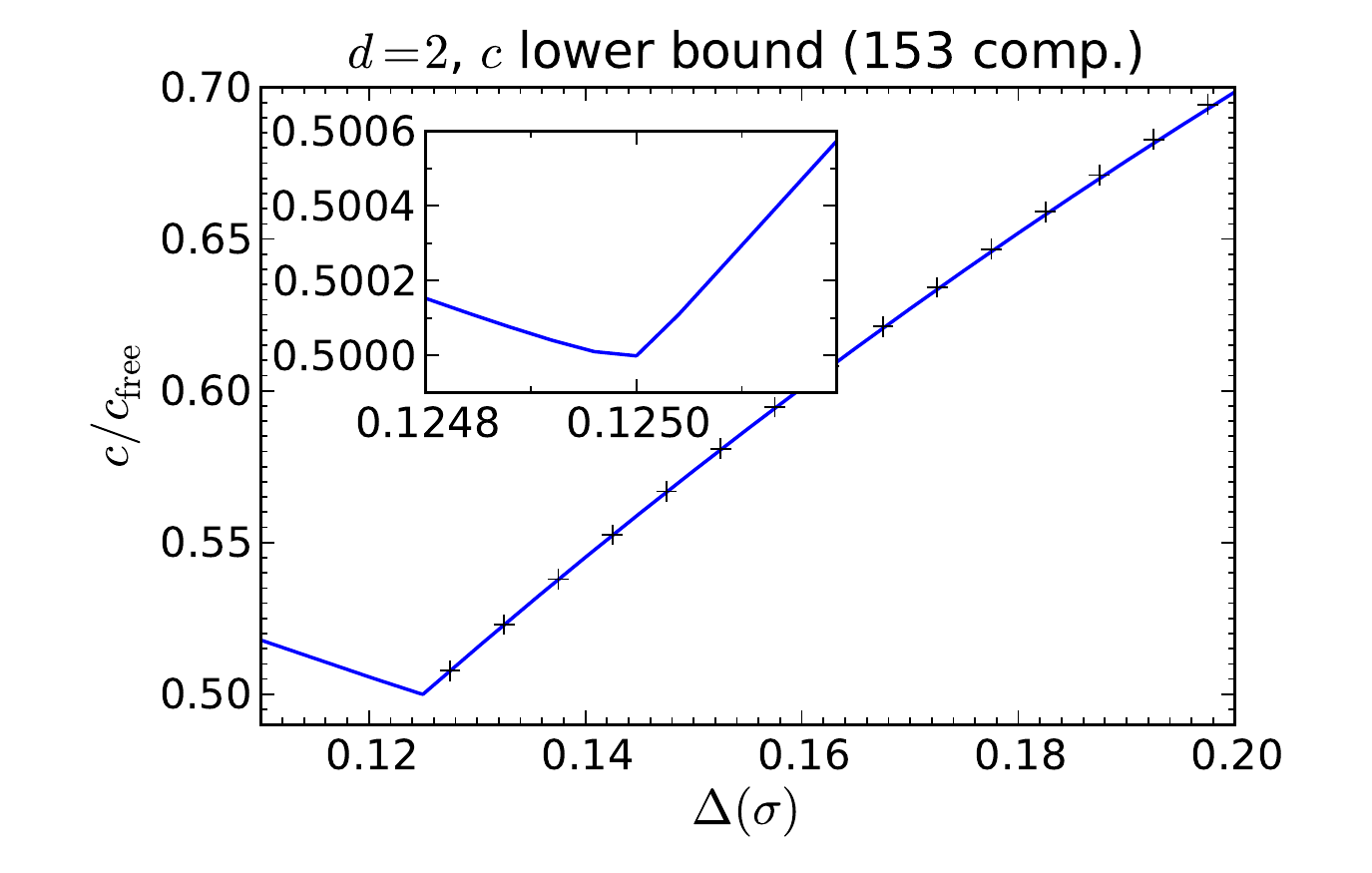} 
\caption{\textit{Blue curve}: a lower bound on $\CT$ in two dimensions similar to Fig.~\ref{sev_ct} but with only $N=153$ components. In the inlay we show the region near $\Delta_\s =\frac{1}{8}$, from which it's clear $\CT$ is minimized, to high precision, at the exact 2d Ising value $\CT=\frac{1}{2}$. \textit{Black crosses}: exact interpolating solution discussed in section \ref{sec:int} below. The interpolating solution is defined for any $\frac18\le\Delta_\s\le \frac12$. The positions of the crosses are chosen arbitrarily to guide the eye; they do not coincide with the positions of the unitary minimal models, of which only two are present in the shown range: $\mathcal{M}(4,3)$ at $\Delta_\s=\frac{1}{8}$ and $\mathcal{M}(5,4)$ at $\Delta_\s=\frac 15$ .}
\label{sev_ct_2d}
\end{center}
\end{figure}

In Figure \ref{sev_ct_2d} we show the lower bound for $\CT$ in 2d, using 153 components. 
Just as in the 3d case (see the discussion in section
\ref{sec-equivalence}), we have to impose an additional constraint
$\Delta_\eps \ge \Delta_\eps^{\text{cutoff}}$ to eliminate spurious minima far away from the Ising point.
The end results are insensitive to $\Delta_\eps^{\text{cutoff}}$ as long as it is sufficiently large. In this study we found $\Delta_\eps^{\text{cutoff}}\approx0.5$
suffices.\footnote{The first study of the $\CT$ lower bound in two dimensions was done in \cite{Vichi-thesis}, section 6.3.2. A peculiarity of the 2d case is that for $\Delta_\eps^{\text{cutoff}}$ close to the unitarity bound, the lower bound on $c$ disappears altogether, allowing solutions to crossing with an arbitrarily small $c$.}

The inset in Figure \ref{sev_ct_2d} shows that by 153 components the minimum seems to have converged, to quite high precision, to the expected
values $(\Delta_\s, \CT)=(\frac{1}{8},\frac{1}{2})$. To be precise, the $\CT$ minimum in this plot is 0.499999. Thus in 2d we can determine $\CT$ with $10^{-6}$ precision at $N=153$, while in 3d we only had $10^{-5}$ precision at $N=231$---the bootstrap analysis seems to converge faster in two dimensions.

A precious feature of the 2d situation is that, unlike in the 3d case, we have a very good guess not only for the physical meaning of the point realizing the overall minimum of $c$---the 2d Ising CFT---but also for the part of the $c$ bound at $\Delta_\sigma>\Delta_\sigma^{\text{2d Ising}}$.
This part of the curve very likely corresponds to the analytic solution to crossing symmetry interpolating between the unitary minimal models, constructed in \cite{Liendo:2012hy} and reviewed in section \ref{sec:int} below. The black crosses in Figure \ref{sev_ct_2d} show that the central charge of the interpolating family follows rather closely our numerical bound. The deviation from the interpolating solution grows somewhat with $\Delta_\s$. At $\Delta_\s=\frac 15$, the interpolating family reaches the second unitary minimal model $\mathcal{M}(5,4)$ of central charge $\frac{7}{10}$. At this point, the deviation from our $N=153$ bound is $1.5\times 10^{-3}$. Although this is less impressive that the $10^{-6}$ agreement at the Ising point, the agreement keeps improving with $N$, so that the exact interpolating family can plausibly be recovered in the $N\to\infty$ limit.

\begin{figure}[htbp]
\begin{center}
\includegraphics[scale=0.6]{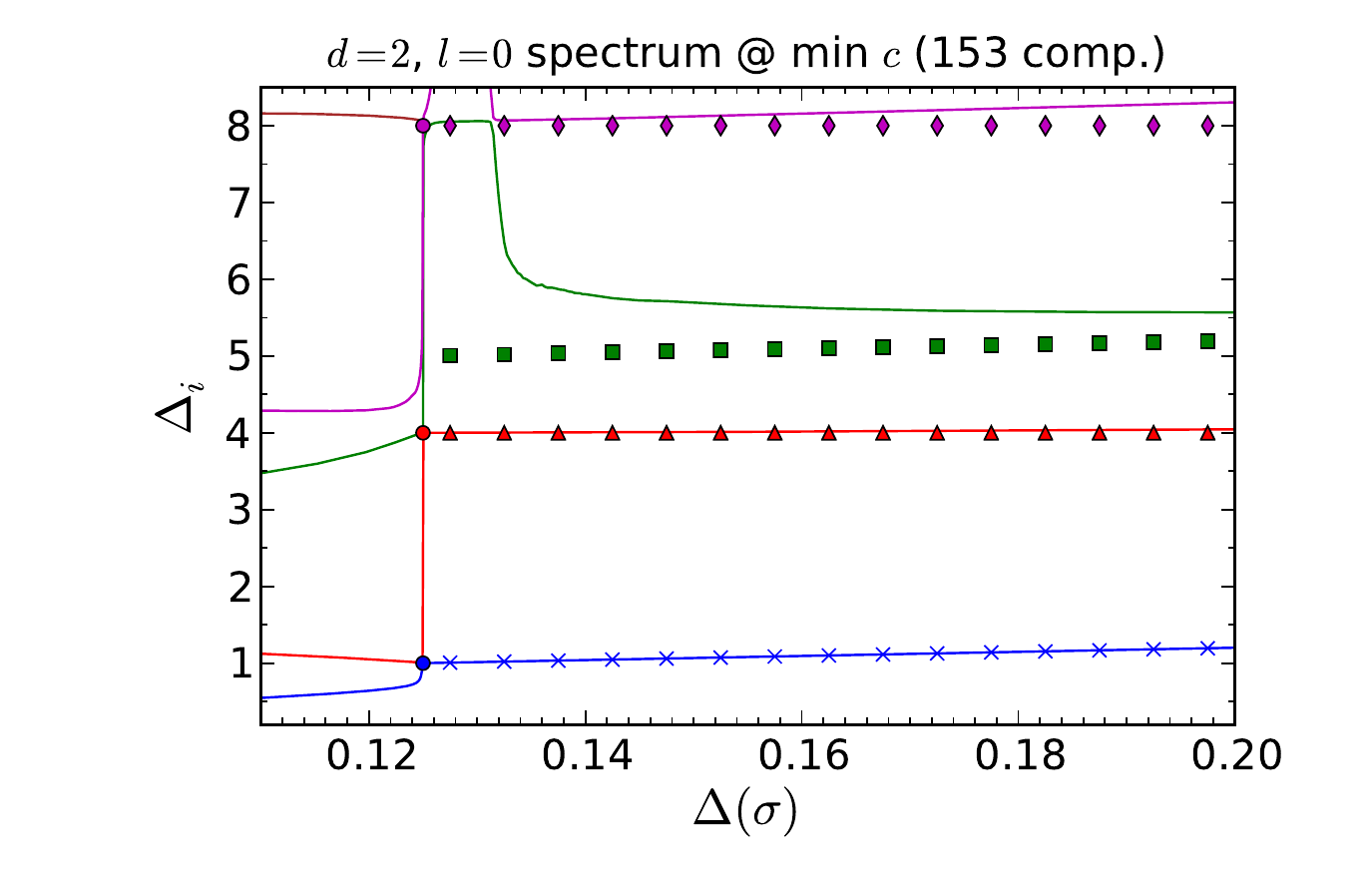}
\hspace{-20pt}
\includegraphics[scale=0.6]{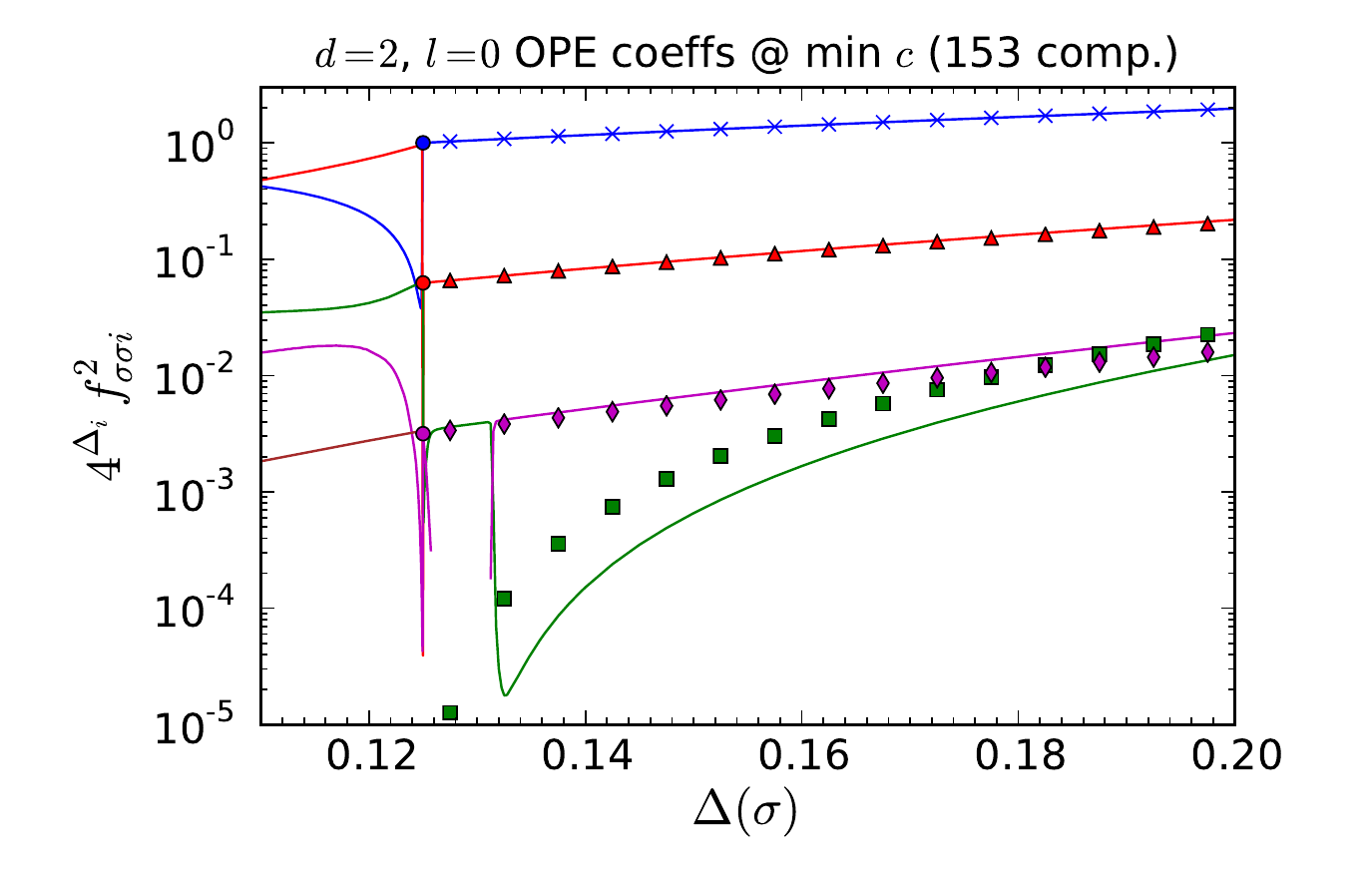}
\caption{\textit{Solid lines}: the spectrum of spin 0 operators with $\Delta<8.5$ and their OPE coefficients, corresponding to the $\CT$ bound at $N=153$.  Note the anomalously small OPE coefficient of the green operator ($\Delta \sim \text{5 - 6}$) which tends to decouple slightly to the right of the Ising point.
\textit{Circles at $\Delta_\sigma=\frac18$}: operators and OPE coefficients in the 2d Ising CFT. \textit{Markers at $\Delta_\sigma>\frac18$} ($\times, \bigtriangleup, \Box, \diamond$): operators and OPE coefficients in the interpolating solution.}
\label{fig-ct-spin0-2d}
\end{center}
\end{figure}

\begin{figure}[htbp]
\begin{center}
\includegraphics[scale=0.6]{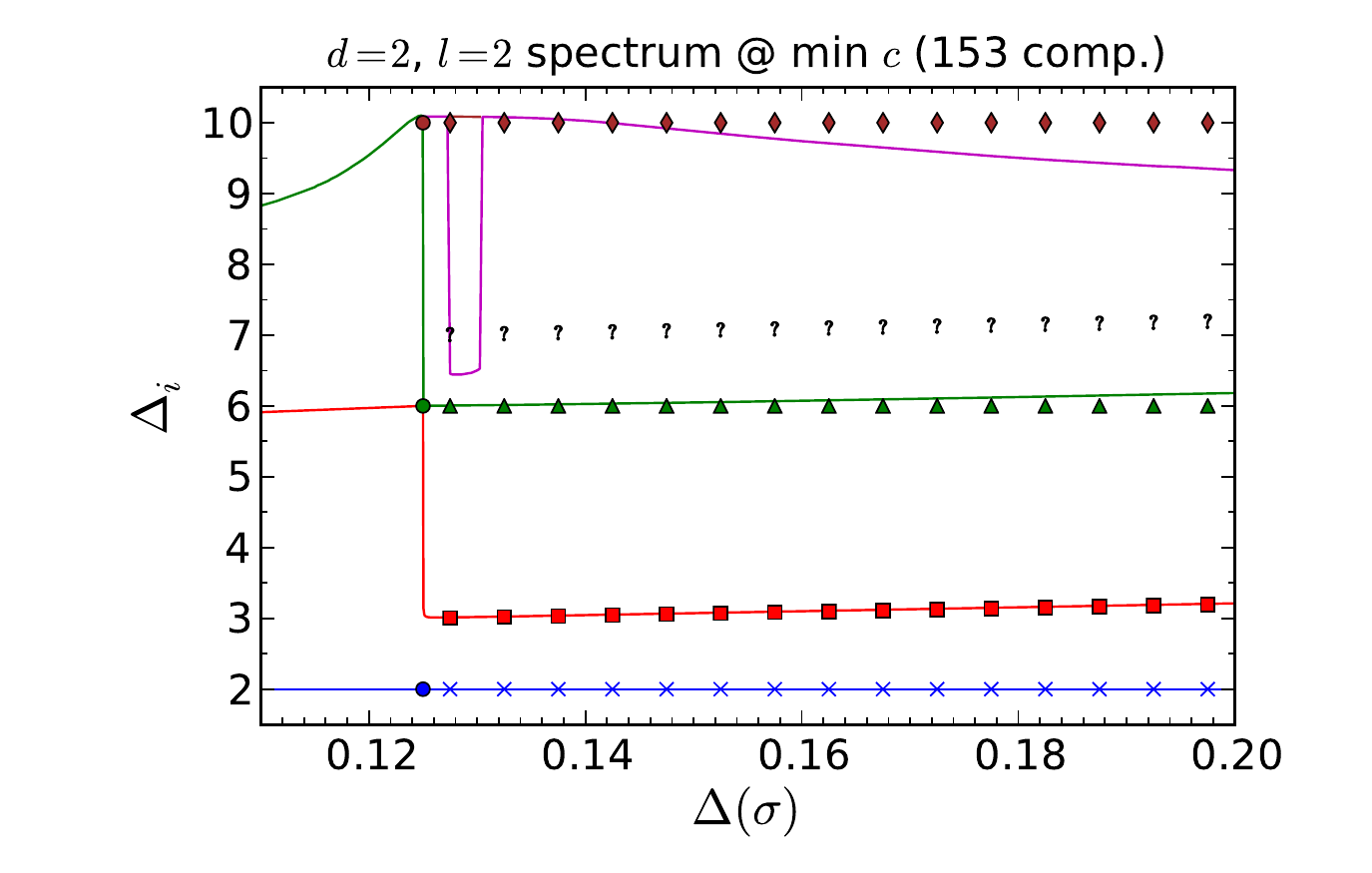}
\hspace{-20pt}
\includegraphics[scale=0.6]{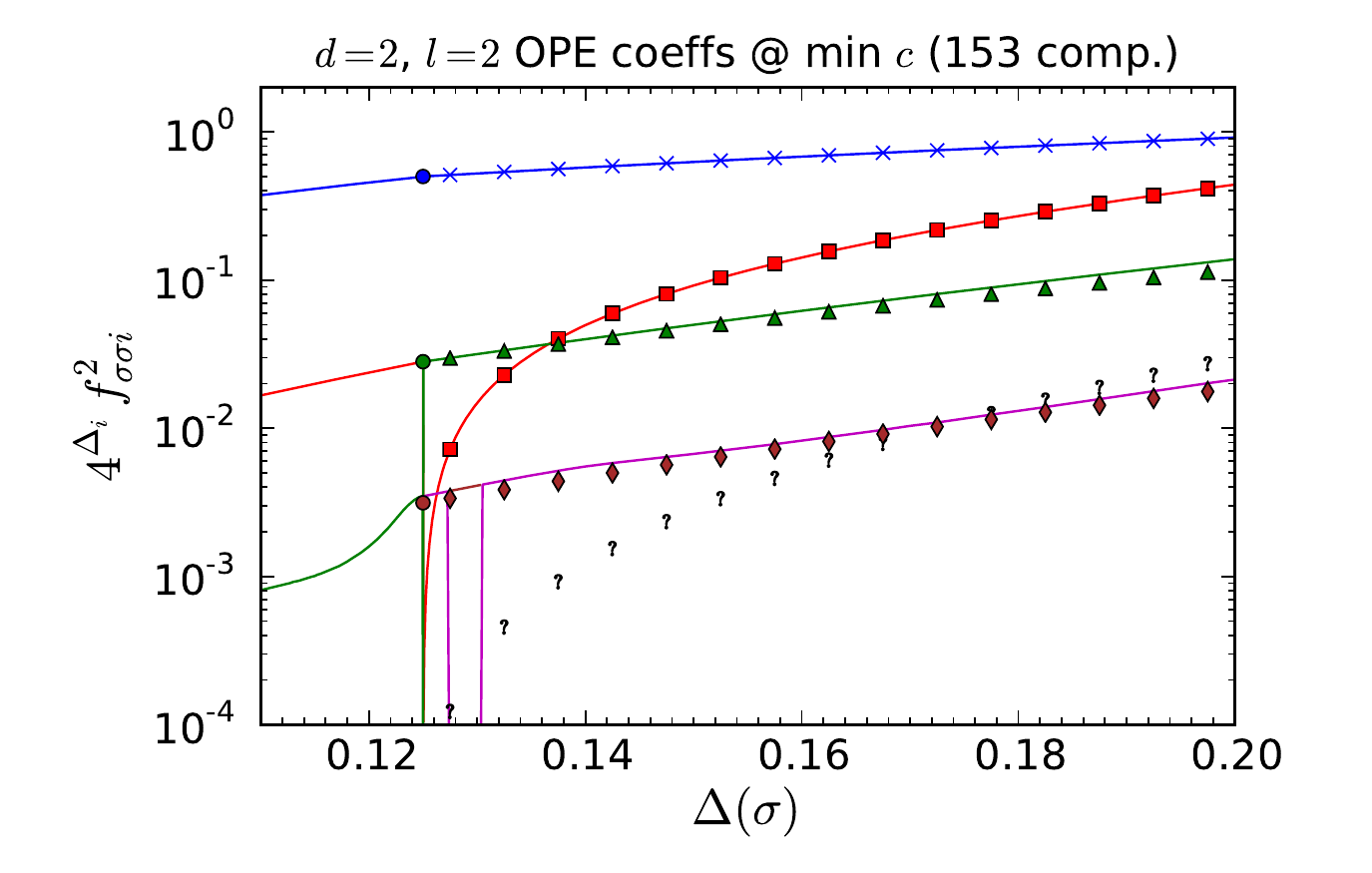}
\caption{\textit{Solid lines}: the spectrum of spin 2 operators with $\Delta<8.5$ and their OPE coefficients, corresponding to the $\CT$ bound at $N=153$. \textit{Circles}: operators and OPE coefficients in the 2d Ising CFT. \textit{Markers}: operators and OPE coefficients in the interpolating solution.}
\label{fig-ct-spin2-2d}
\end{center}
\end{figure}

As in three dimensions, uniqueness of the extremal solution determines the OPE
coefficients and operator dimensions corresponding to the minimal value of
$\CT$ as a function of $\Delta_\s$.  The scalar and spin 2 spectra, as well as
the OPE coefficients, are plotted in Figures \ref{fig-ct-spin0-2d} and
\ref{fig-ct-spin2-2d}, respectively.  We observe the familiar dramatic re-arrangements of the operators at $\Delta_\s \sim 1/8$.

In the same figures, we show the dependence of the low-lying operator dimensions and OPE coefficients of the interpolating family for $\Delta_\s> \frac18$. We see that our numerical solution reproduces well the general features of the interpolating family. In Figure \ref{fig-ct-spin0-2d}, the first, second, and fourth scalars agree very well, together with their OPE coefficients. The third scalar is reproduced less precisely, and in the region close to the Ising point it disappears from our numerical spectrum; this must be due to the fact that it has a very small OPE coefficient and so sensitivity to its dimension is reduced. Indeed, as we discuss in section \ref{sec:int}, in the exact interpolating solution this scalar becomes null and decouples at the Ising point. The tendency of this state to decouple is well reproduced by our numerics. 

Turning to the spin 2 sector in Figure \ref{fig-ct-spin2-2d}, we see that the first four states in the numerical spectrum all correspond to states in the interpolating family.\footnote{A state of dimension $\sim 6.5$ present only in a small interval around $\Delta_\s=0.128$ and with a tiny OPE coefficient is clearly a numerical artifact.} The lowest one is the stress tensor. The second state is decoupling at the Ising point, but it's OPE coefficient is not as small as the coefficient of the decoupling scalar, so it is captured well by the numerics. 

A minor blemish is that one more decoupling spin 2 state (marked with question marks) is altogether missed by our numerics. This is perhaps due to it being close to another state with a larger OPE coefficient. Perhaps the observed slight deviation of the 3rd and the 4th numerical state dimensions can be explained as a distortion induced by trying to compensate for the absence of the ``question mark" state. This story should serve as a warning that it will be difficult to resolve closely spaced high-lying states in the future bootstrap studies.  On the other hand including additional correlators, where the OPE coefficient in question is larger, should allow us to compensate for this.

Comparing Figures \ref{fig-ct-spin0-2d} and \ref{fig-ct-spin2-2d} with the analogous 3d spectrum plots in sections \ref{sec:spin0} and \ref{sec:spin2}, we see several similarities as well as some differences. The two most important similarities are, firstly, rapid spectrum rearrangements at the Ising point and, secondly, the subleading spin 2 operator decoupling when approaching the Ising point from the right. In 3d this decoupling is a mystery, but in 2d it has a natural explanation in terms of the interpolating solution. Maybe also in 3d there exists a continuous family of CFTs saturating the $c$-minimization bound for $\Delta_\s>\Delta_\s^{\text{3d Ising}}$?

One difference between 2d and 3d is that in 2d we saw a scalar decoupling when approaching the Ising point from the right, while no such scalar was observed in 3d.
But the most noticeable difference is in the behavior of the scalar spectrum to the left of the Ising point.  In 2d the lowest
dimension scalar bifurcates to the left of the Ising model while in 3d it is
the subleading scalar which exhibits this behavior. To clarify this puzzle, we carried out a preliminary spectrum study for $c$-minimization in fractional $2<d<3$. We saw that for such $d$ the spectrum to the left of the Ising point is qualitatively similar to the 3d case. In particular, the subleading scalar on the left does not continuously connect to the leading scalar on the right, but curves up and connects to the subleading scalar on the right, as it does for $d=3$. This ``change of topology'' happens even though the extremal spectrum as a whole seems to vary continuously with $d$.

Our last comment concerns the meaning of the extremal numerical solution to the left of the 2d Ising point. Can one find a family of (perhaps non-unitary) CFTs corresponding to this solution, similar to what happens for $\Delta_\s>\frac 18$? We believe that the answer is no, for the following reason.
On the one hand, as can be seen Figure \ref{fig-ct-spin2-2d}, the extremal solution on the left contains a local stress tensor (spin 2, dimension 2 operator). A 2d conformal field theory with a local stress tensor will have a Virasoro symmetry, so let's see if the rest of the spectrum respects it. Unfortunately, the spectrum to the left of the Ising point is not consistent with Virasoro (unlike to the right where it's largely consistent apart from the missing ``question mark" operator). For example, take the two lowest scalars on the left. In the presence of Virasoro symmetry, we would expect to see their spin 2 descendant on level 2, and yet there are no states of such dimension in Figure \ref{fig-ct-spin2-2d}. Perhaps these states are null? However it's easy to check that the dimension of these scalars is inconsistent with them being degenerate on level 2, given the central charge from Figure \ref{sev_ct_2d}.
Thus we really have a contradiction, and any CFT interpretation of the extremal solution on the left is excluded. This solution must thus be unphysical.

\subsection{Interpolating Between Minimal Models}
\label{sec:int}
Ref.~\cite{Liendo:2012hy} constructed a family of analytic solutions to crossing which interpolate between the unitary minimal models $\mathcal{M}(m+1,m)$.
More precisely, they found a family of crossing symmetric four-point
functions $g^{(\De_\s)}(z,\bar z)$ with positive conformal block coefficients
for all $\frac 1 8 \leq \De_\s \leq 1$. These correlators are not
realized in a fully unitary CFT unless $\De_\s$ sits precisely at a minimal
model value (this should be visible as a breakdown of positivity in other
correlators). However, these correlators do satisfy all the constraints
imposed in this work.

As discussed in the previous section, the family of \cite{Liendo:2012hy} likely saturates the $c$-minimization for all $\frac 1 8 \leq \De_\s\leq \frac 1 2$. It also likely saturates the $\Delta_\eps$-maximization bound in the same range.\footnote{In the range
$\frac 1 2 \leq \De_\s \leq 1$, the function $g^{(\De_\s)}$ may still saturate the
bounds on $c$ and $\Delta_\eps$, 
 but this range has not been sufficiently explored and the conclusive numerical evidence is lacking.} 

Because the correlators $g^{(\De_\s)}(z,\bar z)$ provide an important reference for our results, let us review their construction \cite{Liendo:2012hy}.  Minimal model primaries $\f_{r,s}$ are labeled by integers $r,s$.  Let us take $\s=\f_{1,2}$ and $\e=\f_{1,3}$, with scaling dimensions
\be\label{minmodel_line}
\De_\s= \frac 1 2 - \frac{3}{2(m+1)},\qquad \De_\e=2-\frac 4 {m+1}.
\ee
The central charge is
\be
c = 1-\frac{6}{m(m+1)}.
\ee
From this point on, we solve for $m$ in terms of $\De_\s$ and use $\De_\s$ as our independent parameter.  For instance, $\De_\e=\frac 2 3 (4\De_\s+1)$.  The 2d Ising CFT corresponds to $\De_\s=\frac 1 8$.

The operator $\s$ has a null descendant at level $2$, so its correlators satisfy a differential equation
\be
\p{\cL_{-2}-\frac 3 {2(\De_\s+1)}\del_z^2}\<\s(z)\cO_1(z_1)\cdots\cO_n(z_n)\>=0,
\ee
where
\bea
\cL_{-2} &=& \sum_{i=1}^n \p{\frac 1{z-z_i} \ptl_{z_i}+\frac {h_i}{(z-z_i)^2}},
\eea
and $h_i=\De_i/2$ for scalar $\cO_i$. Applying this differential operator to the ansatz
\be
\<\s(x_1)\s(x_2)\s(x_3)\s(x_4)\>= \frac {1}{x_{12}^{2\De_\s}x_{34}^{2\De_\s}}g^{(\De_\s)}(z,\bar z)
\ee
gives a hypergeometric equation that can be solved as
\bea
g^{(\De_\s)}(z,\bar z) &=& G_1(z)G_1(\bar z) + N(\De_\s)G_2(z)G_2(\bar z),
\label{eq:interpolatingsolution}\\
G_1(z) &=& (1-z)^{-\De_\s}{}_2F_1\p{\frac {1-2\De_\s}{3},-2\De_\s,\frac{2(1-2\De_\s)}{3};z}\,,\\
G_2(z) &=& (1-z)^{\frac{1+\De_\s}{3}}z^{\frac {1+4\De_\s}{3}} {}_2F_1\p{\frac{2(1+\De_\s)}{3},1+2\De_\s,\frac{4(1+\De_\s)}{3};z}.
\eea
Crossing symmetry then fixes the normalization $N(\De_\s)$ to be
\be
N(\De_\s) = \frac{2^{1-\frac{8(1+\De_\s)}{3}}\G(\frac 2 3 - \frac {4\De_\s}{3})^2\G(1+2\De_\s)^2 \p{-\cos\p{\frac{\pi(1+4\De_\s)}{3}}+\sin\p{\frac{\pi(1+16\De_\s)}{6}}}}{\pi \G(\frac 7 6 + \frac {2\De_\s}{3})^2}.
\ee
Notice that $N(\De_\s)$ is positive for $0<\De_\s<1$.

The interpretation of this solution is that the only Virasoro primaries appearing in the OPE $\sigma\times\sigma$ are $\mathds{1}$ and $\eps$. This is consistent with the OPE structure of the degenerate fields $\phi_{r,s}$. The two terms in the RHS of \reef{eq:interpolatingsolution} are the holomorphic times anti-holomorphic Virasoro conformal blocks of $\mathds{1}$ and $\eps$, respectively. The function $N(\Delta_\s)$ is the square of the $\sigma\sigma\eps$ OPE coefficient. 

The solution (\ref{eq:interpolatingsolution}) can be further decomposed into $SL(2,\mathbb{C})$ conformal blocks, which are the ones used in our numerical analysis. We can then isolate the squared OPE coefficients of the $SL(2,\mathbb{C})$ primaries in the $\s\x\s$ OPE. The low-lying spectrum of $SL(2,\mathbb{C})$ primaries and their OPE coefficients plotted in Figures \ref{fig-ct-spin0-2d} and \ref{fig-ct-spin2-2d} were obtained this way. These states are Virasoro descendants of $\mathds{1}$ and $\eps$. When we do this computation, we lump together all $SL(2,\mathbb{C})$ primaries of the same dimension and spin.

An interesting feature of the solution (\ref{eq:interpolatingsolution}) is that its expansion into $SL(2,\mathbb{C})$ conformal blocks has positive coefficients for all $\frac 1 8 \leq \De_\s\leq \frac 1 2$. This has been checked to a high level in \cite{Liendo:2012hy}, and we checked it to even higher levels. It seems likely that positivity holds for all levels. This may seem surprising because by the non-unitarity theorem \cite{Friedan:1983xq,Friedan:1986kd} the Verma modules of the $\phi_{r,s}$ operators do contain negative norm states unless $m$ is an integer. What happens is that the norms of these negative norm states always remain much smaller in absolute value that the norms of the many positive norm states present at the same level. Intuitively this is to be expected since these negative norms must vanish at the minimal models, so there is not enough room for them to grow.
Thus the total contribution per level never becomes negative. We checked explicitly that this is precisely what happens in the interval between the first and the second unitary models. In this interval the first negative norm descendant of $\mathds{1}$($\eps$) occurs at level 12(6), respectively.

The situation changes qualitatively for $\Delta_\s<\frac18$, where the four-point function ceases to be unitary. The reason is that in this interval the first negative norm descendant of $\eps$ occurs at level two, and since it's the only descendant on this level there is no room for the cancelation. The operator in question is $\cO_2\equiv (L_{-2}-\frac 3 {2(\De_\s+1)}L_{-1}^2)\e$ of spin 2 and a norm given by
\be
\frac{(1+\De_\s)(2+ 5\De_\s)(8\De_\s-1)}{6(7+4\De_\s)(5+8\De_\s)}.
\label{eq:norm}
\ee
We see explicitly that the norm becomes negative when $\De_\s < \frac 18$, to the left of the Ising point. 

In Figure \ref{fig-ct-spin2-2d}, $\cO_2$'s dependence on $\Delta_\s$ is marked by the red squares. The OPE coefficient goes to zero at the Ising point. This dependence is reproduced well by our numerical solution. Another spin 2 state which decouples at the Ising point can be obtained by acting on $\cO_2$ by $(L_{-2}+\ldots)(\bar L_{-2}+\ldots)$, where $\ldots$ is fixed to get a quasiprimary. This is the ``question mark" state which is missed by our numerics.
Finally, the decoupling scalar in Figure \ref{fig-ct-spin0-2d} is $(\bar L_{-2}-\frac 3 {2(\De_\s+1)}\bar L_{-1}^2)\cO_2 $. Its norm thus goes to zero as the square of \reef{eq:norm}.

\section{Technical Details}

While the computational algorithms used in this paper share many conceptual similarities with our previous work \cite{ElShowk:2012ht}, there are two major technical novelties.

Firstly, we switched to a very efficient representation for conformal block derivatives recently developed in 
\cite{Hogervorst:2013sma,Hogervorst:2013kva,Kos:2013tga}. 
In our previous work \cite{ElShowk:2012ht}, we had to discretize $\Delta$ with a small step and precompute large tables of conformal blocks and their derivatives corresponding to the discretization. The new representation is sufficiently fast to evaluate conformal blocks inside the simplex algorithm, and allows us to dispense with the discretization.  

Secondly, we switched to the ``primal" method instead of the ``dual" method used in prior work (see section \ref{sec-approaching}). Although the two methods are formally equivalent, the primal one has an advantage that at every step of the extremization algorithm we have a valid solution to the crossing symmetry constraint. 
We implemented our own version of Dantzig's primal simplex method algorithm, capable of dealing with a continuum of constraints, and using multiple precision arithmetic with $O(100)$ significant digits. Demands of final accuracy and numerical stability render the standard double precision arithmetic (i.e., 16 significant digits) insufficient for our needs (see note \ref{note:MPFR} for a discussion why).

There are many small subtleties and tweaks which go into the implementation of the above two ideas; they will be described in detail in the rest of this section. It should be noted that the numerical bootstrap field is still rapidly developing and our algorithms will continue to improve dramatically in the foreseeable future. For this reason it doesn't seem worthwhile yet to release general-purpose code for doing these computations.
But we will be happy to provide anyone interested with a current version of our code. 

\subsection{Partial Fraction Representation for Conformal Blocks}
\label{sec:partialfractionsforblocks}

In this section, we describe a representation for conformal blocks that is efficient to compute and well suited for our optimization algorithm.  It is based on the series expansion of \cite{Hogervorst:2013sma} and the idea of rational approximations introduced in \cite{Kos:2013tga}.

\begin{figure}[htbp]
\begin{center}
\includegraphics[scale=1]{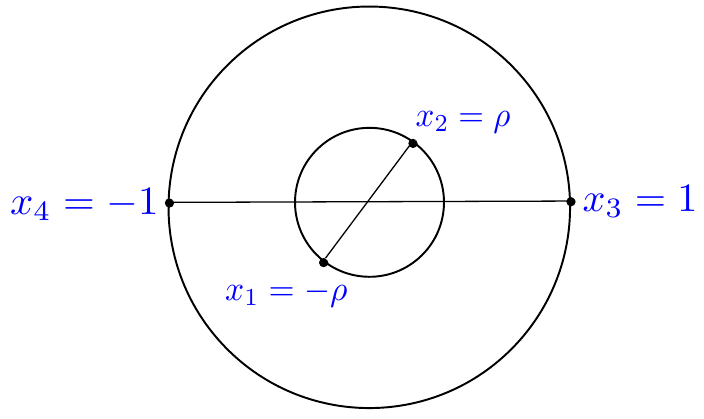}
\caption{Configuration of points for radial quantization in the $\rho$-coordinate \cite{Hogervorst:2013sma}.}
\label{fig:rho-representation}
\end{center}
\end{figure}

Conformal blocks are best studied in the radial coordinates of \cite{Hogervorst:2013sma}.  Via a conformal transformation, we can always place four operators $\s(x_1),\dots,\s(x_4)$ on a two-plane, so that $x_3=1$, $x_4=-1$ lie on the unit circle, and $x_1=-\rho$, $x_2=\rho$ lie on a smaller circle around the origin (Figure~\ref{fig:rho-representation}).  The complex coordinate $\rho=r e^{i\th}$ is related to the usual conformal cross-ratios via
\be
\rho = \frac{z}{(1+\sqrt{1-z})^2}.
\ee

A conformal block is then given by inserting all the states in a conformal multiplet (i.e. a primary $\cO$ and its descendants $\ptl \cO, \ptl^2\cO, \dots$) on a sphere separating $x_1,x_2$ from $x_3,x_4$, in radial quantization,
\be
x_{12}^{-2\De_\s} x_{34}^{-2\De_\s} G_{\De,\ell}(u,v) = \sum_{\a=\cO,\ptl\cO,\dots} \frac{\<0|\s(x_3)\s(x_4)|\a\>\<\a|\s(x_1)\s(x_2)|0\>}{\<\a|\a\>}.
\label{eq:conformalblockassumoverstates}
\ee
By classifying the states $\a$ according to their representations under dilatation and rotations, this sum can be written \cite{Hogervorst:2013sma}
\be
\label{eq:gegenbauerexpansion}
G_{\De,\ell}(r,\eta) = \sum_{n=0}^\oo \sum_j B_{n,j}(\De,\ell) r^{\De+n} \frac{j!}{(2\nu)_j}C^{(\nu)}_j(\eta)
\ee
where $r=|\rho|$, $\eta=\cos\th=\frac{\rho+\bar\rho}{2|\rho|}$, $\nu=\frac{d-2}{2}$, and $C^{(\nu)}_j(\eta)$ are Gegenbauer polynomials.  The coefficients $B_{n,j}(\De,\ell)$ express contributions of descendants of spin $j$ at level $n$ of the multiplet; they are rational functions of the dimension $\De$.  This follows directly from the expression (\ref{eq:conformalblockassumoverstates}) for $G_{\De,\ell}$ as a sum over states: each term in the numerator and denominator is a polynomial in $\De$ that can be computed using the conformal algebra.

For our purposes, we will be interested in computing conformal blocks and their derivatives around the crossing-symmetric point $z=\bar z=1/2$. This corresponds to $r=r_*\equiv 3-2\sqrt 2\approx 0.17$, so the series (\ref{eq:gegenbauerexpansion}) is a rapidly convergent expansion at this point.\footnote{For the conformal blocks corresponding to identical external scalars used here, the expansion is actually in powers of $r^2$ \cite{Hogervorst:2013sma}.} To get a good approximation, we can truncate it at some large but finite value of $n$, with the result
\be
G_{\De,\ell}(r,\eta) \aeq r^\De \frac{P_\ell(r,\eta,\De)}{Q_\ell(\De)},
\label{eq:rationalexpressionforblock}
\ee
where $P_\ell, Q_\ell$ are polynomials in $\De$.  Taking derivatives around the crossing-symmetric point, and expanding the resulting rational function of $\De$ in partial fractions, we can write
\be
\left.\ptl_r^m \ptl_\eta^n G_{\De,\ell}(r,\eta)\right|_{r=r_*,\eta=1}
\aeq
r_*^\De \p{p^{m,n}_\ell(\De)+\sum_i \frac{a^{m,n}_{\ell,i}}{\De-\De_i}},
\label{eq:partialfractionrepresentation}
\ee
where $p^{m,n}_\ell(\De)$ are polynomials and $a^{m,n}_{\ell,i}$ are numerical coefficients. As Table \ref{tab:polepositions} below shows, there are only simple poles in $\Delta$ except if $d=2,4,6,\ldots$ when there are also double poles. Conformal blocks vary continuously in $d$, so the double poles get resolved into a pair of simple poles when $d$ is slightly perturbed away from an even integer.\footnote{We used this feature in our $d=2$ computations presented in section \ref{sec:2dchecks}. Instead of dealing with double poles, we used our generic simple-pole code and ran it at $d=2+10^{-5}$. We checked that using $d=2+10^{-7}$ or even $d=2+10^{-15}$ does not change the results.}

This representation of the conformal blocks in terms of partial fractions has the virtue that once the data $p^{m,n}_\ell$ and $a_{\ell,i}^{m,n}$ have been computed, we can calculate the blocks at any value of the dimension $\De$ extremely rapidly and with very high precision.  This will be crucial in our optimization algorithm.  Now, having established what representation of the conformal blocks we would like to use, let us describe two methods for computing it.  For this work, we have implemented both methods.

\subsubsection{The Casimir Equation}
\label{sec:casimirequationmethod}

The conformal block $G_{\De,\ell}(r,\eta)$ is an eigenvector of the quadratic Casimir of the conformal group.  This implies a differential equation for $G_{\De,\ell}(r,\eta)$ which can be solved iteratively order by order in $r$ \cite{Hogervorst:2013sma}. Using the Casimir equation, one can recursively compute the coefficients $B_{n,j}(\De,\ell)$ starting from the initial conditions:
\beq
B_{0,j}(\De,\ell)=4^\Delta \delta_{j\ell},
\eeq
expressing the fact that the primary itself is the only state on level zero of the multiplet. The $4^\Delta$ fixes the normalization of conformal blocks to be the same as in \cite{ElShowk:2012ht}. Once $B_{n,j}(\De,\ell)$ are known, it is straightforward to obtain $p_\ell^{m,n}, a_{\ell,i}^{m,n}$.

In this work, we did not literally use this method but a slight variation due to \cite{Hogervorst:2013kva}. The point is that Eq.~\reef{eq:gegenbauerexpansion} contains more information than needed: it can be used to recover conformal blocks for any $\eta$ while in the applications here we only need $\eta\approx1$. The variation described below focuses directly on $\eta\approx 1$ without having to deal with Gegenbauer polynomials.

One considers first the conformal block restricted to the ``diagonal" $\rho=\bar\rho$ (i.e.~$\eta=1$):
\beq
\label{eq:diagonal}
G_{\Delta,\ell}(r)\equiv G_{\Delta,\ell}(r,\eta=1)=\rho^\Delta\sum_{n=0}^\oo b_n(\Delta,\ell) r^{n},\quad b_n(\Delta,\ell)=\sum_j B_{n,j}(\De,\ell) .
\eeq
It was shown in \cite{Hogervorst:2013kva} that $G_{\Delta,\ell}(r)$ satisfies a fourth-order ordinary differential equation (which becomes third-order for $\ell=0$). Using this equation, we evaluate $b_n(\Delta,\ell)$ to a very high order ($n=120$ was used in most computations, and a few results were checked at $n=200$) starting from $b_0(\Delta,\ell)=4^\Delta$. The equation has a regular singular point at $\rho=0$, which is why we can evaluate the whole series expansion starting from just one initial condition for the leading term. Knowing $b_n(\Delta,\ell)$, we obtain coefficients $p_\ell^{m,0}, a_{\ell,i}^{m,0}$ in \reef{eq:partialfractionrepresentation}.

To obtain derivatives normal to the diagonal, we then use the quadratic Casimir partial differential equation, solving it \textit{\`a la} Cauchy-Kovalevskaya in a power-series expansion around the diagonal. This idea was already used in \cite{ElShowk:2012ht}, where it was shown that all normal and mixed derivatives of conformal blocks could be reduced in this way to derivatives along the diagonal. 

To be precise, we switch halfway to the variables $a$ and $b$ defined as in \cite{ElShowk:2012ht}, 
\beq
z=(a+\sqrt{b})/2,\quad \bar{z}=(a-\sqrt{b})/2\,,
\eeq
so that $b=0$ is the diagonal, and the crossing symmetric point $z=\bar{z}=1/2$ corresponds to $a=1$, $b=0$. Since the conformal blocks are symmetric under $z\leftrightarrow \bar z$, the expansion around $b=0$ contains integer powers of $b$. So, we first evaluate the diagonal derivatives $\del_r^m G_{\Delta,\ell}$ as described above, then do a change of variables $\rho\to a$ to express the diagonal derivatives $\del_a^m G_{\Delta,\ell}|_{a=1}$, and finally use the Cauchy-Kovalevskaya method, precisely as described in section 4 of \cite{ElShowk:2012ht}, to obtain the mixed and normal derivatives $\del_a^m \del_b^n G_{\Delta,\ell}|_{a=1,b=0}$. For all these derivatives we get a representation of the form \reef{eq:partialfractionrepresentation}. 

\subsubsection{Recursion Relations}
\label{sec:recursionrelationmethod}

An alternate way to compute the partial-fraction representation~(\ref{eq:partialfractionrepresentation}) was developed in \cite{Kos:2013tga}.  Here, we review this method and include additional details about its implementation.  The idea is to use a recursion relation expressing conformal blocks as a sum over poles in $\De$, where the residue at each pole is itself a conformal block:
\bea
h_{\De,\ell}(r,\eta) &\equiv& r^{-\De} G_{\De,\ell}(r,\eta)\nn\\
h_{\De,\ell}(r,\eta) &=& h_{\ell}^{(\oo)}(r,\eta) + \sum_i \frac{c_i r^{n_i}}{\De-\De_i} h_{\De_i+n_i,\ell_i}(r,\eta),
\label{eq:cbrecursionrelation}
\eea
with
\be
\label{eq:asymptoticblock}
h_{\ell}^{(\oo)}(r,\eta) = \frac{\ell!}{(2\nu)_\ell} \frac{C^{(\nu)}_\ell(\eta)}{(1-r^2)^\nu\sqrt{(1+r^2)^2-4r^2\eta^2}}.
\ee
\begin{table}
\centering
\begin{tabular}{c|c|c|cl}
$n_i$ & $\De_i$ & $\ell_i$ & $c_i$\\
\cline{1-4}
$2k$ & $1-\ell-2k$ & $\ell+2k$ & $c_1(k)$  & \quad $k=1,2,\dots$\\
$2k$ & $1+\nu-k$   & $\ell$    & $c_2(k)$  & \quad $k=1,2,\dots$\\ 
$2k$ & $1+\ell+2\nu-2k$ & $\ell-2k$ & $c_3(k)$ & \quad $k=1,2,\dots,\lfloor \ell/2 \rfloor$
\end{tabular}
\caption{The positions of poles of $G_{\De,\ell}$ in $\Delta$ and their associated data.  There are three types of poles, corresponding to the three rows in the table.  The first two types exist for all positive integer $k$, while the third type exists for positive integer $k\leq \lfloor \ell/2\rfloor$.  The coefficients $c_1(k), c_2(k), c_3(k)$ are given in Eqs.~(\ref{eq:poleresidues}).}
\label{tab:polepositions}
\end{table}

The $\De_i$ above are special values of the dimension where a descendant state $|\a\>$ in (\ref{eq:conformalblockassumoverstates}) can become null.  By definition, these degenerate values are always below the unitarity bound~(\ref{eq:unitaritybounds}).  The null descendant $|\a\>$ has dimension $\De_i+n_i$ and spin $\ell_i$, and each pole comes with a numerical factor $c_i$.  For the reader's convenience, we summarize this data in Table~\ref{tab:polepositions} and Eq.~(\ref{eq:poleresidues}):
\bea
\label{eq:poleresidues}
c_1(k) &=& -\frac{k(2k)!^2}{2^{4k-1}k!^4}\frac{(\ell+2\nu)_{2k}}{(\ell+\nu)_{2k}},\nn\\
c_2(k) &=& -\frac{k(\nu+\ell-k)(\nu)_k(1-\nu)_k\p{\frac{\nu+\ell+1-k}{2}}_k^2}{k!^2(\nu+\ell+k)\p{\frac{\nu+\ell-k}{2}}_k^2},\nn\\
c_3(k) &=& -\frac{k(2k)!^2}{2^{4k-1}k!^4}\frac{(1+\ell-2k)_{2k}}{(1+\nu+\ell-2k)_{2k}}.
\eea

Note that each term in the sum over poles $\De_i$ is suppressed by at least $r^{2}$.  Thus, by iterating the recursion relation~(\ref{eq:cbrecursionrelation}), starting with the initial term $h^{(\oo)}_\ell$, we rapidly converge to the correct value of the conformal block.  In practice, it is convenient to take derivatives first and perform this iteration numerically at a given point $(r,\eta)=(r_*,1)$, while keeping $\De$ as a variable.  Specifically, let us define the vector of derivatives
\bea
\bh_\ell(\De) &=&[\ptl_r^0\ptl_\eta^0 h_{\De,\ell}(r_*,1)\quad\ptl_r^1\ptl_\eta^0 h_{\De,\ell}(r_*,1)\quad\cdots\quad]^T\nn\\
 &=& \bh^{(\oo)}_\ell + \sum_i \frac{\bd_{\ell,i}}{\De-\De_i},
 \label{eq:ansatzforHvector}
\eea
where we include all derivatives $\ptl_r^m\ptl_\eta^n$ with $m+n\le 2K$ for some $K$.
The vector $\bh^{(\oo)}_\ell$ is given simply by derivatives of the known function (\ref{eq:asymptoticblock}), which can be computed beforehand.  Multiplication by $r$ is represented on the space of derivatives by a matrix $\bR$, so Eq.~(\ref{eq:cbrecursionrelation}) implies
\be
\bd_{\ell,i} = c_i(\ell) \bR^{n_i} \p{\bh^{(\oo)}_\ell + \sum_j \frac{\bd_{\ell_i,j}}{\De_i(\ell)+n_i-\De_j(\ell_i)}}.
\ee
Iterating this equation numerically, we can compute the residues $\bd_{\ell,i}$.  The number of iterations, together with the number of poles in the ansatz~(\ref{eq:ansatzforHvector}) is roughly equivalent to the order at which we truncate the Gegenbauer expansion~(\ref{eq:gegenbauerexpansion}).  In this work, we keep poles up to roughly $k=100$ and perform 100 iterations.

Finally, to recover the vector of derivatives for $G_{\De,\ell}$ itself, we should multiply by the matrix $\bR^{\De}$.  This matrix takes the form $\bR^\De=r^\De \bS(\De)$, where $\bS(\De)$ is a matrix polynomial in $\De$.  Decomposing $\bS(\De)\bh_\ell(\De)$ into partial fractions then yields the representation (\ref{eq:partialfractionrepresentation}).  For example, the coefficients $\mathbf a_{\ell,i}=(a_{\ell,i}^{m,n})$ in section~\ref{sec:partialfractionsforblocks} are given by
\be
\mathbf a_{\ell,i} = \mathop{\mathrm{Res}}_{\De\to\De_i}\bS(\De)\frac{\bd_{\ell,i}}{\De-\De_i}=\bS(\De_i)\bd_{\ell,i}.
\ee

\subsection{A Customized Simplex Method}
\label{sec:customsimplex}

In this section, we describe our procedure for optimizing over $\cC_{\De_\s}$.  The underlying algorithm is the Simplex Method, due to Dantzig.  We will first review this algorithm, 
and then discuss its specialization to our case of interest. 

\subsubsection{The Primal Simplex Method}
\label{sec:primalsimplex}

The material in this section is standard in the mathematics and computer science literature.  We include it to establish notation and because it may be unfamiliar to physicist readers.  We will essentially follow the presentation of \cite{NumericalRecipes}.

Given vectors $\bc\in \R^n, \bb\in \R^m$ and a matrix $\bA\in \R^{m\x n}$, we would like to minimize the {\it objective function}
\be
\bc\.\bx=c_1x_1+\cdots+c_n x_n\nn
\ee
over $\bx\in \R^n$ such that
\begin{itemize}
\item $x_i\geq 0$, and
\item $\bA\bx=\ba_1 x_1+\cdots +\ba_n x_n=\bb$.
\end{itemize}
We assume $n>m$, so that the space of possible $\bx$'s has positive dimension.

The space of possible $\bx$'s is a convex polytope.  Because of convexity, the minimum we seek is always realized at a vertex of this polytope (though it may be non-unique).  So it suffices to minimize $\bc\.\bx$ over vertices.

At a vertex, as many as possible of the inequalities $x_i\geq 0$ are saturated.  Since $\bx$ lives in $n$-dimensions and is subject to $m$ equality constraints $\bA\bx=\bb$, generically $n-m$ inequalities can be saturated. This leaves $m$ nonzero coordinates $x_{j_1},\dots,x_{j_m}$.\footnote{At non-generic vertices, it's possible that extra inequalities can be saturated, so that some of the $x_{j_i}$ actually vanish.}  The equality $\bA \bx=\bb$ then expresses $\bb$ as a nonnegative sum of the corresponding vectors $\ba_{j_1},\dots,\ba_{j_m}$, which are called {\it basic vectors},
\be
\ba_{j_1}x_{j_1}+\cdots+\ba_{j_m}x_{j_m} = \bb.
\ee
The remaining $\ba_i$ with $x_i=0$ are called {\it nonbasic}.  Specifying a vertex is equivalent to specifying a set of $m$ basic vectors.

The idea of the simplex algorithm is to travel from vertex to vertex along the polytope edges, following the direction of steepest descent.  Let's assume that we have found a vertex of our polytope (we address the question of finding an initial vertex later), and describe how to pass to the next vertex.

Suppose our starting vertex is characterized by $m$ basic vectors $\ba_{j_1},\dots,\ba_{j_m}$ with nonzero coordinates $x_{j_1},\dots,x_{j_m}$.  For convenience, let us partition $\bA$ into an $m\x m$ matrix $\bA_B$ whose columns are basic vectors, and an $m\x(n-m)$ matrix $\bA_N$ whose columns are the remaining nonbasic vectors,
\be
\bA = [\bA_B\,|\,\bA_N]
\ee
We similarly partition $\bx$ into basic coordinates $\bx_B=(x_{j_1},\dots,x_{j_m})^T$ and the remaining nonbasic coordinates $\bx_N$ (all of which vanish at our vertex).

Now consider adjusting some nonbasic coordinate $x_k$ away from zero.  To preserve $\bA\bx=\bb$, we must simultaneously adjust $\bx_B$.  We have
\be
\bA_B \bx_B+\ba_k x_k = \bb
\qquad
\implies
\qquad
\bx_B = \bA_B^{-1}(\bb-\ba_k x_k),
\ee
so that our objective function becomes
\be
\bc_B\.\bx_B+c_k x_k=\bc_B^T\bA_B^{-1}\bb+(c_k-\bc_B^T\bA_B^{-1}\ba_k)x_k.
\ee
To decrease $\bc\.\bx$ as quickly as possible, we should choose $k$ that minimizes the quantity in parentheses, known as the {\it reduced cost},
\begin{align}
\label{eq:minreducedcost}
k_* &=\mathrm{argmin}_k \mathrm{RC}_k ,\nn\\
\mathrm{RC}_k &\equiv (c_k-\bc_B^T\bA_B^{-1}\ba_k).
\end{align}

If the minimum reduced cost is nonnegative, then our starting vertex is already a minimum, and the algorithm terminates.  Assume instead that the minimum reduced cost is negative.  Once we've picked $k_*$, we turn on $x_{k_*}$ as much as possible until one of the original basic coordinates goes to zero, 
\be
x_\ell=(\bA_B^{-1}\bb-\bA_B^{-1}\ba_{k_*} x_{k_*})_\ell=0.
\ee
We choose $\ell$ to be the first index for which this happens (this is commonly known as the ``ratio test").
When $x_\ell$ becomes zero, the result is a new set of $m$ basic vectors where $\ba_\ell$ has been replaced by $\ba_{k_*}$.  This defines a new vertex with strictly smaller objective function $\bc\.\bx$.  By repeating this process, we eventually reach a minimum.

\subsubsection{Choosing an Initial Vertex}
\label{sec:phaseone}

To run the iteration described above, we need an initial vertex.  We can find one by solving an auxiliary optimization problem (typically called ``phase 1" of the algorithm, while the main optimization stage described above is called ``phase 2").  We extend $\bx$ with $m$ additional {\it slack variables} $\bs=(s_1,\dots,s_m)^T$,
\bea
\bx' &=& \left(
\begin{array}{c}
\bs\\
\hline
\bx
\end{array}\right
),
\eea
and extend $\bA$ with {\it slack vectors},
\bea
\bA' &=&  = 
\left.\left(
\begin{array}{cccc}
b_1 & 0 & \dots & 0\\
0 & b_2 & \dots & 0\\
\vdots & \vdots & \ddots &\vdots \\
0 &  0 &\dots & b_m
\end{array}
\right| \bA
\right).
\nn
\eea

These are designed so that we can trivially satisfy the conditions
\be
\bA'\bx'=\bb\quad\textrm{and}\quad x'_i\geq 0\ (i=1,\dots,n+m)
\ee
by choosing the vertex $s_i=1$, $\bx=0$.  Starting from this initial condition, we can now try to minimize
\be
\bc'\.\bx'\equiv \sum_{i=1}^m s_i
\ee
using the algorithm described above.  This objective function is bounded below because the $s_i$'s are nonnegative.
If the minimum is zero, then every $s_i$ vanishes and we've found a good vertex for starting our original optimization problem.  If the minimum is nonzero, then no such vertex exists and the original optimization problem is infeasible.

\noindent\textit{Hot Start}

This is a slightly more involved strategy for an initial vertex search, which may give a considerable speed up, provided that we have a good guess for an initial basis. Let this guess be given by an $n\times n$ ``trial basis" submatrix $\tilde\bA_B$ of $\bA$. The trial coordinates are given by 
\be
\tilde\bx_B = \tilde \bA_B^{-1} \bb\,.
\ee
If all of them are positive, then our guess was perfect, and we can go straight to phase 2. Otherwise, we extend the matrix $\bA$ by the \textit{negatives} of all columns in $\tilde\bA_B$ whose trial coordinates are negative. If the initial guess was reasonably good, there will be only a few such vectors. The added vectors plus all the remaining vectors of $\tilde \bA_B$ form a good basis for the extended problem. We then form the objective function given by the sum of the coordinates of all the added vectors, and attempt to minimize it to zero. 

\subsubsection{Adaptation to Optimization over $\cC_{\De_\s}$}
\label{sec:specializationofprimalsimplex}

Let us describe the space $\cC_{\De_\s}$ in the language of the previous sections.  Crossing symmetry Eq.~\reef{eq:crossingsymmetry} can be written
\bea
\sum_{\De,\ell} p_{\De,\ell} F^{\De_\s}_{\De,\ell}(u,v) &=& - F^{\De_\s}_{0,0}(u,v),
\eea
where
\bea
\label{eq:definitionofF}
F^{\De_\s}_{\De,\ell}(u,v) &\equiv& u^{-\De_\s}G_{\De,\ell}(u,v)-v^{-\De_\s}G_{\De,\ell}(v,u)
\eea
and $F^{\De_\s}_{0,0}$ corresponds to the unit operator.

As explained in section~\ref{sec-approaching}, in practice we work with the spaces $\cC_{\De_\s}^{(N)}$, given by truncating the crossing symmetry constraint to $N$ derivatives around the crossing-symmetric point $z=\bar z =1/2$.  Let us define vectors of derivatives $\bF_{\De,\ell}^{\De_\s}$ with components
\bea
\label{eq:DSDders}
(\bF^{\De_\s}_{\De,\ell})^{(m,n)}&=& \left.\ptl_z^m\ptl_{\bar z}^n F_{\De,\ell}^{\De_\s}(u,v)\right|_{z=\bar z = 1/2},
\eea
where $u=z\bar z$, $v=(1-z)(1-\bar z)$ as usual.  Because $F_{\De,\ell}^{\De_\s}(u,v)$ is antisymmetric under $u\leftrightarrow v$, these derivatives are nonzero only if $m+n$ is odd.  Further, since $u$ and $v$ are invariant under $z\leftrightarrow \bar z$, it suffices to take $m\geq n$.  Demanding $m+n\leq 2K$ for integer $K$, we have $N=\frac{K(K+1)}2$ nontrivial components for $\bF_{\De,\ell}^{\De_\s}$. This is the truncation parameter $N$ used to label plots throughout the paper. For example $N=231$ corresponds to $K=21$.

The vectors $\bF^{\De_\s}_{\De,\ell}$ are related by a linear transformation to the derivatives of conformal blocks discussed in section~\ref{sec:partialfractionsforblocks}.  We first transform from derivatives in $r,\eta$ coordinates (section~\ref{sec:recursionrelationmethod}) to $z,\bar z$ coordinates, and then multiply by an additional matrix to account for multiplication by $u^{-\De_\s}$ and $v^{-\De_\s}$ in Eq.~(\ref{eq:definitionofF}). 

Alternatively, we can work with the $a,b$ coordinates from section \ref{sec:casimirequationmethod} and define
\bea
(\bF^{\De_\s}_{\De,\ell})^{(m,n)}&=& \left.\del_a^m\ptl_b^n F_{\De,\ell}^{\De_\s}(u,v)\right|_{a=1, b=0},
\eea
where $m,n\ge0$ arbitrary, $m$ is odd (otherwise the derivative vanishes), and $m+2n\le 2K-1$. Working with these derivatives or with \reef{eq:DSDders} corresponds to a different choice of basis in the same $N$-dimensional space and gives the same constraints.

After taking derivatives, the crossing equation takes the familiar form
\be
\sum_i \ba_i x_i = \bb,\qquad x_i\geq 0,
\ee
with the identifications
\bea
\begin{array}{ccc}
i &\leftrightarrow& (\De,\ell),\\
x_i & \leftrightarrow & p_{\De,\ell},\\
\ba_i &\leftrightarrow& \bF^{\De_\s}_{\De,\ell},\\
\bb   &\leftrightarrow& -\bF^{\De_\s}_{0,0}.
\end{array}
\eea
$p_T$-maximization now resembles the linear program studied in the previous section, with objective function $\bc\.\bx=-p_{d,2}$.

The key difference between optimization over $\cC_{\De_\s}^{(N)}$ and a typical linear program is that the number of vectors $\bF_{\De,\ell}^{\De_\s}$ is continuously infinite, whereas before we only considered a finite collection of vectors $\ba_i$.  Such optimization problems are called a {\it semi-infinite programs}.  In the simplex algorithm, finding the minimum reduced cost Eq.~(\ref{eq:minreducedcost}) now requires minimizing over infinitely many vectors.

For concreteness, consider a vertex with $N$ basic vectors
\be
\bF^{\De_\s}_{\De_1,\ell_1}, \dots, \bF^{\De_\s}_{\De_N,\ell_N}.
\ee
We call the dimensions and spins $\{(\De_1,\ell_1),\dots,(\De_N,\ell_N)\}$ the ``spectrum" at this vertex.  When we turn on a nonbasic coefficient $p_{\De,\ell}$, the reduced cost is now a function of $\De$ and $\ell$,
\be
\mathrm{RC}_{\De,\ell} = c_{\De,\ell}-\bc_B^T \bA_B^{-1} \bF_{\De,\ell}^{\De_\s},
\ee
where
\bea
\bA_B &=& (\bF_{\De_1,\ell_1}^{\De_\s}\ \cdots\ \bF_{\De_N,\ell_N}^{\De_\s}),\\
\bc_B &=& (c_{\De_1,\ell_1}\ \cdots\ c_{\De_N,\ell_N})^T,
\eea
and $c_{\De,\ell}$ is the coefficient of $p_{\De,\ell}$ in the objective function.

Suppose $c_{\De,\ell}$ is nonzero only for finitely many $(\De,\ell)$.  For example, in $p_T$-maximization, we have $c_{d,2}=-1$, while $c_{\De,\ell}=0$ otherwise.  If there are any such $(\De,\ell)$ which are not already in the basis, we can simply compute the reduced cost for these special $(\De,\ell)$. This is not even needed in $p_T$-maximization, since the stress tensor will always be in the basis.\footnote{Except possibly at one intermediate step after the hot start, see section \ref{sec:MFThotstart}.} The challenge is scanning over the infinite $(\De,\ell)$ for which $c_{\De,\ell}$ vanishes.  In this case, we must minimize
\be
\label{eq:rc}
\mathrm{RC}_{\De,\ell}= -\bc_B^T \bA_B^{-1}\bF_{\De,\ell}^{\De_\s}
\ee
over all dimensions and spins $(\De,\ell)$ which can appear in the spectrum.

By the results of section~\ref{sec:partialfractionsforblocks}, $\mathrm{RC}_{\De,\ell}$ can be arbitrarily well approximated by a partial fraction expansion in $\De$,
\be
\label{eq:wemustminimizeapartialfractionthingy}
\mathrm{RC}_{\De,\ell} \approx r_*^{\De}\p{q_\ell(\De) + \sum_i \frac{s_{\ell,i}}{\De-\De_i}},
\ee
where $r_*=3-2\sqrt 2\approx 0.17$. The polynomials $q_\ell$ and residues $s_{\ell,i}$ are given by dot products of $-\bc_B^T\bA_B^{-1}$ with the vectors $\bp_\ell(\De)$ and $\ba_{\ell,i}$ appearing in Eq.~(\ref{eq:partialfractionrepresentation}).  This expression for $\mathrm{RC}_{\De,\ell}$ has several nice properties.  Firstly, since $r_*\ll 1$, it's clear that $|\mathrm{RC}_{\De,\ell}|$ falls off quickly with $\De$ so that it suffices to minimize over a finite range of dimensions $\De$.\footnote{The simplex algorithm can proceed as long as we always find a negative reduced cost at every step.  Thus we can rescale $\mathrm{RC}_{\De,\ell}$ by any positive function of $\De$, and we will still eventually find the correct optimum.  However, different rescalings cause the search to proceed in different ways.  For example, if we strip off the factor $r_*^\De$ from $\mathrm{RC}_{\De,\ell}$, the simplex algorithm will proceed along a different path, favoring larger values of $\De$ in the intermediate steps. By the exponential decoupling theorem of \cite{Pappadopulo:2012jk}, low-lying operators will dominate the constraints of crossing symmetry, and it is more practical to start the search by exploring the low dimensions first. We have found that the normalization in Eq.~(\ref{eq:wemustminimizeapartialfractionthingy}) realizes this requirement in practice, and leads to the fastest solution times. This is because this normalization is natural from the point of view of the OPE convergence estimates of \cite{Pappadopulo:2012jk}.}  Secondly, we can evaluate it efficiently on a computer for any $\De$.

Our modification of the simplex algorithm is to approximate the reduced cost by Eq.~(\ref{eq:wemustminimizeapartialfractionthingy}) and minimize over $\De$ using an efficient univariate minimization algorithm, described in the next section.  
In initial bootstrap studies (using out-of-the-box solvers), the strategy was to discretize the allowed dimensions $\De$ \cite{Rattazzi:2008pe}.  This meant minimizing the reduced cost by scanning over every dimension in this discrete set---an expensive operation.  Precision errors could also be introduced by the discretization. Our new algorithm evades these difficulties.\footnote{An alternative way to avoid discretizing $\De$ is to use semidefinite programming \cite{Poland:2011ey, Kos:2013tga}, see Appendix~\ref{app:semidefinitecomparison}.}

\subsubsection{A Strategy for Reduced Cost Minimization}
\label{sec:slavasearch}

To minimize the reduced cost, we must minimize functions of the form (\ref{eq:wemustminimizeapartialfractionthingy})
over dimensions $\De$ satisfying the unitarity bound~(\ref{eq:unitaritybounds}).  We will restrict to $\ell\leq \ell_\mathrm{max}$ for some large $\ell_\mathrm{max}$ (typically 40 or 50).\footnote{\label{note:lightcone}This truncation of spins is an approximation.  By the unitarity bound, operators with large spin have large dimension, and one can show that the contribution of large dimension operators to a four-point function is exponentially suppressed at $z=\bar z = 1/2$ \cite{Pappadopulo:2012jk}.  However, if we investigate the constraints of crossing symmetry near the lightcone ($z\to 0$ with $\bar z$ fixed), then it is known that large spin operators play an important role \cite{Fitzpatrick:2012yx,Komargodski:2012ek}.  It will be interesting to explore this regime in future bootstrap studies.}  Further, since $\mathrm{RC}_{\De,\ell}$ falls of exponentially quickly with $\De$, it suffices to minimize over $\De$ in some finite interval $[\De_\mathrm{unitarity},\De_\mathrm{max}]$.  We typically take $\De_{\mathrm{max}}=50$ or the sliding cutoff $\De_{\mathrm{max}}=50+\ell$. We checked that our results are insensitive to varying these cutoffs.

Thus, we have the problem of minimizing a smooth univariate function $f(\De)$ over an interval $\De\in[a,b]$.  There are many possible strategies.  Here, we present one that works well in practice, though we sacrifice some rigor for speed.  This is because the simplex algorithm can proceed as long as we find {\it some} negative minimum at each step.  In other words, it's not absolutely necessary to find the true minimum reduced cost every time.  However, we benefit if our algorithm manages to find the true minimum most of the time.\footnote{When the simplex algorithm terminates, we do want to be sure of the true minimum.  In this case, we can either apply a more rigorous minimization strategy or simply decrease the tolerance parameter $\e$ in Eq.~(\ref{eq:goodinterval}).}

Our strategy proceeds in four steps:
\begin{enumerate}
\item Recursively divide $[a,b]$ into smaller intervals where the shape of $f'$ is well understood.

\item Find which of these small intervals could contain a local minimum of $f$.

\item Compute the local minima using Newton's method.

\item Pick the overall minimum among the local minima and the endpoints $a,b$.
\end{enumerate}

Let us describe these steps in detail.  For an interval $[x,y]$, define a quadratic fit for $f'$ around the midpoint $z=\frac{x+y}{2}$,
\be
f'_\mathrm{fit}(t)\equiv f'(z)+f''(z)(t-z)+\frac 1 2  f'''(z) (t-z)^2.\\
\ee
Call the interval $[x,y]$ ``good" if this fit agrees closely with the true values at the endpoints,
\bea
\label{eq:goodinterval}
\frac{|f'(x)-f'_\mathrm{fit}(x)|}{|f'(x)|+|f'_\mathrm{fit}(x)|} \leq \e
\quad\textrm{and}\quad
\frac{|f'(y)-f'_\mathrm{fit}(y)|}{|f'(y)|+|f'_\mathrm{fit}(y)|} \leq\e,
\eea
where $\e$ is a small parameter (typically $0.05$).

In step 1, we check whether the interval $[a,b]$ is good.  If not, we split it into two intervals $[a,\frac{a+b}{2}]$ and $[\frac{a+b}{2},b]$ and recursively check and split each interval.  We stop when we've completely partitioned $[a,b]$ into good intervals.\footnote{Each time we split an interval, we can re-use our calculation of $f'$ at the endpoints and midpoint in the next level of recursion.}

In step 2, the good intervals which can contain a local minimum are those where $f'$ is negative on the left endpoint and positive on the right endpoint.  This criterion is sufficient but not necessary for the existence of a zero of $f'$ (further, it's possible that a single interval could contain multiple zeros).  This is where we sacrifice rigor for speed.  Most computation time is spent evaluating $f$ and its derivatives.  Thus, it is fruitful to reduce the number of evaluations even if that requires making assumptions about the shape of $f'$.

Once we have the zeros of $f'$ isolated into intervals, we can use Newton's method to compute them with high precision (step 3).\footnote{In fact, we use a hybrid of Newton's method and binary search which is guaranteed to stay within the interval and find a zero.}  This is the least computationally intensive part of the algorithm, and it is easy to compute within an error of $10^{-30}$ or smaller.

\subsubsection{Hot Start from the Mean Field Theory Spectrum}
\label{sec:MFThotstart}

In section \ref{sec:phaseone} we described a general hot start strategy for the initial vertex search. This strategy works particularly well in our problem of $p_T$-maximization, because a good initial guess for a spectrum can be obtained from Mean Field Theory (MFT). Recall that in MFT the $\sigma\times\sigma$ OPE contains operators
\beq
\s(\del\del\ldots\del)_{\text{spin }\ell}(\del^2)^n\s
\eeq 
of even spin $\ell$ and dimension $\Delta=2\Delta_\sigma+\ell+2n$.
A trial spectrum for our problem can be obtained by truncating this infinite spectrum to $N$ operators. We found that the best truncation strategy is to keep the $N$ lowest dimension operators (preferring low twist operators whenever the dimension is equal). This strategy gives the smallest number of auxiliary variables and the fastest solution time. 

It may happen that the basis obtained by hotstarting from the MFT spectrum does not contain the stress tensor (since the MFT spectrum does not have it). Then one intermediate step is required after hot start and before the main stage of the algorithm, to bring the stress tensor into the basis.

Without hot start, the initial vertex search takes about the same time as the subsequent $p_T$-maximization. Hot start from the MFT spectrum speeds up the initial vertex search by an order of magnitude or more, thus reducing the total solution time by about a factor 2.

\subsubsection{Linear Functionals}
\label{sec:functionals}

In this section we clarify the relationship between the primal simplex method described above and the linear functionals discussed in previous bootstrap studies.  At each stage in the simplex algorithm, the reduced cost can be written
\be
\mathrm{RC}_{\De,\ell}=-\bc_B^T\bA_B^{-1}\bF_{\De,\ell}^{\De_\s}\equiv\Lambda[F_{\De,\ell}^{\De_\s}],
\ee
where $\Lambda$ is a linear functional.  In the case of $p_T$-maximization, all components of $\bc_B$ vanish except for $c_{d,2}=-1$.  Further, the columns of $\bA_B$ are precisely the basic vectors $\bF_{\De_i,\ell_i}^{\De_\s}$.  Acting with $\Lambda$ on these basic vectors, we find
\be
\Lambda[F_{\De_i,\ell_i}^{\De_\s}]=-\bc_B^T\bA_B^{-1}\bF_{\De_i,\ell_i}^{\De_\s}=-c_{\De_i,\ell_i}=
\left\{
\begin{array}{cl}
1 & \textrm{if }(\De_i,\ell_i)=(d,2),\\
0 & \textrm{otherwise}
\end{array}
\right.
\ee
In other words, $\Lambda[F_{\De,\ell}^{\De_\s}]$ has a zero when $(\De,\ell)$ is in the spectrum (excluding the stress-tensor).

\begin{figure}[htbp]
\begin{center}
\includegraphics[scale=1]{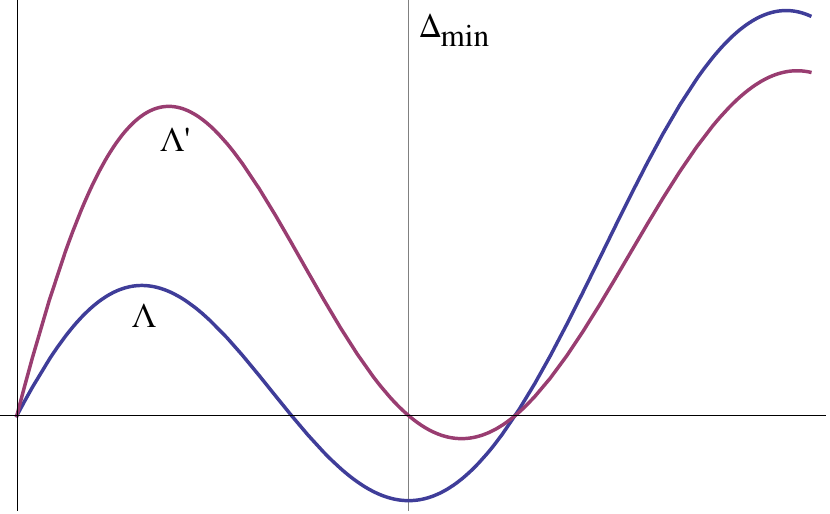}
\caption{The change in the functional $\Lambda\to\Lambda'$ after a single step of the simplex algorithm.  We first minimize $\Lambda[F_{\De,\ell}^{\De_\s}]$ to find $\De_\mathrm{min},\ell_\mathrm{min}$.  Swapping the corresponding vector into the basis results in a new functional with a zero at $\De_\mathrm{min},\ell_\mathrm{min}$.  After many steps, the minima move up to zero and the functional becomes nonnegative.}
\label{fig:functionalexample}
\end{center}
\end{figure}

If $\Lambda[F_{\De,\ell}^{\De_\s}]$ has a negative minimum, then the simplex algorithm instructs us to swap the corresponding vector into the spectrum, raising the functional up to zero there (Figure~\ref{fig:functionalexample}).  By repeating this process, we ``push up" all the minima of $\Lambda[F_{\De,\ell}^{\De_\s}]$, until we obtain a nonnegative functional $\Lambda_*$ and the algorithm terminates.\footnote{The algorithm actually does not terminate after a finite number of steps but instead converges quickly to a solution.  One can terminate the algorithm by hand when the minimum reduced cost is sufficiently close to zero, $\mathrm{min}(\mathrm{RC}_{\De,\ell})\geq -\de$ for small $\de$.  Some of our plots (Figures~\ref{fig-3dkinklarge} (right), \ref{fig-ctmin-old} (right), \ref{fig-difference}, \ref{fig-nmax13-spin0}, and \ref{fig-nmax13-spin2}) were produced with $\de=10^{-60}$.  We have found that a much less conservative criterion $\de\lesssim 10^{-12}$ suffices to reach the optimal spectrum with reasonable precision.  Another criterion is to terminate if $p_T$ is reduced by less than $\delta$ in the last $M$ iterations (we used $M=1000$, $\delta=10^{-15}$). The point is that the minimum reduced cost is not a perfect predictor for the actual reduction in $p_T$, and it may also fluctuate significantly from one step to the other. Thus looking at the change in $p_T$ integrated over many iterations may give a better idea about the progress of the algorithm. } 

The zeros of $\Lambda_*$ are precisely the dimensions and spins present in the optimal spectrum.  This establishes the equivalence of the primal method and the dual (or ``extremal functional") method discussed in \cite{Poland:2010wg, ElShowk:2012hu}.   In each step of the primal method, we have a solution to crossing symmetry and a functional of indefinite sign.  In each step of the dual method, we have a nonnegative functional but no solution to crossing.  Both methods terminate with a solution to crossing and a nonnegative functional.

Nonnegativity, together with continuity in $\De$, implies that $\Lambda_*[F_{\De,\ell}^{\De_\s}]$ actually has a {\it double zero} at generic $(\De,\ell)$ in the optimal spectrum.  ($\Lambda_*$ can have a single zero if it occurs at the end of an interval of allowed $\Delta$, for instance at the unitarity bound.)  At intermediate stages in the simplex algorithm, this manifests as pairs of dimensions that approach each other as we converge to the optimal solution.\footnote{\label{note:MPFR}These nearby pairs of dimensions are responsible for making the matrix $\mathbf{A}_B$ nearly-degenerate. In technical language, its condition number decreases as the algorithm progresses. This is a well-known potential numerical instability of the semi-infinite programming problems, see \cite{reemtsen_numerical_1998}, Figure 3. It is one of the reasons why we have to work with the multiple precision arithmetic---to avoid large rounding errors when inverting the poorly-conditioned matrix $\mathbf{A}_B$. Another reason for using multiple precision is not related to the ``spectrum doubling"---it comes from the fact that for large $N$ there is a huge disparity in size between low and high-order derivative components in the vectors $\bF_{\De,\ell}^{\De_\s}$ composing the matrix $\mathbf{A}$. Computing the minimum reduced cost, Eq.~\reef{eq:rc}, is an arithmetic operation which mixes these components. Since the minimum reduced cost may become tiny at the final steps of the algorithm, we have to perform its computation at a sufficient number of digits to be able to determine it accurately.} The functional dips negative between each pair and is forced up to zero as they squeeze together.  Thus, although our basis at each stage contains $N$ vectors, we end up with between $N/2$ and $N$ operators in the optimal spectrum.

At this point, let us address the claim in section~\ref{sec:spaceofunitarycrossingsymmetric4ptfunctions} that $\cC_{\De_\s}$ is infinite dimensional.  For each $N$ we explored, the space of possible spectra is infinite dimensional.  Even if we consider only extremal spectra, we have an $N$-dimensional space of possible objective functions to optimize, and for each objective function we generically observe a different spectrum of dimensions $\De_i$.  Together, these optimal four-point functions span an infinite dimensional space.  It would be extremely surprising if this phenomenon ceased to be true at sufficiently large $N$.

\subsubsection{$\De_\e$-Maximization}
\label{sec:deltaemaximization}

To optimize a nonlinear objective function like $\De_\e$, we must combine the simplex algorithm with an additional search.  For example, in $\De_\e$-maximization we start by assuming all scalars have dimension $\De\geq \De_\mathrm{min}$ for some $\De_\mathrm{min}$.  Subject to this assumption, we perform a ``phase 1" optimization (described in section~\ref{sec:phaseone}) to determine if a feasible spectrum exists.  If one exists, we raise $\De_\mathrm{min}$ and repeat.  If not, we lower $\De_\mathrm{min}$ and repeat.

In general, the most efficient way to find the maximal value of $\De_\mathrm{min}$ where a feasible spectrum exists is to perform a binary search.  To determine $\De_\e$ within accuracy $\de$ then requires $O(\log_2(1/\de))$ runs of the simplex algorithm.  This contrasts with $p_T$-maximization, which requires only two runs (phase 1 to determine an initial vertex, and a second run to perform the optimization).  In addition, individual $\Delta_\eps$-maximization runs start taking more and more steps (hence longer and longer time) to find a feasible solution as $\De_\mathrm{min}$ approaches the boundary of the feasible region. On the other hand, $p_T$-maximization always works with a feasible spectrum and is free from such a slowdown. For these reasons,  $p_T$-maximization is preferable.

\subsection{Implementation and Running Time Details}

A typical sequence of steps in a $p_T$-maximization bootstrap computation is as follows. One first specifies $N$ and runs code which computes and stores conformal block derivative expansion \reef{eq:partialfractionrepresentation}
(a table of $a_{\ell,i}^{m,n}$ and of $p_{\ell,i}^{m,n}$ polynomial coefficients) using one of the two methods from section \ref{sec:partialfractionsforblocks}. This is not computationally intensive. For example, our Mathematica code takes 12 minutes on an 8 core iMac to produce and store the 90MB table for $N=231$ at the 64-digit precision using the method from section \ref{sec:casimirequationmethod}, expanding up to $n=120$ in Eq.~\reef{eq:diagonal}.

One then picks a $\Delta_\s$ and runs separate code which loads the above table, produces from it a similar $\Delta_\s$-specific table for the derivatives of the functions $\bF_{\De,\ell}^{\De_\s}$, and carries out the simplex algorithm described in the previous section. This code outputs the maximal attained value of $p_T$ for the given $\Delta_\s$, and the corresponding solution to crossing. These computations \textit{are} intensive, and would be too slow to perform in Mathematica. We have two independently developed versions of this code, one in Python\footnote{We use Cython \cite{Cython} to link the MPFR multiple precision arithmetic library \cite{MPFR}, and to compile more computationally intensive parts of our Python code.} and another in C++. 

We will give performance details for the Python code (C++ code performance is similar). Hotstarting the $p_T$-maximization from the MFT spectrum makes phase 1 practically negligible, with a running time of a few minutes at $N=231$. Phase 2 takes about 55 hours at $N=231$, about 24 hours at $N=190$, and about 6 hours at $N=153$. These times are for an average single-core process on our clusters (composed of standard $\sim$3GHz machines with a $\sim4$GB memory limit per core; our computations require $<1$GB), and for $\Delta_\sigma$ in the transition region near the 3d Ising point. This region near the 3d Ising point appears to be the most time-consuming, while away from it the $p_T$-maximization concludes even faster. Also in 2d, the $p_T$-maximization is about factor two faster for the same $N$. 

In a given computation, we may spawn $O(100)$ single-core Python processes for a number of $\Delta_\s$ values in an interval of interest. We did not keep careful track of the total CPU time used to produce the results of this paper, but we estimate it as {2 - 3} single-core CPU years. In comparison, the less precise Monte Carlo computations of \cite{Hasenbusch2010} took about 30 CPU years, and would need about 1000 CPU years to get the leading critical exponents at the accuracy that we achieved here.


\section{Summary and Discussion}

In this work we performed a precision study of the 3d conformal bootstrap in the vicinity of the solution corresponding to the 3d Ising model at criticality. We also performed a detailed comparison with the analogous results in 2d. The goals were to both perform a high precision determination of operator dimensions as well as to gain insights into why the 3d Ising solution is special in the space of unitary solutions to crossing symmetry. We believe that we have succeeded on both of these fronts. 

First, using the bootstrap combined with the conjecture that the 3d Ising CFT minimizes the central charge we have determined the leading operator dimensions $\{\De_\s, \De_\e\}$ in the 3d Ising CFT at a level that is {2 - 3} times more precise than the previous record determinations, by Monte Carlo techniques. Moreover, we have made new precise determinations of the central charge $c$, the OPE coefficient $f^2_{\s\s\e}$, the second $\bZ_2$-even scalar dimension $\De_{\e'}$ and its OPE coefficient $f^2_{\s\s\e'}$, as well as the second $\bZ_2$-even spin 2 dimension $\De_{T'}$ and OPE coefficient $f^2_{\s\s T'}$. Using our techniques we additionally obtain reasonable estimates for all low dimension ($\Delta \lesssim 13$) scalar and spin 2 operators in the $\bZ_2$-even spectrum. Moreover, we showed that the same procedure in 2d can accurately reproduce the known spectrum of the 2d Ising CFT to a high degree of precision. It is now a challenge for other methods to test our very precise new 3d predictions. The new approach to lattice radial quantization being developed in~\cite{Brower:2012vg} seems to be promising in this respect. The entanglement renormalization methods \cite{2009arXiv0912.1651V} are also known to compute very well 2d Ising operator dimensions and OPE coefficients; would they be competitive for 3d? 

Second, we observed that by following the boundary of the region allowed by crossing symmetry, there is a dramatic rearrangement of the operator spectrum as one crosses the ``kink" associated with the Ising model. This rearrangement shows compelling numerical evidence that the solution corresponding to the 3d Ising model contains fewer operators than a generic solution to crossing symmetry, i.e.~that certain operators decouple as one approaches the Ising point. Such behavior is similar to what happens in 2d, where the decoupling of operators can be understood in terms of null states of the Virasoro symmetry. While there is no clear symmetry interpretation of the decoupled states in 3d, we believe that this gives an important characterization of why the 3d Ising solution is special, going beyond the conjecture that the central charge is minimized. 

Could it be that the critical 3d Ising model is, after all, exactly solvable? While at present we do not have a good idea of how to turn our numerical results into a systematic search for an exact solution, we hope that our results will stimulate theoretical thought in this direction.

An alternative point of view could be that the 3d Ising CFT is not exactly solvable, yet we found a very efficient method to solve it numerically, much better than any previously known technique. Both possibilities are interesting, and the future will show which one is true.

While this fundamental dilemma may have to await its resolution for some time, we can see several research directions which can be attacked immediately. It is clearly important to better understand the interpretation of the decoupling states. For this a precision study of the interpolating solution between 2d and 3d, going beyond our recent results~\cite{El-Showk:2013nia} would be useful. It would also be good to perform a more careful analysis of higher spin operators, where one can compare with MC determinations as well as verify ``Nachtmann's theorem"~\cite{Nachtmann:1973mr} and recent analytic results for the asymptotic behavior at large spin~\cite{Fitzpatrick:2012yx,Komargodski:2012ek} (see section \ref{sec:higher}).  

It is also important to extend our analysis beyond the single correlator $\<\s\s\s\s\>$, where e.g.~the correlator $\<\s\s\e\e\>$ would allow us to study the $\bZ_2$-odd spectrum. Work in this direction is ongoing \cite{Kos:future}. More generally, including higher dimension external operators should boost sensitivity to higher dimension exchanged operators, since the exponential decoupling sets in later in this case (see \cite{Pappadopulo:2012jk}). It would also be useful to more systematically study constraints from different regions of cross ratio space, where e.g.~it is expected that going to the lightcone limit should boost sensitivity to high spin operators with low twist (see note \ref{note:lightcone}).

One could also incorporate additional symmetries into precision studies of the bootstrap. It would be interesting to perform a similar precision analysis of the spectrum of the $O(N)$ vector models for different values of $N$, extending the bootstrap results of~\cite{Kos:2013tga}. In the XY universality class ($N=2$) there exists an $8\sigma$ discrepancy between the lattice determination of the $\nu$ critical exponent and the direct experimental measurement at the $\lambda$-point of ${}^4$He (see \cite{Vicari:2007ma} for a review). Borel-resummed perturbation theory methods are not sufficiently precise to say who's right and who's wrong. It seems likely that the bootstrap should be able to do so.

It would also be interesting to consider other discrete symmetries (e.g., $\bZ_n$, $n\ge3$) via the bootstrap. One could also do similar spectrum studies in 4d or other higher dimensions, where one could also look for evidence of transitions in the solutions living on the boundary. In fact, similar bounds in 4d $\mathcal{N}=1$ supersymmetric theories show evidence of a ``kink"~\cite{Poland:2011ey}, and it would be very interesting to study the corresponding transition in detail. More generally, similar spectrum studies can be performed in systems with various amounts of supersymmetry across different dimensions. 

It should be clear that this paper brings us one step closer to the dream of the conformal bootstrap---that one can solve theories using only basic inputs about symmetries, even in more than 2 dimensions. An extremely powerful approach is gradually emerging, which will give us precision access to strongly-coupled physics in systems where perturbation theory and $1/N$ expansions break down. Many of these systems have beautiful realizations in condensed matter and statistical physics. For some of us, this experimental connection is the best justification of this research program. For loftier thinkers, bootstrap results may provide a window into the world of strongly-coupled quantum gravity via the AdS/CFT correspondence. There is something for everyone in the bootstrap!

\section*{Acknowledgements}
We are grateful to M.~Hasenbusch, M.~Henkel, D.~Mouhanna and E.~Vicari for the useful communications concerning their work. We are grateful to B.~van Rees for the discussions of the interpolating solution. In addition, we thank N.~Arkani-Hamed, C.~Beem, A.L.~Fitzpatrick, G.~Fleming, H.~Ooguri, H.~Osborn, J.~Kaplan, E.~Katz, F.~Kos, J.~Maldacena, J.~Penedones, L.~Rastelli, N.~Seiberg, and A.~Zhiboedov for related discussions.

S.R. is grateful to the Samara Chernorechenskaya Scientific Center for their hospitality. S.R., D.S.D., and D.P. are grateful to KITP for their hospitality.  We would also like to thank the organizers and participants of the Back to the Bootstrap 3 conference at CERN.

This research was supported in part by the National Science Foundation under Grant No.~PHY11-25915.  The work of S.E. was partially supported by the French ANR contract
05-BLAN-NT09-573739, the ERC Advanced Grant no. 226371 and the ITN
programme PITN-GA-2009-237920.
M.P. is supported by DOE grant DE-FG02-11ER41742.
A.V is supported by DOE grant DE-AC02-05CH1123.
The work of D.S.D. is supported by DOE grant number DE-SC0009988.  

Computations for this paper were run on National Energy Research Scientific Computing Center supported by the Office of Science of the U.S. Department of Energy under Contract No.~DE-AC02-05CH1123; on the CERN cluster; on the Aurora and Hyperion clusters supported by the School of Natural Sciences Computing Staff at the Institute for Advanced Study; on the Omega cluster supported by the facilities and staff of the Yale University Faculty of Arts and Sciences High Performance Computing Center; on the TED cluster of the Chemistry Department and High Energy Theory group at Brown University; and the Kelvin cluster at the C.E.A. Saclay funded by the European Research Council Advanced Investigator Grant ERC–AdG–228301. S.E. would like to thank D. Kosower for providing access to the Kelvin cluster.

\appendix

\section{Comparison to Semidefinite Programming}
\label{app:semidefinitecomparison}

An alternative optimization algorithm that has proved useful in bootstrap studies is semidefinite programming \cite{Poland:2011ey,Kos:2013tga}.  Like our algorithm presented here, semidefinite programming avoids discretizing the possible operator dimensions $\De$.  The starting point is an approximation for linear functionals in terms of polynomials times positive functions,
\be
\left.\sum_{m,n}a_{mn}\ptl^m_r \ptl^n_\eta G_{\De,\ell}(r,\eta)\right|_{r=r_*,\eta=1} \approx \frac{r_*^\De}{Q_\ell(\De)}\sum_{m,n}a_{mn}P^{m,n}_\ell(\De),
\label{eq:rationalapproxofblocks}
\ee
where $P_{\ell}^{m,n}(\De)$ are polynomials and $Q_\ell(\De)=\prod_i (\De-\De_i)$.  This approximation follows from Eq.~(\ref{eq:partialfractionrepresentation}).  Positivity of the polynomial $\sum_{m,n}a_{mn}P^{m,n}_\ell(\De)$ can be encoded in terms of $r\x r$ positive semidefinite matrices, where 
\be
r=\max_{m,n}\left\lceil \frac 1 2 \deg P^{m,n}_\ell(\De)\right\rceil=\left\lceil K+\frac 1 2 q\right\rceil.
\ee
Here, the number of derivatives satisfies $m+n\leq 2K$, as in section~\ref{sec:specializationofprimalsimplex}, and $q$ is the number of poles $\De_i$ included in the rational approximation (\ref{eq:rationalapproxofblocks}).

The most popular semidefinite program solvers used in the bootstrap are {\tt SDPA} and its arbitrary precision version {\tt SDPA-GMP} \cite{SDPA}.  Their performance scales differently from the performance of our algorithm.  In practice, the running time increases slowly as the number of crossing relations $n_C$ is increased, but quickly as the degrees of the polynomial approximations $r$ ($K$ or $q$) are increased.  (By contrast, our algorithm is relatively insensitive to the number of poles $q$.) This makes {\tt SDPA} a good choice for studying theories with global symmetries, where we have different crossing relations for each tensor structure which can appear in a four-point function \cite{Rattazzi:2010yc}.  (For example, theories with $O(N)$ symmetry have $n_C=3$ for a four-point function of vectors.  Theories with $SU(N)$ symmetry have $n_C=6$ for a four-point function of anti-/fundamentals.)  The overall dimension of the space of linear functionals is
\be
d_\Lambda = n_CN = n_C\frac{K(K+1)}{2}.
\ee
In \cite{Poland:2011ey}, this was taken as high as $n_C=6$ and $K=11$, so $d_\Lambda=396$.  Each optimization for a problem of this size takes approximately 48 hours.

In this work, we have been interested in studying a single crossing relation $n_C=1$ and exploring as many derivatives as possible.  We have been able to reach $K=21$, so that $d_\Lambda=231$.  Although this value of $d_\Lambda$ is smaller than what was achieved with {\tt SDPA-GMP} in the case of global symmetries, the number of derivatives $K=21$ could be difficult to match with {\tt SDPA-GMP}.  As the number of derivatives $K$ is increased, one must also increase the number of poles $q$ to maintain numerical stability.  Further, one should tune the {\tt SDPA-GMP} parameters to ensure the solver uses good initial data and termination criteria.  It will be important to explore whether this can be done in the future.

{\tt SDPA-GMP} uses a primal-dual solution method, so in principle it could be used for precision spectrum studies similar to what we do here.  This will be interesting to explore in future work.

\end{document}